\documentclass[remotesensing,article,accept,moreauthors,pdftex]{Definitions/mdpi} 

\newcommand{\omTilde}{{\tilde{\omega}}}
\newcommand{\OmTilde}{{\tilde{\Omega}}}
\newcommand{\fTilde}{{\tilde{f}}}

\newcommand{\V}{\mathbf{v}}

\newcommand{\A}{\mathbf{A}}
\newcommand{\B}{{\mathbf{B}}}
\newcommand{\x}{{\mathbf{x}}}

\newcommand{\E}[1]{\Big\langle\, #1 \,\Big\rangle}
\newcommand{\e}[1]{\big\langle\,#1 \,\big\rangle}
\newcommand{\norm}[1]{| #1 |}
\newcommand{\sinc}[1]{\text{sinc}#1}
\newcommand{\NOm}{{|\OmTilde|}}
\usepackage{scalerel}
\usepackage[final]{changes}
\firstpage{1} 
\pubvolume{1}
\issuenum{1}
\articlenumber{0}
\pubyear{2021}
\copyrightyear{2021}
\externaleditor{Academic Editor: Cornelis Weemstra, Nori Nakata and Aurélien Mordret 
} 
\datereceived{} 
\dateaccepted{} 
\datepublished{} 
\hreflink{https://doi.org/} 

\Title{Seismic Interferometry from Correlated Noise Sources}

\TitleCitation{Seismic Interferometry from Correlated Noise Sources}

\Author{{Daniella Ayala-Garcia} 
 $^{1,}$*, Andrew Curtis $^{2}$ and Michal Branicki $^{1,3}$}

\AuthorNames{Daniella Ayala-Garcia, Andrew Curtis and Michal Branicki}

\AuthorCitation{Ayala-Garcia, D.; Curtis, A.; Branicki, M.}

\address{%
$^{1}$ \quad School of Mathematics, The University of Edinburgh, Edinburgh, \added{{EH9 3FE}}
, UK; m.branicki@ed.ac.uk\\
$^{2}$ \quad School of GeoSciences, The University of Edinburgh, Edinburgh, \added{{EH9 3FD}}, UK; Andrew.Curtis@ed.ac.uk\\
$^{3}$ \quad 
The Alan Turing Institute for Data Science, London, \added{{NW1 2DB}}, UK \\
}

\corres{Correspondence: c.d.ayala-garcia@sms.ed.ac.uk}

\abstract{ It is a well-established principle that cross-correlating seismic observations at different receiver locations can yield estimates of band-limited inter-receiver Green's functions. This principle, known as \added{Green's function retrieval or} seismic interferometry, is a powerful technique that can transform noise into signals which enable remote interrogation and imaging of the Earth's subsurface. In practice it is often necessary and even desirable to rely on noise already present in the environment. Theory that underpins many applications of ambient noise interferometry assumes that the sources of noise are uncorrelated in \deleted{space and} time. However, many real-world noise sources such as trains, highway traffic and ocean waves are inherently correlated \deleted{both} in space and time, in direct contradiction to the \added{these}\deleted{current} theoretical foundations. Applying standard interferometric techniques to recordings from correlated energy sources makes the Green's function liable to estimation errors that so far have not been fully accounted for theoretically nor in practice. We show that these errors are significant for common noise sources, always perturbing \deleted{and sometimes} \added{or entirely} obscuring the phase one wishes to retrieve. Our analysis explains why stacking may reduce the phase errors, but also shows that in commonly encountered circumstances stacking will not remediate the problem. This analytical insight allowed us to develop a novel workflow that significantly mitigates effects arising from the use of correlated noise sources. Our methodology can be used in conjunction with already existing approaches, and improves results from both correlated and uncorrelated ambient noise. Hence, we expect it to be widely applicable in \deleted{real life} ambient noise studies.}


\begin{document}

\section{Introduction}

 The critical zone and Earth's crust are constantly monitored across ecological and geological disciplines due to their importance to terrestrial life, for industrial applications and for advancing science. Remote sensing methods are commonly deployed, in which physical energy fields are recorded on or above the Earth's surface, and are used to infer structure and properties of the subsurface. In particular, seismic interferometry is a powerful technique which transforms previously discarded data, such as seismic energy from earthquakes or from the ambient background noise field, into useful signals that remotely illuminate subsurface Earth structures \citep{campillo2003long, wapenaar2006green, curtis2006seismic}.  The origin of seismic interferometry can be traced back to the seminal work of Claerbout \cite{claerbout1968synthesis} who showed that the reflection response of a horizontally layered medium could be estimated from the autocorrelation of its transmission response. Seismic, or more generally wavefield interferometry has since become a rapidly evolving field of research \citep{rickett1999acoustic, weaver2001ultrasonics,derode2003estimate,slob2007electromagnetic}, leading to fundamental advances in our ability to image the Earth's crust at global \citep{ruigrok2008global,nishida2009global}, regional \citep{shapiro2005high,nishida2008three, arroucau2010new} and industrially relevant scales \citep{bakulin2006virtual,bakulin2007virtual,halliday2010interferometric}. 
 
 In general, interferometric methods rely on cross-correlating, convolving or deconvolving  pairs of recorded signals to extract information about the medium. The goal is to estimate signals which would have been acquired if the receivers \citep{hong2006tomographic,curtis2009virtual} or sources~\citep{campillo2003long, wapenaar2004retrieving, wapenaar2006green} been deployed at different locations or at different times \citep{curtis2010source,curtis2012benefit,entwistle2015constructing,chen2020empirical}. Theoretically, the response of a medium at a given location to an impulsive source at a different location is described by Green's functions which are associated with specific equations describing the wave dynamics. The purpose of seismic interferometry is usually to estimate various approximations to these Green's functions.

\begin{figure}[H]
\includegraphics[width=0.6\textwidth]{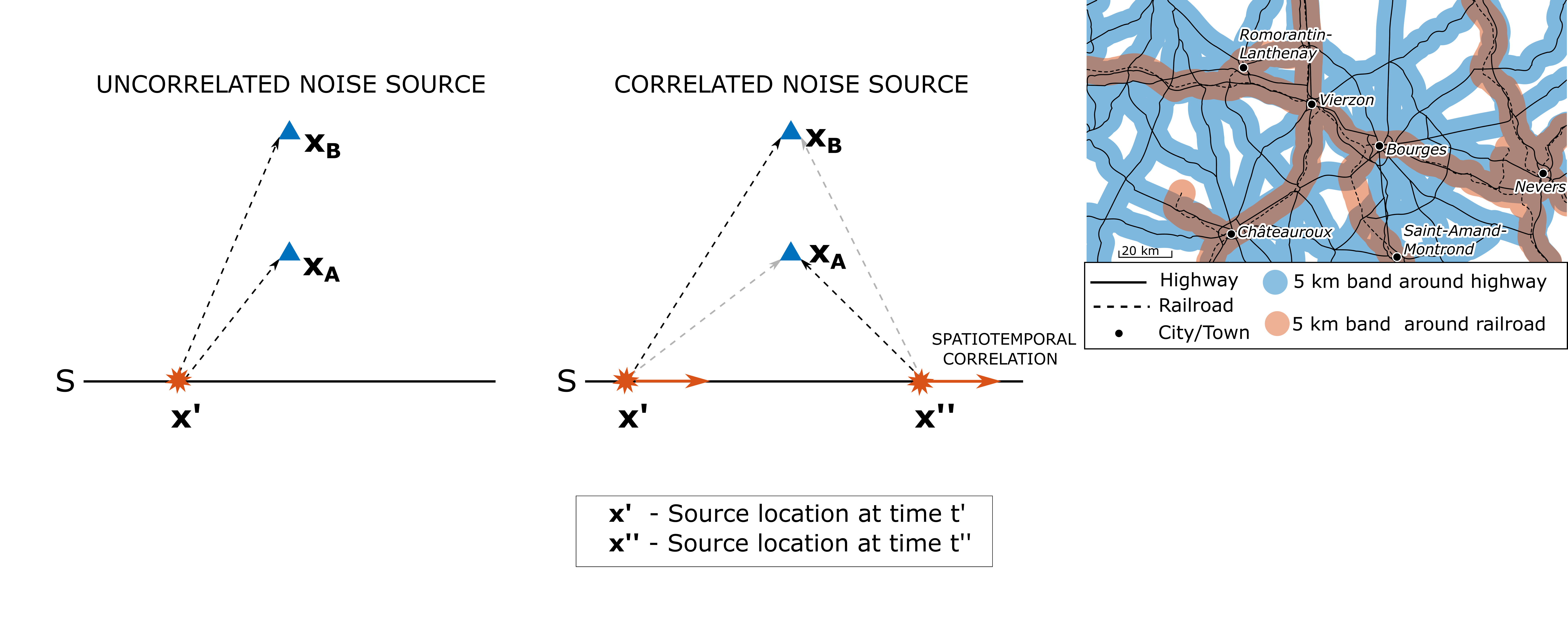}
\caption{Geometry of a boundary $S$ of uncorrelated, impulsive sources (see Equation~\eqref{DirectInt}) versus a time-ordered train of impulsive sources  (see Equation~\eqref{CorrInt}). Blue triangles denote receivers and source locations are represented by  orange stars. Orange arrows in the source boundary indicate the presence of a  spatiotemporal correlation of sources along the boundary. The top-right panel shows the main roads (solid black) and railways (black dashes) within an area of about 9000 km\textsuperscript{2} around Bourges, France.  A zone of 5 km on either side of each road (blue) and railway (orange) highlights the prevalence of near-field traffic noise. The bottom panel shows oblique waves breaking against the coast of Skeleton Bay, Namibia. These are common examples of correlated noise sources.
}\label{geometryCorrVSUncorr}
\end{figure}

Two theoretical assumptions that underpin many interferometric methods are that recorded energy comes from sources that are distributed isotropically in space around the receivers, and that the time series emitted by the sources are statistically uncorrelated \added{in time} between pairs of sources. The latter assumption precludes sources that are spatio-temporally correlated. However, it is often necessary or even desirable to rely on noise sources that are already present in the environment. This is especially true in areas of the Earth where environmental constraints preclude the use of active artificial seismic sources, or when we wish to illuminate large volumes of the Earth that require more powerful sources than can be generated artificially. Sources of freely available ambient noise abound in the Earth and interferometry has been successfully performed by cross-correlating earthquake codas (the long tail of energy that is recorded after the initial impulses from first-arriving seismic waves) which are assumed to act as an approximately diffuse, reverberating wavefield approaching receivers from all directions \citep{campillo2003long}, or using recorded wavefields assumed to come from isotropic noise fields \citep{curtis2006seismic, wapenaar2006green,wapenaar2011seismic,nicolson2012seismic}. Many known physical noise sources that contribute to ambient noise are in motion, e.g., storm sources, ocean waves or wind~\cite{ardhuin2011ocean}, \deleted{and} \added{in addition to seismic energy deriving from} man-made activities such as shipping~\cite{sabra2005arrival} or noise from traffic. Recently it has been recognized that highway and/or railway traffic can comprise a dominant component of the ambient noise field, and interferometry has been applied to highway traffic noise \citep{halliday2008seismic_a,nakata2011shear,behm2013love}, railway noise \citep{dales2020virtual, brenguier2019train, quiros2016seismic, pinzon2021humming,liu2021retrievability}, and noise generated from waves breaking along coastlines \citep{gerstoft2006green}, as illustrated in \mbox{Figure \ref{geometryCorrVSUncorr}}. 
However, all these sources of ambient noise are inherently correlated in time, which is in direct contradiction to the theoretical assumptions outlined above. \added{Although different methodologies for mitigating non-ideal source distributions have been considered, for example in~\cite{curtis2010directional,fichtner2016generalised,van2012interferometric, van2015retrieving}, the effects of statistical correlation in ambient noise sources were not considered, particularly in the important case of correlation induced by motion of the source. The impact that source motion has on cross-correlational interferometry was investigated in~\cite{sabra2010influence}~but under the standard assumption of statistically uncorrelated ambient noise.} Importantly,  the extent of the error in the interferometric estimates of the Green's function due to the assumption of uncorrelated noise sources has so far not been quantified, and no general methods to reduce these errors have been published.

In this paper, we present a unified theory of  inter-receiver seismic interferometry that encompasses correlated and uncorrelated noise sources in both the near and far fields, including the case of moving sources. Our theoretical framework allows us to develop a novel workflow that mitigates the spurious effects arising from the use of correlated noise sources, leading to the retrieval of the Green's function from a short-time recording of a single correlated moving noise source.  Moreover, our framework provides a systematic insight into the mechanisms through which the correlation in the sources induces errors in the estimates of the Green's function and its phase. We quantify these errors and show that they have a significant effect for commonly used noise sources, with spurious effects that obscure the estimates of phase (and hence of wave travel times). Our novel method is based on an appropriate randomization of the recorded traces which allows for an accurate interferometric retrieval even from a single moving energy source. Furthermore, our analysis explains why stacking multiple traces may reduce errors due to noise correlation in the interferometric estimates, but also highlights limitations of this approach, and identifies potentially commonly occurring circumstances in which it will fail.  Our unified methodology is applicable to both correlated and uncorrelated ambient noise, and is particularly useful in cases where stacking does not improve the signal to noise ratio in the~retrieval.

In what follows we begin by briefly presenting the theoretical principles of standard seismic interferometry for the case when individual recordings from static impulsive sources are available, as well as for an ambient noise scenario. Subsequently, we highlight the challenges in retrieving the interferometric inter-receiver Green's function from ambient noise. The important  case of performing interferometric retrieval from wavefields generated by correlated moving sources is considered next, and the corrupting effect due to the Doppler spread on the quality of the interferometric retrieval is discussed. We then review the standard ensemble average approach in ambient noise interferometry, which is referred to as {\it stacking}. 
Potential pitfalls of the retrieval through the stacking average are discussed in terms of the correlation structure of the source signature. Moreover, an analytical decomposition of the ambient noise retrieval is derived which, based on the characteristic correlation length of  the source signature, separates the retrieval into a term comprising predominantly coherent contributions that correctly approximate the inter-receiver phase, and the crosstalk term associated with spurious incoherent contributions. These analytical findings motivate the development of a novel and versatile technique, referred to as the \emph{random windowing} approach.  In contrast to the standard stacking average, the random windowing average does not rely on spatial localization of the correlation structure of the source signature, and it can be used for interferometric phase retrieval from short recordings of a correlated noise signal. We conclude by outlining a pseudo-algorithm for carrying out the random windowing retrieval and present an example with the resulting phases retrieved successfully from both correlated and uncorrelated ambient noise.

\section{Theoretical Background}\label{theor_sec}
It is a well-established principle that cross-correlating seismic observations recorded at different receiver locations yields seismic responses that under certain conditions, provide useful estimates of the inter-receiver Green's function of linear wave equations. This principle has been expressed in a variety of ways, often via the acoustic or elastic representation theorems, and using either the convolution, deconvolution or cross-correlation of recorded seismic traces. For our purposes, a cross-correlational Green's function representation theorem is used in line with \cite{wapenaar2006green,schuster2009seismic}, in a regime of volume-injection rate impulsive~sources. 

Consider a medium of density $\rho$ and sound speed $c$, and {let} 
 $\hat{G}(\x,\x',\omega)$ denote the frequency-domain response of the medium to an impulsive source located at $\x'$ and recorded at location~$\x$. Throughout,  $\hat{\cdot}$  denotes quantities in the frequency domain, and $\cdot^*$ denotes complex conjugation. Given a pair of receiver locations $\x_\A$ and $\x_\B$ we assume an empirical estimate $\hat{g}(\x_\A,\x_\B,\omega)$ of the inter-receiver Green's function $\hat{G}(\x_\A,\x_\B,\omega)$ can be represented as

\begin{equation}\label{DirectInt}
 \hat{g}(\x_\A,\x_\B,\omega) = \frac{2}{\rho c}\int_S \hat{G}^*(\x_\A,\x',\omega)\hat{G}(\x_\B,\x',\omega)d\x'\,,
\end{equation}
\noindent
as in Wapenaar and Fokkema~\citep{wapenaar2006green}, where each $\x'$ is an impulsive source location along the boundary $S$ (see leftmost panel in Figure \ref{geometryCorrVSUncorr}). The integrand in Equation \eqref{DirectInt} is equivalent to the convolution of the time-reversed trace recorded at $\x_\A$ with the trace recorded at $\x_\B$, and this operation corresponds to a cross-correlation in the time {domain} 
 (this statement holds exactly for recordings ${G}(\x,\x',t)$ defined for all time). Please note that $\hat g$ in Equation \eqref{DirectInt} is proportional to the spatial average of the integrand in the right-hand side of Equation~\eqref{DirectInt} over $S$; this interpretation will be helpful below. 

The main theoretical requirements for Equation \eqref{DirectInt} to provide a good estimate of the inter-receiver Green's function are: (i) that the medium is lossless, (ii) that the source boundary $S$ encloses the receivers and that energy is emitted equally from all directions, (iii) that $S$ is sufficiently far from the receivers for the recorded energy flux to be emitted approximately perpendicular to the boundary, and (iv) that individual recordings are acquired independently for each impulsive source location $\x'$ within $S$. Under these idealized conditions, the empirical estimate $\hat{g}(\x_\A,\x_\B,\omega)$ provides a good approximation to the homogeneous inter-receiver Green's function \citep{wapenaar2006green}, i.e., $\hat{g}(\x_\A,\x_\B,\omega) = \hat{G}(\x_\A,\x_\B,\omega)$\deleted{$\pm$}\added{$-$}$\hat{G}^*(\x_\A,\x_\B,\omega)$  or, equivalently $g(\x_\A,\x_\B,t) = G(\x_\A,\x_\B,t)$\deleted{$\pm$}\added{$-$}$G(\x_\A,\x_\B,-t)$ in the time domain. \mbox{Wapenaar et al.~\cite{wapenaar2011seismic}} discuss spurious effects that are introduced when the above theoretical assumptions are violated. However, in practice  artefacts due to the violation of these theoretical assumptions are usually assumed to be sufficiently weak to allow for retrieval of useful empirical estimates of the Green's function. In particular, seismic interferometry has been shown empirically to be applicable in regimes where the medium illumination is one-sided~\mbox{\citep{halliday2008seismic_b, wapenaar2011seismic, mehta2007improving, snieder2004extracting}}. Hence, while the boundary $S$ in Equation~\eqref{DirectInt} is typically assumed to be a closed contour which is sufficiently distant from the receivers, we assume $S$ to be a straight source-line as depicted in the left insets of Figure \ref{geometryCorrVSUncorr}. It is worth noting that while the amplitudes of empirical Green's function estimates are acknowledged to be unreliable, the phases of the Green's function recovered through such a procedure are assumed to be well approximated (see for example, \cite{wapenaar2006green,thorbecke2008analysis}, and remarks under Equation \eqref{WapenaarRetrieval} below).

The main practical obstacle in the context of ambient noise interferometry stems from the fact  that the medium's response to impulsive sources is not readily available. Hence,  the information required for constructing the integrand in  Equation~\eqref{DirectInt} cannot be obtained directly. Instead, Wapenaar and Fokkema~\cite{wapenaar2006green} propose the ambient noise interferometric relationship \deleted{given by}\added{defined as}
\deleted{$\hat{h}(\x_\A,\x_\B,\omega) = \frac{2}{\rho c}\,\hat{p}^*(\x_\A,\omega)\,\hat{p}(\x_\B,\omega)\,,$}

\begin{equation}\label{AmbientInt}
\hat{h}(\x_\A,\x_\B,\omega) := \frac{2}{\rho c}\,\hat{p}^*(\x_\A,\omega)\,\hat{p}(\x_\B,\omega)\,,
\end{equation}
where $\hat{p}(\x,\omega)$ denotes ambient noise recorded at a receiver location $\x$ for frequency $\omega$. It is useful (and theoretically appropriate) to consider each recording $\hat{p}(\x,\omega)$  as a random field  which is obtained via the Fourier transform of a realization of a stochastic process (a random mechanism) generating {${p}(\x,t)$} (these statements hold for $t,\omega\in \mathbb{R}$, we do not delve into the accuracy of finite-time approximations of the Fourier transform).
In such a framework  source locations $\x'\in S$   are no longer required to be known explicitly. Instead, the source characteristics  are  implicitly present in the ambient noise representation in Equation~(\ref{AmbientInt}) and they are  accounted for, in principle, in the ambient noise recordings  represented via 

\begin{equation}\label{p}
\hat{p}(\x,\omega) = \int_S \hat{F}(\x',\omega)\hat{G}(\x,\x',\omega)d\x'\,,
\end{equation}
\noindent
where $\hat{F}(\x,\omega)$ characterizes the unknown {\it source signature}. Wapenaar and Fokkema~\cite{wapenaar2006green} showed that provided that $\hat{p}(\x_\A,\omega)$ and $\hat{p}(\x_\B,\omega)$ are obtained from  recordings of uncorrelated noise sources, taking the empirical average over a large number ($N\gg 1$) of realizations of $\hat{h}(\x_\A,\x_\B,\omega)$ in~Equation~\eqref{AmbientInt}, referred to hereafter as {\it stacking}, leads to a good {approximation} 
(such an approximation is increasingly accurate for $N, t\rightarrow \infty$ provided that ${h}(\x_\A,\x_\B,t)$ is ergodic which seems to be a reasonable assumption in practice). of the phase of the inter-receiver Green's function $\hat{G}(\x_\A,\x_\B,\omega)$.  However, as discussed in the subsequent sections, the stacking average is likely to produce unsatisfactory results in the presence of correlated noise sources  even when long-time recordings are available. 

In this paper, we delve deeper into the relationship between  the retrievals $\hat{g}(\x_\A,\x_\B,\omega)$ and $\hat{h}(\x_\A,\x_\B,\omega)$ in order to identify conditions under which the ambient noise retrieval based on Equation~\eqref{AmbientInt}  provides a satisfactory approximation of the retrieval Equation~(\ref{DirectInt}), and to develop techniques that allow us to achieve this for both uncorrelated and correlated noise sources. 

To understand the need for the (statistical) averaging over multiple realizations of the ambient noise recordings in the retrieval (\ref{AmbientInt}) for some fixed boundary $S$, substitute the integral expression \eqref{p} for the noise recordings at receivers $\x_\A$ and $\x_\B$ into Equation~\eqref{AmbientInt} to obtain

\begin{equation}\label{CorrInt}
\hat{h}(\x_\A,\x_\B,\omega) =  \frac{2}{\rho c}\int_S \int_S \hat{F}^*(\x',\omega)\hat{F}(\x'',\omega)\,\hat{G}^*(\x_\A,\x',\omega)\hat{G}(\x_\B,\x'',\omega)d\x''\,d\x'\,.
\end{equation}

Please note that the above formula represents a product of two independent integrals which are re-written in a form suitable for the subsequent analysis; to simplify notation, we skip the explicit dependence of $\hat F$ in Equation~(\ref{CorrInt})  on the particular recording of ambient noise, but this remains implicit. The geometry corresponding to the retrieval based on, respectively, Equations~(\ref{DirectInt}) and (\ref{CorrInt}) is sketched on the left side of Figure \ref{geometryCorrVSUncorr}. In contrast to  $ \hat{g}(\x_\A,\x_\B,\omega)$ in Equation~(\ref{DirectInt}), where averaging over the source locations $\x'\in S$ leads to the cancellation of all phases except for the inter-receiver phase, the ambient noise retrieval $\hat{h}(\x_\A,\x_\B,\omega)$ in~Equation~\eqref{CorrInt}  does not represent a spatial average over $S$. Consequently, the retrieval in Equation~(\ref{CorrInt})  is dominated by acausal and spurious information contained in the integrand for $\x'\ne\x''$ (see middle panel of Figure \ref{geometryCorrVSUncorr}).  

It turns out that a reliable phase retrieval of the inter-receiver Green's function $\hat{G}(\x_\A,\x_\B,\omega)$ from Equation~\eqref{CorrInt} relies on an appropriate statistical average which couples the source characteristics  $\hat{F}(\x',\omega)$ and $\hat{F}(\x'',\omega)$ for all $\x',\x''\in S$ to generate an integral kernel which is localized in the neighborhood of $\x'=\x''$. As discussed in subsequent sections, the suitable  statistical average depends on   the nature of the noise. In particular, for correlated noise sources the (stacking) average over multiple realizations of \mbox{Equation~(\ref{AmbientInt})} is largely unsatisfactory regardless of the amount of available data, and a more versatile averaging approach must be used. \added{Please note that while a constant medium density and propagation speed have been assumed, the analysis presented still holds when these two quantities are space-dependent. } 
  
In light of the above discussion, it is clear that the non-averaged relationship in Equation \eqref{CorrInt} is dominated by contributions from incongruous source locations, and these may lead to spurious contributions that completely obscure the phase of the inter-receiver Green's function. Consider, for example, the interferometric retrieval from a simulated sequence of sources which generate seismic impulses in a time-ordered fashion; hereafter, we refer to such a sequence as a \emph{train of sources}. The train of sources is not to be confused with a  {\it set of impulsive sources} recorded one-by-one and used to calculate the retrieval Equation~(\ref{DirectInt}); these two cases will be compared in what follows. In both cases we assume that the signal is recorded  by a pair of receivers deployed in the vicinity of source-line $S$. To simulate the wavefield generated by trains of sources, we derived an analytical two-dimensional model which is described in Appendix \ref{pDetails}. Figure \ref{CorrVSUncorrFig} illustrates the situation where the interferometric trace retrieved via Equation~\eqref{CorrInt} from a train of sources differs significantly from the  retrieval obtained from individually recorded impulsive sources, according  to  Equation~\eqref{DirectInt}. Both retrievals are based on sources on the same boundary $S$ and recorded over the same time interval, except that the impulses in the train of sources are generated in a time-ordered sequence of physical locations which span $S$.

\begin{figure}[H]
\includegraphics[width=0.7\textwidth]{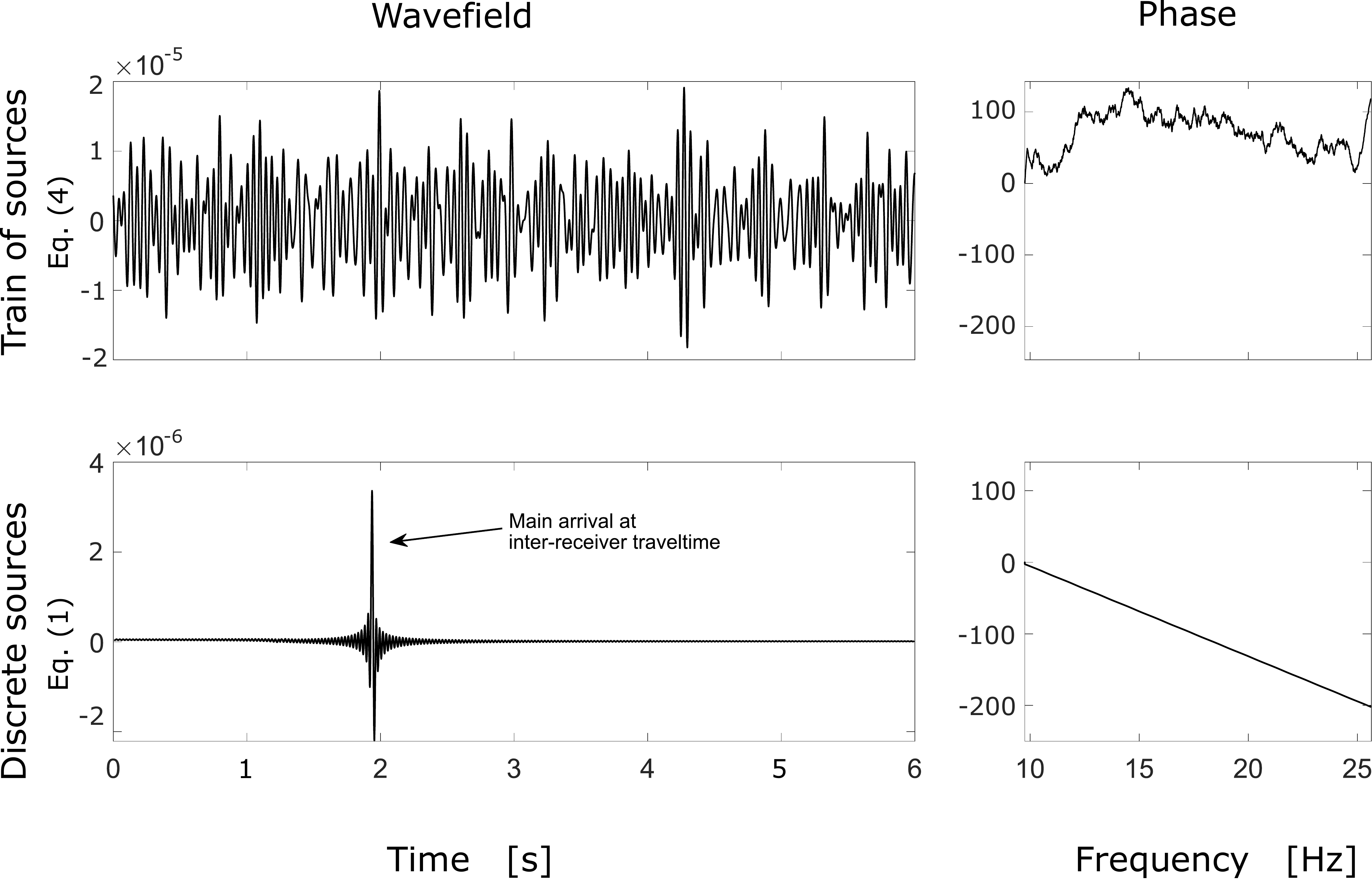}
\caption{{Comparison of} 
 the interferometric estimate of the Green's function from a simulated train of sources as per Equation \eqref{CorrInt} (\textbf{top}), versus the estimate from a boundary of static point sources (\textbf{bottom}) recorded individually, according to Equation \eqref{DirectInt}. The left side panels show the waveforms in time, while the panels to the right show the corresponding estimated phases (unwrapped). The source train moves along the boundary line at 25 m/s (90 km/h) and emits sound in a frequency range between 10 and 25 Hertz. The receiver pair is oriented perpendicularly to the source boundary and 2000 m away from each other. Sound speed in the medium was set at 1000 m/s. The receiver closest to the source boundary was 400 m away.}\label{CorrVSUncorrFig}
\end{figure}

When the sources are individually recorded (bottom of Figure \ref{CorrVSUncorrFig}) the inter-receiver Green's function is retrieved accurately with the main arrival found around 2 s, which  corresponds to the inter-receiver travel time. In the top panels, the inter-receiver arrival is completely obscured by spurious arrivals, and the phase does not reproduce the reference estimate in the bottom panels.

In addition to the above issues, if the noise source is in motion, one must  take into account both the spatiotemporal correlations of the noise sources, as well as  the Doppler effect which shifts the energy emitted by the source at each frequency to a range of other recorded frequencies, even if the source is monochromatic. The instantaneous frequency $\breve{\omega}(\x,t')$ of the train of sources recorded at location $\x$ and time $t'$ can be derived analytically from Equation \eqref{p} when it is used to represent a continuously moving source, resulting in the following expression (see the derivation of Equation~\eqref{dopplerApp} in Appendix~\ref{pDetails})

\begin{equation}\label{doppler}
\breve{\omega}(\x,t') = \frac{\omTilde}{1-\frac{|\V |}{c}\cos\theta_\x(t')}\,,
\end{equation}
\noindent
where $\omTilde$ is the frequency emitted by the source, and $\theta_\x(t')$ is the angle between the line from the instantaneous source location at time $t'$ to the receiver location $\x$, and the source-line $S$. Equation \eqref{doppler} agrees with the well-known expressions for the Doppler shift for the instantaneous frequency when the source is moving directly towards, or away from, the receiver. However, it is more general in that it quantifies the Doppler spread in the recorded frequency in terms of the relative source-receiver position angle $\theta_\x(t')$, when the source has an azimuthal velocity component. It is clear from Equation \eqref{doppler} that for a pair of distinct receivers $\x_\A$ and $\x_\B$, the corresponding relative position angles of the source $\theta_{\x_\A}(t')$ and $\theta_{\x_\B}(t')$ will be different; this means that the same source emitting the single frequency $\tilde\omega$ will be recorded with a different frequency at each location.

\begin{figure}[H]
\includegraphics[width=0.7\textwidth]{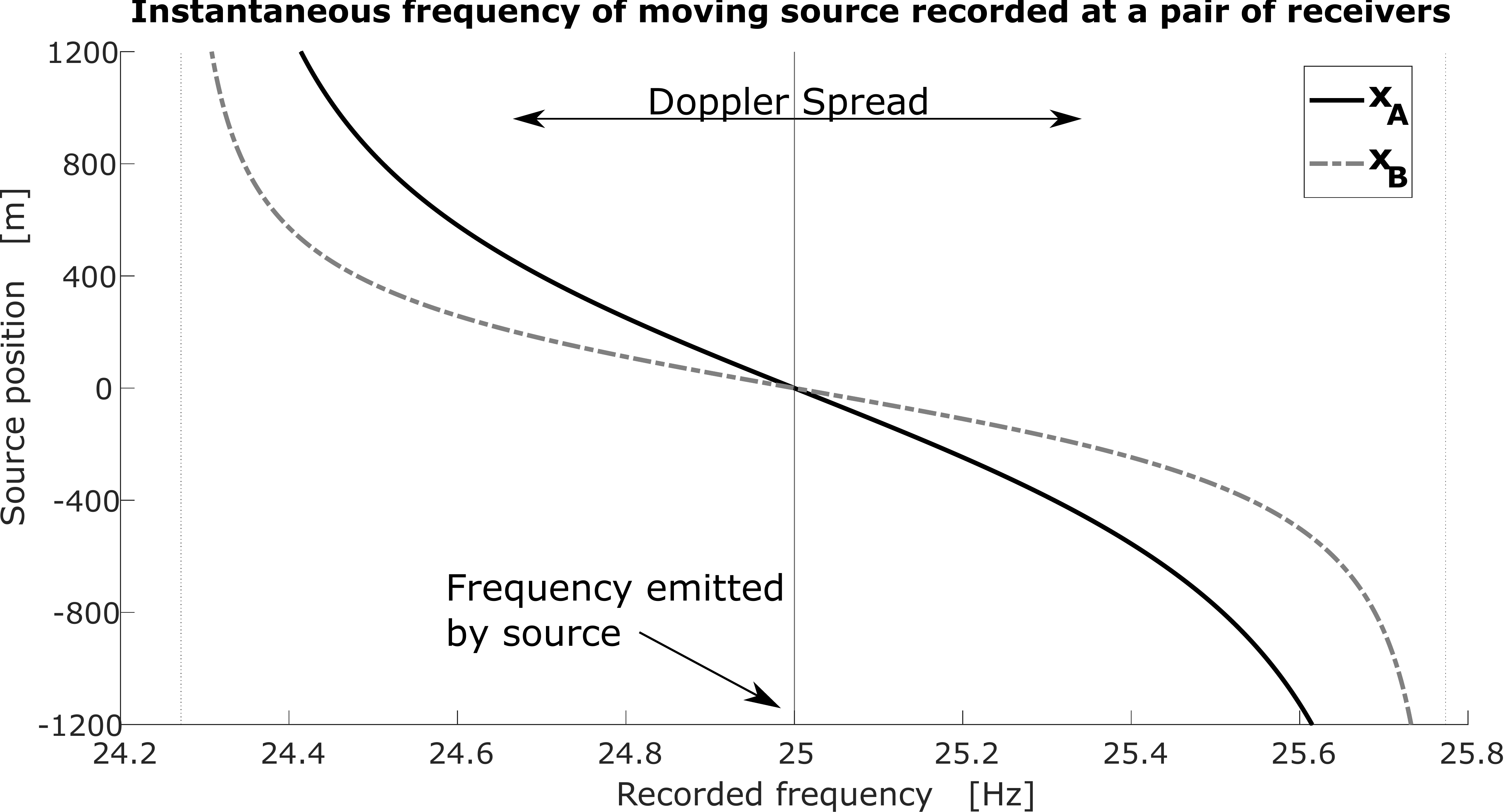}
\caption{{Relationship between} 
 source location and instantaneous frequency recorded at two different receiver locations $\x_\A$, and $\x_\B$. The simulated source was travelling at 30 m/s (108 km/h) and the medium speed was set at $1000$\,m/s. For illustration purposes the source was set to be monochromatic, emitting noise at a fixed frequency of 25 Hz. The limiting values for the Doppler spread, given by $25/(1\pm \norm{\V}/c)$ Hz, are marked with vertical dotted lines.}\label{DopplerFig}
\end{figure}

 Figure \ref{DopplerFig} illustrates the relationship between the instantaneous location of a moving noise source and the recorded frequency $\breve{\omega}$ at each receiver $\x_\A$ and $\x_\B$. As expected, the recorded frequency is higher when the source is moving towards the receivers, and lower as it moves away. Clearly, as can be deduced from Equation~(\ref{doppler}) and observed in Figure~\ref{DopplerFig}, the only (instantaneous) location of the source that  will lead to recording the true emitted frequency occurs when the source is collinear with the pair of receivers, and the line through $\x_\A$ and $\x_\B$ is perpendicular to the boundary $S$; depending on the geometry of the problem, such a location need not exist. Thus, in general, the frequency emitted by a moving noise source and recorded at the receivers will differ from the true frequency $\tilde \omega$, and the frequency bias will be different at $\x_\A$ and $\x_\B$ at all instantaneous locations except when the source is collinear with the receiver pair. 
 
This has important implications for seismic interferometry, since recordings at any particular frequency will usually have been emitted at different source locations. For example, in Figure \ref{DopplerFig} the spectrum value recorded at 24.5 Hz would have been emitted by a source that was 800 m along the boundary for receiver $\x_\A$, but only 400 m for receiver $\x_\B$. Given that  interferometry is performed frequency by frequency, this inconsistency in the recorded frequency is conceptually equivalent to introducing further spurious contributions from spatially incongruous source locations. Moreover, 
 in the case of a more realistic broadband source the spread and consequent interaction among different frequencies becomes compounded, with multiple source locations along the boundary $S$ introducing  spatially incongruous contributions for every recorded frequency. Hence, the Doppler shift associated with moving noise sources further compounds the spurious effects inherent to ambient noise interferometry if we use the retrieval (\ref{AmbientInt}). 
 
Equation \eqref{doppler} shows that the Doppler shift is amplified for higher frequencies, as well as increasing source train speeds. In addition, larger inter-receiver distances, as well as near-field recordings,  will increase the discrepancies  in frequencies recorded $\x_\A$ and $\x_\B$, resulting in further deterioration of the interferometric retrieval. \added{Some of the potential obstacles for retrieving inter-receiver Green's functions from sources in motion have been considered theoretically by Sabra in \cite{sabra2010influence}, where the inconsistency induced by the differential Doppler effect at the receivers is mitigated via a stationary phase approximation, consistent with the discussion above, but under the assumption that the ambient noise sources are statistically uncorrelated in time. More recent work by Liu et al.}~\cite{liu2021retrievability}, applies standard ambient noise interferometry (\ref{AmbientInt}) to an expression for the wavefield generated by a train, and the potential retrievable wave types are discussed using intuitive ray-path arguments which are then tested on train recordings. However, the explicit role of the source speed in the retrieval, and systematic methods for mitigating these effects have not been addressed (and the Doppler effect is not considered in the latter study). 
 
In summary,  the crosstalk terms associated with the integrand in Equation~(\ref{AmbientInt}) for $\x'\ne \x''$ emerge naturally in systematic ambient noise interferometry analysis but their effects  are commonly removed from considerations by imposing the assumption that the ambient noise field is uncorrelated; such an assumption greatly simplifies the theoretical retrieval procedure but it is often unrealistic. This is the case, for example, when performing traffic or railway noise interferometry.
Moreover, ambient noise interferometry (\ref{AmbientInt}) using moving point sources must contend with the Doppler effect in addition to the unavoidable correlations between the noise sources.

We discuss  consequences of these issues in detail in the following section, where we point out  that the delta-correlated assumption is unnecessarily restrictive, and that excluding near-diagonal terms reduces uncertainty in the retrieved amplitudes.  We show in detail how the success of the standard stacking procedure depends on the propertie\deleted{d}\added{s} of the source signature and on the source speed in cases where the noise sources are in motion. Moreover, we show how this standard stacking technique may fail, and propose an alternative workflow in Section~\ref{newtech} which proves to be successful at isolating terms from the incongruous crosstalk terms, leading to the extraction of the underlying inter-receiver phase regardless of the presence of correlation in the ambient noise field. Finally, for the scenario where the presence of correlation is caused by motion of the source, the workflow presented implicitly mitigates spurious contributions due the Doppler spread as explained~above.

\section{When and Why Stacking Works}\label{stacking_sec}

A commonly used averaging procedure in ambient noise interferometry, referred to as stacking, relies on the arithmetic average over several interferometric retrievals $\hat{h}(\x_\A,\x_\B,\omega)$ in Equation~(\ref{AmbientInt})---e.g.,  \cite{wapenaar2006green}.
Here, we define  \emph{stacking} more formally as the operation consisting of taking the statistical average of functions evaluated on an  ensemble of  recordings of the ambient noise field over some time interval; we denote such an ensemble average by $\langle \,\cdot\, \rangle$.  Below, we formally link the interferometric retrieval \eqref{AmbientInt} to the retrieval~\eqref{DirectInt} and determine conditions leading to the likely failure of the stacking average retrieval when the ambient noise field is correlated. 

Application of  the stacking average operator to  the ambient noise retrieval  represented via Equation~\eqref{CorrInt} leads to 

\begin{equation}\label{AverageAmbient}
\e{\hat{h}(\x_\A,\x_\B,\omega)} =  \frac{2}{\rho c}\int_S \int_S\e{\hat F^*\hat F}(\x',\x'',\omega)\,\hat{G}^*(\x_\A,\x',\omega)\hat{G}(\x_\B,\x'',\omega)d\x'\,d\x''\,,
\end{equation}
\noindent
where 

\begin{equation}\label{stack_av}
\e{\hat F^*\hat F}(\x',\x'',\omega):=\big\langle\hat{F}^*(\x',\omega)\hat{F}(\x'',\omega)\big\rangle 
\end{equation}
\noindent
is referred to as the {\it kernel of the stacking average} or simply as the {\it stacking average kernel}. This is determined by the correlation structure of the noise source characteristics {$\hat F$}.
Recall that for brevity of notation we consistently skip the explicit dependence of $\hat F$ in~Equation~(\ref{AverageAmbient})  on the particular recording of  ambient noise, and that for any fixed $\x$ and $\omega$, the source signature $\hat F(\x,\omega)$ should be thought of as a random variable, since it is associated with the Fourier transform of the ambient noise recording $p(\x,t)$. The stacking average  $\langle\,\cdot\,\rangle$ is  linear by construction and  we assume that the order of integration is irrelevant (Fubini's theorem is satisfied).  Please note that the averaging in  $\e{\hat{h}(\x_\A,\x_\B,\omega)}$ effectively couples the two integrals which contribute to  the single-recording ambient noise retrieval  ${\hat{h}(\x_\A,\x_\B,\omega)}$ in Equation~(\ref{CorrInt}). The formula in Equation~\eqref{AverageAmbient} formally highlights the link between  $\hat{g}(\x_\A,\x_\B,\omega)$ in Equation~(\ref{DirectInt}), the stacked ambient noise retrieval $\e{\hat{h}(\x_\A,\x_\B,\omega)}$, and the  properties of the stacking average kernel $\e{\hat F^*\hat F}$.  Clearly, the properties of the kernel of the stacking average control the discrepancy between   $\e{\hat{h}(\x_\A,\x_\B,\omega)}$ and ${\hat{g}(\x_\A,\x_\B,\omega)}$, which one is interested in minimizing.

In particular, the common interferometric assumption that the ambient noise field is uncorrelated leads to the stacking average  kernel in Equation~(\ref{AverageAmbient}) of the form 

\begin{equation}\label{sourceDelta}
\e{\hat F^*\hat F}(\x',\x'',\omega) \propto \delta(\x'-\x'')|\hat{s}(\omega)|^2\,,
\end{equation}
\noindent
where $\hat{s}(\omega)$ is the power spectrum of the ambient noise source. \added{There is a further implicit assumption in Equation \eqref{sourceDelta} that the power spectrum of the source does not depend on the spatial variables.} Clearly, if Equation~(\ref{sourceDelta}) holds, evaluating the integrals in Equation~(\ref{AverageAmbient}) leads to 

\begin{equation}\label{WapenaarRetrieval}
\e{\hat{h}(\x_\A,\x_\B,\omega)}\propto \hat{g}(\x_\A,\x_\B,\omega)|\hat{s}(\omega)|^2\,,
\end{equation}
\noindent
where $\hat{g}(\x_\A,\x_\B,\omega)$ represents  the retrieval from individually recorded discrete sources defined by Equation \eqref{DirectInt}. The result in Equation~\eqref{WapenaarRetrieval} implies  that even in the idealized scenario when Equation~(\ref{sourceDelta}) holds, ambient noise interferometry does not generally allow for accurate retrieval of the amplitude of the Green's function  due to the presence of the noise power spectrum $|\hat{s}(\omega)|^2$, and the lack of individual source recordings which  could  allow for a source-specific spectrum correction. On the other hand, the estimated phase remains unaffected by the presence of the real factor $|\hat{s}(\omega)|^2$ in Equation~(\ref{WapenaarRetrieval}). Note also that in practice, one has to consider empirical averages in Equations~(\ref{AverageAmbient})--(\ref{sourceDelta}) based on a finite number of ambient noise recordings instead of the abstract stacking in terms of the statistical average over an `infinite' number of recordings implied by the operator $\langle\,\cdot\,\rangle$. In such a case, if  the relationship in Equation~(\ref{sourceDelta}) holds in the limit of averaging over several samples tending to infinity, it can be shown that the acausal contributions from off-diagonal, incongruent source locations (i.e., when  $\x'\neq\x''$) decay with the number of samples and Equation~(\ref{WapenaarRetrieval}) holds asymptotically; we postpone these more technical details to another publication. A related derivation of a bound on the signal to noise ratio can be found in Appendix B of \cite{van2006interferometric}.

The requirement that ambient noise be delta-correlated in space is restrictive  and it is often unrealistic to assume such an approximation  in applications. As pointed out in~\cite{van2014surface}, most sources in practice are, unsurprisingly, neither fully correlated nor fully uncorrelated. Although some advances have been made in the context of simultaneously acting sources \citep{wapenaar2012relation,van2015retrieving} based on deconvolution techniques to tackle the resulting underdetermined systems, further investigation of such a deconvolution setup is required. Nevertheless, the relationship in Equations~(\ref{sourceDelta}) and (\ref{WapenaarRetrieval}) is desirable, as it allows for the retrieval of correct phases of the associated Green's functions from~Equation~(\ref{AverageAmbient}).
Therefore, in what follows we consider approaches which resemble the relationship in~Equation~(\ref{sourceDelta}) but without the requirement for the spatially uncorrelated noise sources. The remainder of this section is devoted to the analysis of the stacking average through the properties of the stacking average kernel $\e{\hat F^*\hat F}$ in Equation~\eqref{stack_av}. A new approach, which exploits a different averaging kernel and applies to both correlated and uncorrelated noise sources, is proposed and analyzed in Section~\ref{newtech}.

To gain a theoretical insight into the ambient noise retrieval,  we decompose Equation~\eqref{CorrInt} as 

\begin{equation}\label{hDecomp}
\hat{h}(\x_\A,\x_\B,\omega)=\hat{g}_{\epsilon}(\x_\A,\x_\B,\omega) + \hat{x}_{\epsilon}(\x_\A,\x_\B,\omega)\,,
\end{equation}
where  $\epsilon \geqslant 0 $ and 

\clearpage
\end{paracol}
\nointerlineskip
\begin{equation}\label{geps}
\hat{g}_{\epsilon}(\x_\A,\x_\B,\omega)=\frac{2}{\rho c}\int\limits_{\x'\in S} \int\limits_{\substack{\x''\in S\\[.1cm] \norm{\x'-\,\x''}\,<\,\epsilon}}\hat{F}^*(\x',\omega)\hat{F}(\x'',\omega)\,\hat{G}^*(\x_\A,\x',\omega)\hat{G}(\x_\B,\x'',\omega)\,d\x''\,d\x'\,,
\end{equation}
\begin{paracol}{2}
\switchcolumn
\noindent and

\begin{equation}\label{xeps}
\hat{x}_{\epsilon}(\x_\A,\x_\B,\omega)=\frac{2}{\rho c}\int\limits_{\x'\in S} \int\limits_{\substack{\x''\in S\\[.1cm] \norm{\x'-\,\x''}\,\geqslant \,\epsilon}}\hat{F}^*(\x',\omega)\hat{F}(\x'',\omega)\,\hat{G}^*(\x_\A,\x',\omega)\hat{G}(\x_\B,\x'',\omega)\,d\x''\,d\x'\,.
\end{equation}

{The} term $\hat{g}_{\epsilon}(\x_\A,\x_\B,\omega)$, referred  to hereafter as the {\it $\epsilon$-diagonal term}, accounts for the contributions from sources such that $\norm{\x' - \x''}<\epsilon$, and one might consider sufficiently small $\epsilon\geqslant 0$, in order to minimize the contribution from acausal arrivals from incongruous source locations in  $\hat{g}_{\epsilon}$. The decomposition in Equation~\eqref{hDecomp}  holds for any choice of  $\epsilon\geqslant 0$ but  taking $\epsilon\rightarrow 0$ does not necessarily remove all acausal arrivals from $\hat{g}_{\epsilon}$ for moving noise sources, as explained in the discussion in Section~\ref{theor_sec}. 
However, the aim of the above decomposition is to find an appropriate choice of the {\it spatial correlation length cut-off}~$\epsilon$ so that $\hat{g}_{\epsilon}$ is dominated by the coherent terms, whereas $\hat{x}_{\epsilon}$ is dominated by the acausal contributions due to incongruous source locations. Please note that  the decomposition in Equation (\ref{hDecomp})  could be applied to the integral with respect to $\x'$ instead of $\x''$ without loss of generality.

Finally,  we consider the stacking average of the decomposition in Equation~(\ref{hDecomp}) given~by 

\begin{equation}\label{meanhDecomp}
\e{\hat{h}(\x_\A,\x_\B,\omega)}=\e{\hat{g}_{\epsilon}(\x_\A,\x_\B,\omega)}+ \e{\hat{x}_{\epsilon}(\x_\A,\x_\B,\omega)}\,
\end{equation}
\noindent
with the aim of linking the characteristics of the stacking average kernel $\e{\hat F^*\hat F}(\x',\x'',\omega)=\big\langle\hat{F}^*(\x',\omega)\hat{F}(\x'',\omega)\big\rangle$ to the spatial correlation length cut-off $\epsilon$ in the above decomposition. Please note that in this section no specific assumptions have been made about the source or its spectrum, as the representation \eqref{p} used to arrive at Equations \eqref{hDecomp} and \eqref{meanhDecomp} is general.

\subsection*{Interferometric Retrieval from Spatially Correlated Noise Due to a Moving Source}
To show the utility of the $\epsilon$-diagonal decomposition (\ref{hDecomp}) in interferometric analysis, we consider a specific but common example of a broadband source train travelling at speed~$\V\neq0$ and emitting sound within  the frequency band $\big[\omTilde_{\min},\,\omTilde_{\max}\big]$, such that the phases for different emitted frequencies are random and independent from each other (we address correlation of the phase across different frequencies below), i.e., for each frequency $\omTilde$ emitted by the source, there is a corresponding random phase shift $\theta_\omTilde$ uniformly distributed between 0 and $2\pi$, and we assume that the phase shifts $\theta_{\omTilde}$ and $\theta_{\omTilde'}$ are independent whenever $\omTilde\neq\omTilde'$. In such a case we find (see Equation \eqref{EKernel} for details) that  the stacking average kernel in Equation~\eqref{stack_av} is given by 

\begin{equation}\label{kernelMean}
\e{\hat F^*\hat F}(\x',\x'',\omega) = \cos\left(\bar \omTilde\displaystyle \frac{\norm{\x''-\x'}}{\norm{\V}} \right)\sinc\left( {\textstyle\frac{1}{2}}\Delta_\omTilde\,\frac{\norm{\x''-\x'}}{\norm{\V}} \right)\vert\hat{s}(\omega)\vert^2,
\end{equation}
\textls[-15]{which depends on the train speed $\V\neq0$, the mean emitted frequency $\bar \omTilde:=\frac{1}{2}(\omTilde_{\max}+\omTilde_{\min})$ and the bandwidth $\Delta_\omTilde:=\omTilde_{\max}-\omTilde_{\min}$, and where  $\sinc(x) := \sin(x)/x$. The model and the resulting stacking average kernel \eqref{kernelMean} implicitly take account of the Doppler effect (see model derivations in Appendix~\ref{pDetails}). Expression \eqref{kernelMean} for the kernel holds for non-monochromatic sources travelling at speeds $\V\neq0$, and for sources whose spectrum can be represented in the form Equation~\eqref{generalSpectrum}. In the limit of $\Delta_\omTilde\to0$, i.e., when the source is monochromatic with frequency $\omTilde$, this kernel reduces to $\e{\hat F^*\hat F}(\x',\x'',\omega) = \cos(\omTilde(\added{\vert}\x''-\x'\added{\vert})\added{\vert\hat{s}(\omega)\vert^2)}/\norm{\V}$\deleted{$\vert\hat{s}(\omega)\vert^2$} which is no longer localized. See remarks under Equation~\eqref{broadband_eq} in \mbox{Appendix~\ref{pDetails}}} concerning a more general model that allows for appropriate localization. \added{For simplicity, we consider the parametrization of the straight boundary $S$ given by $\x = (\,x,\,0\,)$, where $x=0$ corresponds to the location where the line through the receiver pair intersects the boundary $S$, so that $\x' = (\,x',\,0\,)$ and similarly $\x'' = (\,x'',\,0\,)$. Using this parametrization we have that $\norm{\x'-\x''} = \norm{x'-x''}$ in Equation \eqref{kernelMean}. For further details see Equation \eqref{timeSpaceLink_eq} and accompanying remarks in Appendix~\ref{UncorrelatedPhases}.}

The movement of the noise source allows us to parameterize any locations $\x',\x''\in S$ via the travel time so that \deleted{$\x' = |\V|t'$ and $\x'' = |\V|t''$} \added{$\x' = (\,|\V|t'\,, 0\,)$ and $\x'' = (\,|\V|t'',\, 0\,)$}. The analytical expression for the stacking average kernel associated with a broadband correlated train of sources allows one to identify the approximate correlation length beyond which the envelope of the kernel (\ref{kernelMean}) decays rapidly. The unwanted long-range correlations contributing to the kernel away from the neighborhood of $\x'=\x''$, $\x',\x''\in S$, represent the main contribution to  $\big\langle\hat{h}(\x_\A,\x_\B,\omega)\big\rangle$ in Equation~(\ref{meanhDecomp}) from the acausal arrivals.  Based on Equation~(\ref{kernelMean}), we set the spatial correlation length cut-off $\epsilon$ to twice the value of the first zero of the envelope of the stacking average  kernel (see the bottom panel of Figure \ref{KernelMeanFig}), which is given by 

\begin{equation}\label{epsilon}
\epsilon = \frac{\norm{\V}}{\omTilde_{\max}-\omTilde_{\min}}\,,
\end{equation}
\noindent
for non-monochromatic sources travelling at speeds $\V\neq0$. Figure \ref{KernelMeanFig} shows  the numerically approximated stacking average kernel  $\e{\hat{F}^*\hat{F}}(\x',\x'',\omega)$  and its exact form~\eqref{kernelMean}. The rate of decay of the source kernel in terms of the spatial discrepancy $\norm{\x'-\x''}$ is controlled by the  envelope of the $\sinc(x)$ factor in Equation~\eqref{kernelMean}. As noted before, the spatial discrepancy  can be represented in terms of the travel times so that $\norm{\x'-\x''}=\norm{t'-t''}|\V|$, where $\V$ is the train speed.
Although the choice of the spatial correlation cut-off length $\epsilon$ has so far been somewhat arbitrary, the main principle in its determination is to use the structure of the stacking average kernel $\e{\hat{F}^*\hat{F}}$ in order to approximately identify the correlation length beyond which the crosstalk terms dominate the correlation structure represented by $\e{\hat{F}^*\hat{F}}$.  

Please note that while the decomposition \eqref{hDecomp} and its stacking average Equation~(\ref{meanhDecomp}) are given in the frequency domain to  highlight the interaction between incongruous source locations, the corresponding stacking average decomposition in the time domain is given~by 

\begin{equation}\label{hDecompTime}
h(\x_\A,\x_\B,t) = g_{\epsilon}(\x_\A,\x_\B,t) +x_{\epsilon}(\x_\A,\x_\B,t)\,,
\end{equation}
 \noindent
and its stacking average is given by 

\begin{equation}\label{stack_hDecompTime}
\e{h(\x_\A,\x_\B,t)} = \e{g_{\epsilon}(\x_\A,\x_\B,t)} +\e{x_{\epsilon}(\x_\A,\x_\B,t)}\,.
\end{equation}

The above time-domain decompositions  hold due to the linearity of the inverse Fourier transform;  as is standard in the interferometric literature, we assume throughout that $h$ is square-integrable in $t\in \mathbb{R}$ (so that the Fourier transform is well-defined on  $h\in L^2(\mathbb{R})$) and that $\hat F$ and $\hat G$ are such that Fubini's theorem holds throughout. Thus, the diagonal band decompositions of $h$ and $\hat h$ are dual to each other and they can be considered~interchangeably.

The formulation of the $\epsilon$-diagonal decomposition in the time domain (i.e., representation (\ref{hDecompTime})) lends itself directly to the analysis based on time series recordings provided that a suitable spatial correlation length cut-off $\epsilon$ can be determined. When the noise source is in motion, as in the case considered here,  the spatial correlation length cut-off~$\epsilon$ has  a temporal counterpart $T_\epsilon = \epsilon/|\V|$ for $\V\neq0$.
In Section~\ref{newtech}, we will focus on a procedure for recovering the $\epsilon$-diagonal band structure, analogous to that in $\langle\hspace{.02cm} g_\epsilon \rangle$ in~Equation~(\ref{stack_hDecompTime}),  from a single short recording when the value of the spatial correlation length cut-off $\epsilon$ cannot be estimated from the stacking average kernel.   First, however, we outline the general properties of the stacking average and its potential pitfalls when retrieving the inter-receiver interferometric Green's function. To this end we consider two different classes of noise sources, and we compare the standard stacked retrieval  $h(\x_\A,\x_\B,t)$ and its stacking average $\e{h(\x_\A,\x_\B,t)}$ with retrievals  based on isolating the $\epsilon$-diagonal band terms  $g_{\epsilon}(\x_\A,\x_\B,t)$ and $\e{g_{\epsilon}(\x_\A,\x_\B,t)}$ in the decomposition Equations~(\ref{hDecompTime}) and (\ref{stack_hDecompTime}), respectively,  using  an `informed' choice of the spatial correlation length cut-off $\epsilon$ given by~Equation~(\ref{epsilon}).

\end{paracol}
\nointerlineskip
\begin{figure}[H]
\widefigure
\includegraphics[width=0.8\textwidth]{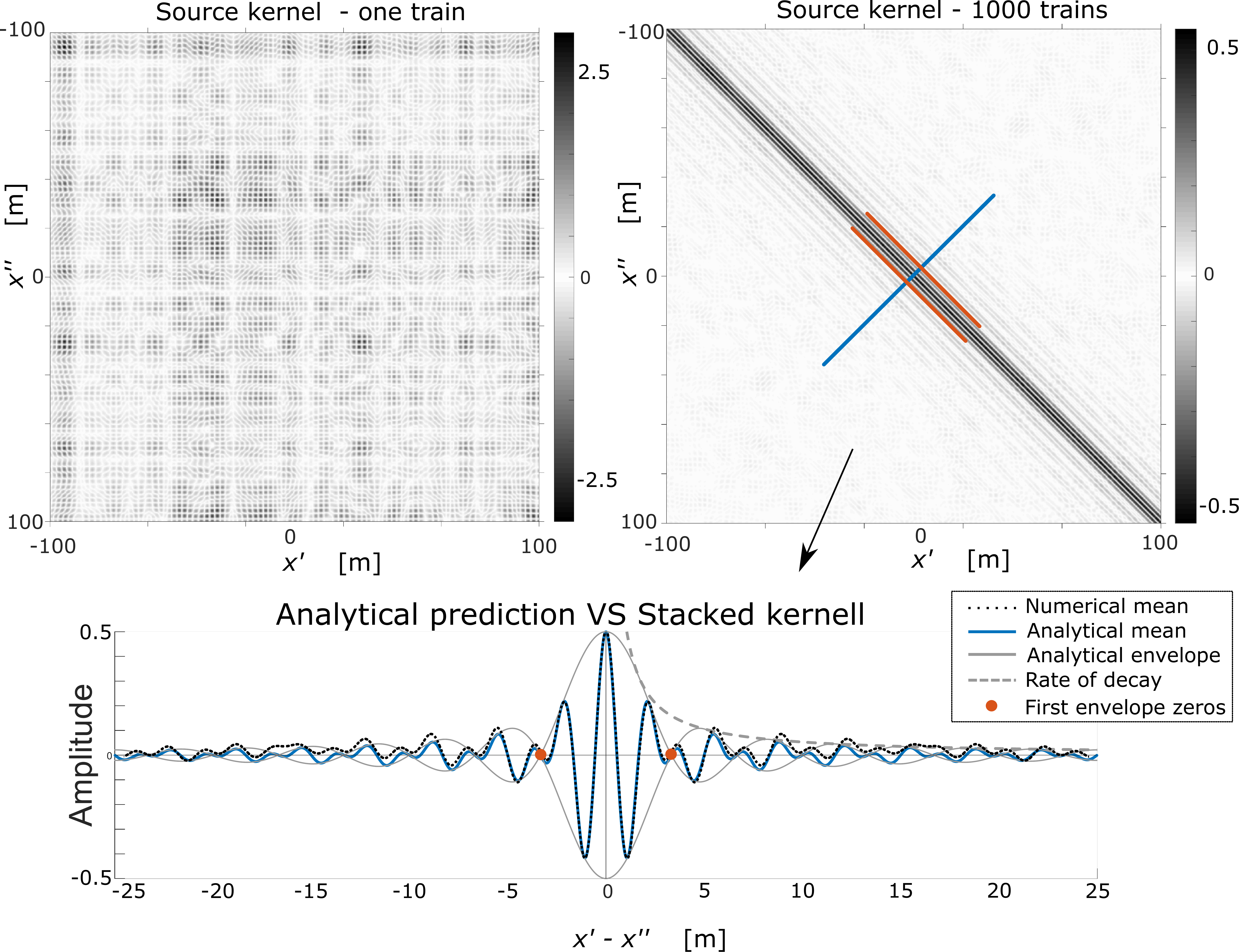}
\caption{ \textls[-15]{{Example of the} 
 structure of the source kernel $\hat{F}^*(\x',\omega)\hat{F}(\x'',\omega)$  and the stacking average kernel $\langle \hat F^*\hat F\rangle(\x',\x'',\omega)=\big\langle\hat{F}^*(\x',\omega)\hat{F}(\x'',\omega)\big\rangle$ derived in Equation~(\ref{kernelMean}). Top left panel shows the spatial part of source kernel~\eqref{kernelMean} (that is, without including the frequency-dependent term $\norm{\hat{s}(\omega)}^2$) as a function of the location of a single train of sources recorded at each receiver. The simulated train travels at 90~km/h and its spectrum ranges from 10 Hz to 25 Hz. The top-right panel shows the corresponding kernel after stacking 1000 source train recordings each five minutes long. The bottom panel compares the numerical anti-diagonal section in black dashes, extracted from the numerical mean in the top-right panel, with the prediction given by Equation \eqref{kernelMean} in blue, as well as its analytical envelope \added{in solid gray} and \added{the} rate of decay in \added{dashed gray}, in terms of the \added{signed} spatial discrepancy \deleted{$\norm{\x'-\x''}$}\added{$x'-x''$, where the parametrization $\x' = (x',\,0)$, $\x'' = (x'',\,0)$ has been used.}\deleted{ in solid and dashed gray, respectively.} The rate of decay is derived from the envelope of the $\sinc$ factor in Equation \eqref{kernelMean}. The orange dots highlight the first pair of zeros of the analytical envelope, capturing the energy around the mean. The radius $\epsilon$ will be taken to be the distance between these zeros.}}\label{KernelMeanFig}
\end{figure}
\begin{paracol}{2}
\switchcolumn

In Figure \ref{DiagDecompFig} we use the spatial cut-off radius $\epsilon$ determined in Equation~(\ref{epsilon}) and  apply the $\epsilon$-diagonal decomposition procedure laid out in Equation~(\ref{hDecompTime}). First, we assume that individual recordings are statistically independent. In particular, for a noise source with a broadband spectrum this constraint translates to the requirement that the frequencies associated with the random source signature $\hat F$ be independent. To generate this figure, a train of sources travelling at 90~km/h along a straight boundary and emitting sound between 10 Hz and 25 Hz was simulated to generate the ambient noise recordings $p(\x_\A,t)$ and $p(\x_\B,t)$. The receiver pair was oriented perpendicularly to the source-line, with an inter-receiver distance of 2000~m and the closest receiver 400~m from the source-line. The medium speed was set to 1000~m/s. The ambient noise recordings were transformed to the frequency domain and $\hat h(\x_\A,\x_\B,\omega)$ was computed according to Equation \eqref{AmbientInt}. Then the $\epsilon$-decomposition was calculated for a single run to generate the insets in the top row. The frequency-domain data were used to extract the phase in the top-right panel of \mbox{Figure \ref{DiagDecompFig}}, as well as transformed back to the time domain to generate the time series $h(\x_\A,\x_\B,t)$, $g_\epsilon(\x_\A,\x_\B,t)$, and $x_\epsilon(\x_\A,\x_\B,t)$ in the top left panel of the same Figure. This procedure was repeated 1000 times and averaged to calculate $\langle\hat h(\x_\A,\x_\B,\omega)\rangle$, as well as its $\epsilon$-decomposition with the associated phases (right) and waveforms (left) in the bottom row of this figure.
\begin{figure}[H]
\includegraphics[width=0.7\textwidth]{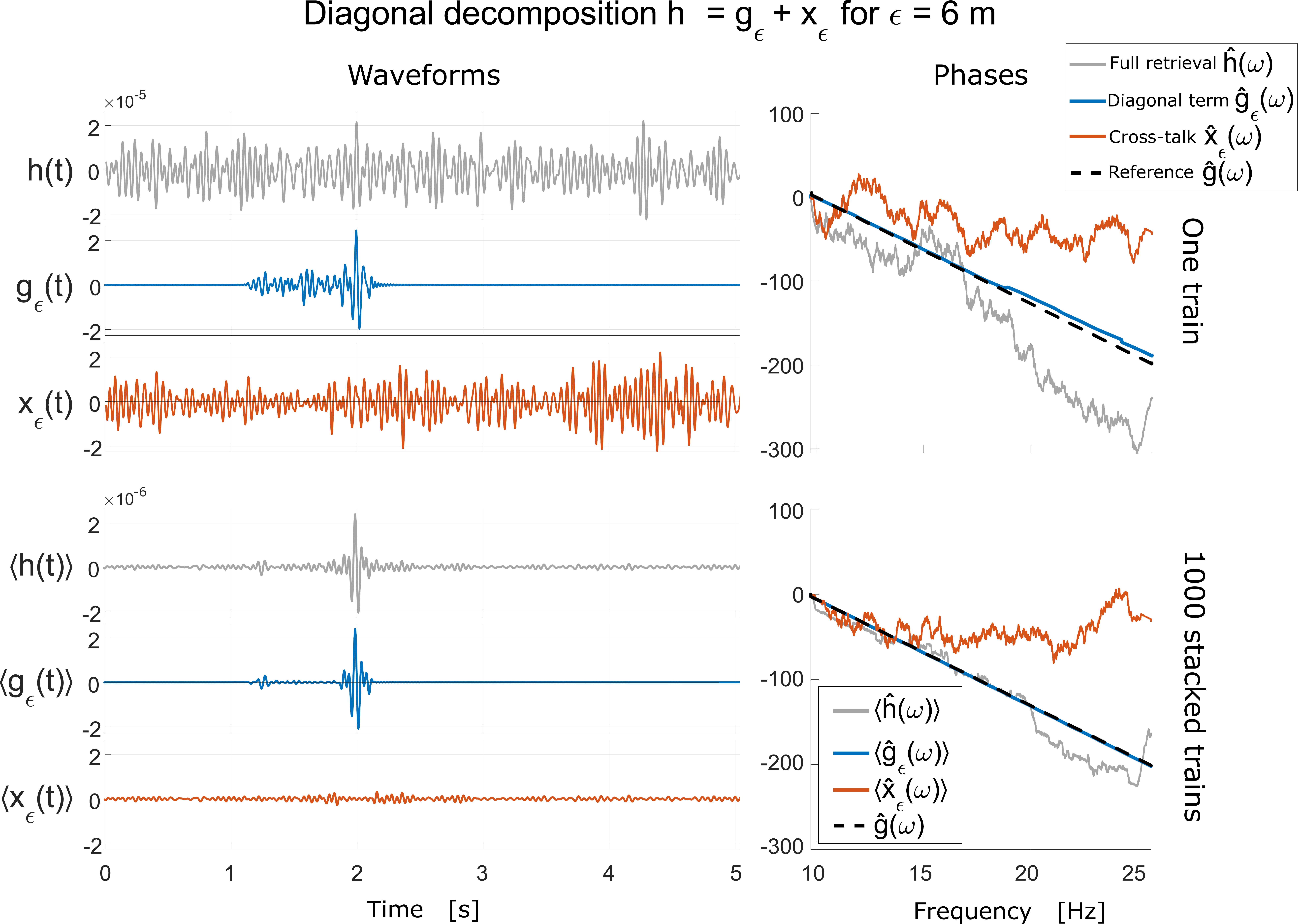}
\caption{{Example of the} 
 $\epsilon$-diagonal decomposition procedure Equation~(\ref{hDecompTime}) and its stacking average version Equation~(\ref{stack_hDecompTime}) for a train travelling at 90\,km/h with the emitted spectrum between 10\,Hz and 25\,Hz. The receiver pair was placed perpendicularly to the road, with an inter-receiver distance of 2000~m and a distance of 400~m from the source-line to the closest receiver. The medium speed was set at 1000~m/s. The decomposition is performed for $\epsilon =  6$\,m ($T_\epsilon = 0.25$\,s) determined from Equation~(\ref{epsilon}) for a single five-minute recorded source train (\textbf{top}) and for a stacked retrieval over 1000~five-minute train source recordings (\textbf{bottom}). The left panels show the waveforms of the components Equations~(\ref{hDecompTime}) and (\ref{stack_hDecompTime}), and the right panels show the corresponding phases in the frequency domain obtained from the frequency representation of the retrieval given by \mbox{Equations~(\ref{hDecomp}) and (\ref{meanhDecomp}).} The retrieval $h(t)$ and its stacking average $\e{h(t)}$ are shown in gray, and their components $g_\epsilon(t)$, $\e{g_\epsilon(t)}$,  and $x_\epsilon(t)$, $\e{x_\epsilon(t)}$, are shown in blue and orange, respectively. The phase plots include the phase of the inter-receiver Green's function (dashed black) estimated from discrete sources, as prescribed by Equation~\eqref{DirectInt} as a  reference.
}\label{DiagDecompFig}
\end{figure}

In this setup, and given the stacking average kernel in Equation~(\ref{kernelMean}), the retrieval of the inter-receiver Green's function is shown in gray in both the temporal and the spatial domains, with the $\epsilon$-diagonal  terms $g_{\epsilon}$, $\hat g_{\epsilon}$  in blue and the crosstalk $x_{\epsilon}$, $\hat x_{\epsilon}$ shown in orange. According to Equation~\eqref{epsilon}, we chose $\epsilon = 6$ m which corresponds to a time window length $T_\epsilon\approx 0.25$\,s. The resulting spectra were transformed to the temporal domain via \deleted{IFFT} \added{inverse Fourier transform} to show the effect of the decomposition on the retrieved waveforms; the components of Equation \eqref{hDecompTime} are shown in the left panels of Figure~\ref{DiagDecompFig}. The decomposition was applied to a single five-minute recording of a source train (top), and to the averaged waveform resulting from stacking 1000 such recordings (bottom). The phases corresponding to each waveform are shown in the panels to the right in Figure \ref{DiagDecompFig}.

This  illustrates  the improvement in the retrieval using the decomposition \eqref{hDecomp}  which satisfactorily  separates the causal contributions to the inter-receiver phase from the spurious~crosstalk.

Please note that the phase of $\langle\hat{h}(\omega)\rangle$ in the bottom right panel of Figure \ref{DiagDecompFig} (in blue) typifies the phase retrieved from the standard ambient noise interferometry approach which relies on averaging over stacked independent recordings of ambient noise  of the form \eqref{AmbientInt}. Although stacking can clearly improve the quality of the retrieved phase of the inter-receiver Green's function, significant discrepancies remain especially in  the higher frequency range of the spectrum due to the increased influence of the Doppler shift even for many stacked recordings.  It should be noted that sources travelling at different speeds and overlapping with each other may help improve the rate of convergence of the standard stacking to the correct phase. On the other hand, the phase of the $\epsilon$-diagonal term $\hat g_\epsilon(\omega)$  shows a much  better agreement with the uncorrelated reference estimate even for a single short recording  (blue phase in top-right panel of Figure \ref{DiagDecompFig}). Application of the stacking average to the $\epsilon$-diagonal term $\langle \hat g_\epsilon(\omega)\rangle$ leads to a further but slight improvement over the standard stacking applied directly to $\hat h$ which is evident  in the higher frequency range of the spectrum. This increased uncertainty for higher frequencies is to be expected given the presence of the Doppler shift.  

The above experiment indicates that although there is some flexibility in the choice of the spatial correlation length cut-off  $\epsilon$, it is possible to carry out the $\epsilon$-diagonal  decomposition~\eqref{hDecomp} in a way which on average, largely separates the desired contributions from coherent sources and  the crosstalk between incongruous source locations into two distinct terms $\hat g_{\epsilon}(t)$ and $\hat x_{\epsilon}(t)$ respectively. 
In the case of correlated moving noise sources, a sufficiently small correlation length cut-off $\epsilon$ results in attributing a significant  proportion of the coherent noise contributions which should be captured by $\hat g_{\epsilon}$  relative to the `crosstalk' term $\hat x_{\epsilon}$; the situation is reversed for a sufficiently large $\epsilon$.  If we were to evaluate these integrals explicitly in order to estimate the true phase, the challenge lies in determining an optimal value of the spatial correlation length cut-off $\epsilon$ that minimizes the error in the phase retrieval in the absence of  knowledge of the spatial correlation structure contained in the stacking average kernel $\big\langle\hat{F}^*\hat F\big\rangle(\x',\x'',\omega)$ in Equation~(\ref{stack_av}).

It is worth noting that a successful phase retrieval of the inter-receiver Green's function based on the empirical average of stacked recordings fundamentally relies, via the law of large numbers, on the  statistical independence of the individual recordings.  In particular, for a noise source with a broadband spectrum this constraint translates to the requirement that the frequencies associated with  the random source signature $\hat F$ are independent; otherwise, the empirical average of the stacked retrieval $\frac{1}{N}\sum_{n=1}^N\hat{h}^{(n)}(\x_\A,\x_\B,\omega)$ might still converge as $N\rightarrow \infty$ but  there are no guarantees that the resulting limit $\big\langle\hat{h}(\x_\A,\x_\B,\omega)\big\rangle$ will lead to the  recovery of the true inter-receiver phase. 
Figure \ref{DiagDecompDependentFig} shows one  such example, where the setup and parameters are identical to  those in Figure \ref{DiagDecompFig}, except that the phases of the frequencies emitted by the sources are no longer uncorrelated.  Such a scenario would be likely to occur, for example, due to an  interaction between sleepers on a rail track and the wheels of the train \citep{brenguier2019train}. Specifically, for each frequency $\omTilde$ emitted by the source, the random phases associated with each emitted frequency are such that for a pair of emitted frequencies $\omTilde$ and $\omTilde'$, $\norm{\omTilde-\omTilde'}\ll1$, the phase shifts satisfy $\theta_\omTilde = \theta_\omTilde' + r $, where $r$ is uniformly distributed between 0 and $2\pi$, so that the phase shifts $\theta_{\omTilde}$ and $\theta_{\omTilde'}$ are no longer~uncorrelated. 

The gray trace in the middle panel of Figure~\ref{DiagDecompDependentFig} shows the averaged waveform in the time domain obtained by  stacking 1000 five-minute long recordings.  In contrast to the result of the equivalent procedure shown in Figure \ref{DiagDecompFig} (bottom left), stacking the recordings  does not converge (or does not converge sufficiently quickly) to the correct inter-receiver arrival, and the crosstalk term (orange) in the middle panel does not decay. The corresponding phases are shown in the right panel of Figure~\ref{DiagDecompDependentFig} with the phase estimated from stationary discrete sources (see~Equation \eqref{DirectInt}) in dashed black  as a reference. However, the phase of the inter-receiver signal is still successfully isolated by the $\epsilon$-diagonal decomposition (\ref{hDecomp}), as indicated by the blue line, showing a good agreement with the phase of the inter-receiver Green's function estimated from discrete sources, as per Equation \eqref{DirectInt}.
\begin{figure}[H]
    \includegraphics[width =0.74\textwidth]{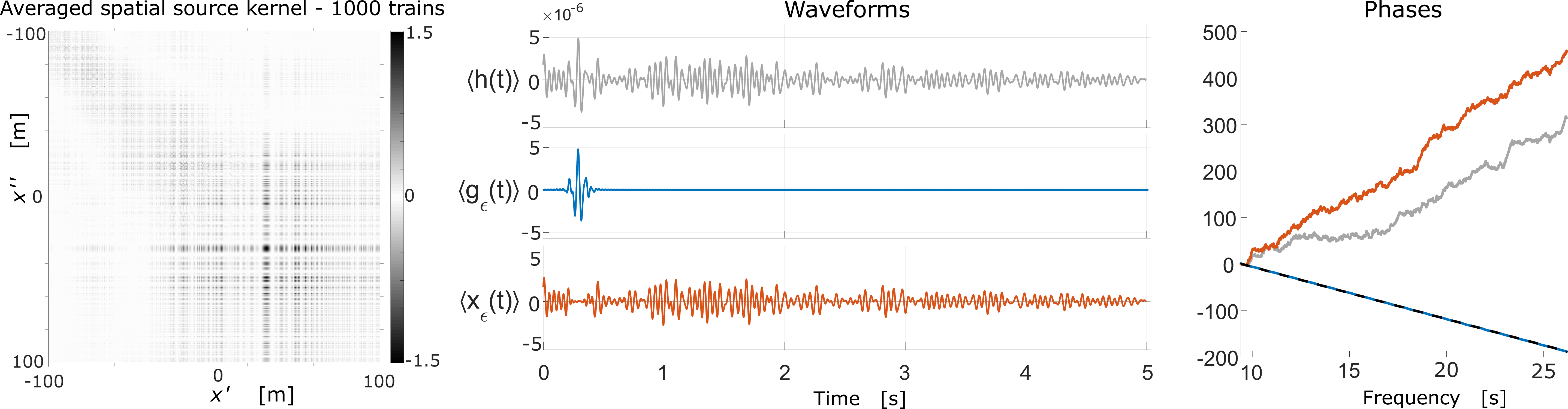}
    \caption{{Example of a} 
 source signature from which the interferometric Green's function estimate cannot be retrieved by stacking. The left  panel shows the numerical/empirical  approximation of the stacking average kernel $\langle \hat F^*\hat F\rangle(\x',\x'',\omega)$ in Equation~(\ref{stack_av})  estimated by stacking 1000 trains of sources travelling at 90 km/h with an emitted spectrum between 10\,Hz and 25\,Hz. The receivers were placed perpendicularly to the source-line, with an inter-receiver distance of 200~m and the receiver closest to the source-line at a distance of 400~m. Away from the diagonal, non-zero elements persist. The middle panel shows the stacked waveform $\e{h(t)}$ and its $\epsilon$-diagonal decomposition into $\e{g_\epsilon(t)}$ and  $\e{x_\epsilon(t)}$. The right panel  shows the corresponding unwrapped phases for the full term $\e{\hat h(\omega)}$ (gray) and its decomposition into the $\epsilon$-diagonal term $\e{\hat g_\epsilon(\omega)}$  (blue) and the crosstalk term and $\e{\hat x_\epsilon(\omega)}$ (orange), as well as the phase of the inter-receiver Green's function (dashed black) estimated from discrete sources as per Equation \eqref{DirectInt} as a reference. In this case, the stacking average kernel $\big\langle\hat{F}^*\hat F\big\rangle(\x',\x'',\omega)$ in Equation~(\ref{stack_av}) is not localized and the value of the correlation length cut-off $\epsilon$ was chosen according to the approach described in Section~\ref{newtech}. The phase retrieval from $\e{\hat g_\epsilon(\omega)}$ and ${\hat g_\epsilon(\omega)}$ is similar and not shown. }\label{DiagDecompDependentFig}
\end{figure}

Clearly,  in most real-world applications it is not possible to gauge the retrieval against some reference benchmark, which leaves the retrieved results and  the convergence trends open to subjective interpretation. Moreover, even if enough recorded data are available to allow for stacked averaging but the stacking average kernel $\big\langle\hat{F}^*\hat F\big\rangle(\x',\x'',\omega)$ in~\mbox{Equation~(\ref{stack_av}) } is not spatially localized, there is no a priori criteria for  estimating the  appropriate value for the spatial correlation length cut-off $\epsilon$ needed in the decomposition (\ref{hDecomp}) (compare the example in Figure \ref{DiagDecompDependentFig} with that illustrated in  Figures \ref{KernelMeanFig} and~\ref{DiagDecompFig}).   
The key point of these considerations is that in principle it remains possible to estimate  the correct phase of the underlying Green's function from a single short recording of a correlated ambient noise source, and that the cause of the noise correlation is not relevant in such considerations.

In summary, in this section we discussed a constructive example of the interferometric retrieval of the inter-receiver Green's function which was based on an abstract decomposition of the retrieval~(\ref{AmbientInt}) as $\hat h = \hat g_\epsilon+\hat x_\epsilon$, as in Equations \eqref{hDecomp} or (\ref{hDecompTime}), and based on  an  `informed' choice of the spatial correlation length cut-off~$\epsilon$ so that the undesirable crosstalk terms are largely contained in the $\hat x_\epsilon$ term. This construct, driven by analytical insight and illustrated in Figures~\ref{KernelMeanFig}--\ref{DiagDecompDependentFig},  suggests that it should be possible to carry out such a decomposition and subsequently obtain the  phase estimate of the inter-receiver Green's function if a suitable correlation cut-off length $\epsilon$ can be estimated. Importantly, this approach would allow one to sidestep the classical stacking average  approach and avoid its  potential pitfalls. In the following section we propose a workflow that facilitates the extraction of the interferometric phase estimate from a single short recording of a passing train (i.e., an instance of a moving correlated noise source). In fact, the choice of the correlation length cut-off $\epsilon$ in the example in Figure \ref{DiagDecompDependentFig} was determined via  a procedure described in the next section, since in the setup of Figure \ref{DiagDecompDependentFig}  the stacking average  kernel $\e{\hat{F}^*\hat{F}}(\x',\x'',\omega)$ was  not spatially localized due to implicit correlations between the noise sources.

\section{The Random Windowing Technique}\label{newtech}
Based on the analytical insight outlined in the previous sections, we consider the problem of estimating the phase of the inter-receiver Green's function from a single short recording of a train of sources, without resorting to the standard averaging over a stacked ensemble  of recordings. Nevertheless, the retrieval based on this procedure effectively results in the extraction of the $\epsilon$-diagonal term in the decomposition $\hat{h}(\x_\A,\x_\B,\omega)=\hat{g}_{\epsilon}(\x_\A,\x_\B,\omega) + \hat{x}_{\epsilon}(\x_\A,\x_\B,\omega)$ introduced in Equation~(\ref{hDecomp}), where $\epsilon$ is determined without the explicit knowledge of the stacking average kernel  $\e{\hat{F}^*\hat{F}}(\x',\x'',\omega)$ and the need for it to be localized. Importantly, the technique applies to both correlated and uncorrelated noise~sources.

Consider  a single time-domain recording of a train of sources at a receiver pair $\x_\A$ and $\x_\B$ and denote these recordings by, respectively, $p(\x_\A,t)$ and $p(\x_\B,t)$ whose Fourier transform leads to the retrieval $\hat h(\x_\A, \x_\B,\omega) = \hat p^*(\x_\A,\omega)\hat p(\x_\B,\omega)$ in Equation~(\ref{AmbientInt}), representing the interferometric approximation of the inter-receiver Green's function $\hat G(\x_\A, \x_\B,\omega)$. In the previous section the correlation length cut-off $\epsilon$ had to be estimated from the stacking average kernel $\e{\hat{F}^*\hat{F}}(\x',\x'',\omega)$ in Equation~(\ref{AverageAmbient}), assuming that the stacking average had  a localized  kernel (e.g., Figure~\ref{KernelMeanFig}) which is not always the case for correlated noise sources (e.g., \mbox{Figure~\ref{DiagDecompDependentFig}}). We now determine an optimal spatial/temporal window size that minimizes the effect of acausal arrivals in the interferometric retrieval from a single short recording. The approach relies on the use of an appropriate randomization technique, termed {\it random windowing}, in order to introduce incoherence into the acausal part of the phase in the retrieval $\hat h(\x_\A, \x_\B,\omega)$ in the available recording and mitigate the influence of the crosstalk terms, largely contained in $\hat{x}_{\epsilon}(\x_\A,\x_\B,\omega)$ of the $\epsilon$-diagonal decomposition of $\hat{h}(\x_\A,\x_\B,\omega)$,  which hinder the accurate retrieval of the Green's function $\hat G(\x_\A, \x_\B,\omega)$.

\subsection{General Setup}\label{rnd_wind_fr}

First, consider the interferometric retrieval $ h(\x_\A, \x_\B,t)$ in Equation~(\ref{AmbientInt}) obtained from recorded time series data within an  interval/window of duration  $T>0$ and centered at some randomly drawn time~$t_n$. For a pair of receiver locations $\x_\A$ and $\x_\B$ restrict the recorded time series $p(\x_\A,t)$ and $p(\x_\B,t)$ to that temporal window (see Figure \ref{RandomWindFig}). The random windowing method introduced below can be applied to recorded time series generated by arbitrary noise sources.  However, in what follows we consider correlated noise generated by a source moving with speed $\V\neq 0$. This setup is relevant in applications, and it also aids understanding of the theoretical underpinning of the random windowing method, since it allows one to link the windowing of the time series to the spatial windowing, as illustrated in Figure~\ref{RandomWindFig}. For time series generated by arbitrary noise sources, $\V$ does not correspond to a physical velocity, and stands for an abstract parameter that allows for the space time transformation. Next, for a given average source velocity, the window of duration  $T$ is linked to a spatial window which we define for a receiver location $\x$ as

\begin{equation}\label{Swindow}
 S(\x,\V,T,t_n) = \Big\{\x'\in S : \x' = \V t' \text{ for } \norm{t'-t_n}<{\textstyle \frac{1}{2}}T \Big\},
\end{equation}
\noindent
and which corresponds to the segment of the (spatial) source boundary traversed by the source moving at speed $\V$ from the origin at time zero, and recorded at a receiver location $\x$ within  the time window of duration $T$ centered at $t_n$. As illustrated in the right panel of Figure \ref{RandomWindFig}, these boundary segments  will differ for each receiver, depending on the inter-receiver distance, the distance to the source-line, and the speed of the source. The theoretical spatial location of $S(\x,\V,T,t_n)$ for each receiver can be calculated explicitly using Equation~\eqref{tEmission} and scaling by the source speed (or by its estimated  mean), as described in Appendix~\ref{pDetails}. 
 The frequency-domain representation of the time-windowed data at each receiver can be expressed as
 
	\begin{equation}\label{windowedp}
	\begin{split}
		\hat{p}(\x_\A,\omega\,;\,T,t_n) &= \int_{S(\x_\A,\V,T,t_n)} \hat{F}(\x',\omega)\hat{G}(\x_\A,\x',\omega)d\x'\,,\\
		\hat{p}(\x_\B,\omega\,;\,T,t_n) &= \int_{S(\x_\B,\V,T,t_n)} \hat{F}(\x',\omega)\hat{G}(\x_\B,\x',\omega)d\x'\,,
	\end{split}
    \end{equation}
    \noindent
    where $S(\x,\V,T,t_n)$ is given by Equation~\eqref{Swindow}. Please note that explicit knowledge of the signals integrated along boundary sections is not required in practice, as the windowing procedure is performed directly on the full recorded time series. A source travelling at speed $\norm{\V}$ will span a distance no greater than $\norm{\V}T$ during a time window of duration  $T$, so that for any random time window location $t_n$ the corresponding spatial location $\x_n = \norm{\V}t_n$ satisfies $\norm{\x'-\x_n}\leqslant\norm{\V}T$. Restricting the time window size to $T$ will implicitly restrict the spatial locations contributing to the cross-correlation terms, and it will limit the crosstalk between incongruous source locations.
 Choosing the set of source locations $\x'$ such that $\norm{\x'-\x_n}\leqslant \norm{\V}T$ allows us to write a recording Equation~\eqref{windowedp} in a manner consistent with representation~\eqref{p}; namely

\begin{equation}\label{windowedpInd}
	\hat{p}(\x,\omega\,;\,T,t_n) := \int_S \hat{F}(\x',\omega)\hat{G}(\x,\x',\omega)\,\chi_{\left\{\norm{\x'-\x_n}\leq T\norm{\V}\right\}}(\x')\,d\x'\,,
\end{equation}
\noindent
where  $\chi_A(\x)$ is the indicator function of a set $A$; i.e., $\chi_A(\x)=1$ if $\x\in A$ and  $\chi_A(\x)=0$ if $\x\notin A$.
Then, applying the ambient noise retrieval Equation~\eqref{AmbientInt} to the time-windowed recordings (see Figure~\ref{RandomWindFig}) for a window of duration  $T$ centered at $t_n$ results in the windowed~retrieval.

 \begin{align}\label{windowedh}
 \hat{h}(\x_\A,\x_\B,\omega\,;\,T,t_n) :=& \frac{2}{\rho c}\,\hat{p}^*(\x_\A,\omega\,;\,T,t_n)\,\hat{p}(\x_\B,\omega\,;\,T,t_n)\notag\\[.2cm]
 =&\frac{2}{\rho c}\int_S\int_S   \hat{F}^*(\x',\omega)\hat{F}(\x'',\omega)\hat{G}^*(\x,\x',\omega)\hat{G}(\x,\x'',\omega)\,\notag\\[.2cm]
&\hspace{1.4cm}\times\chi_{\left\{\norm{\x'-\x_n}\leq T\norm{\V}\right\}}(\x')\,\chi_{\left\{\norm{\x''-\x_n}\leq T\norm{\V}\right\}}(\x'')\,d\x'd\x'',
\end{align}
where the second equality is derived by substituting Equation \eqref{windowedpInd} for each recording, and recasting the resulting product of integrals in a manner consistent with the form of \mbox{Equation \eqref{CorrInt}}. Please note that as in Equation~(\ref{CorrInt}),  the term $\hat{F}^*(\x',\omega)\hat{F}(\x'',\omega)$ is still present in~Equation~\eqref{windowedh}, and one could consider applying the standard stacking to \mbox{Equation \eqref{windowedh}}. However, if  the resulting averaged/stacked source kernel $\langle \,\hat{F}^*(\x',\omega)\hat{F}(\x'',\omega)\,\rangle$ is not sufficiently localized (as discussed in Section \ref{stacking_sec}), there are no guarantees that we will retrieve the correct inter-receiver phase. Instead, we consider a new averaging  operator which relies on averaging over randomized  window locations $t_n$. We denote this operator by $\langle\,\cdot\,\rangle_T$ and define it for a scalar function $\gamma:\mathbb{R}\rightarrow \mathbb{R}$  as 

\begin{equation}\label{operatorT}
\big\langle\gamma\,\big\rangle_T := \int_\mathbb{R} \gamma(\tau)f_T(\tau)d\tau,
\end{equation}    
where $f_T$ is the probability density function associated with the choice of locations $t_n$ which is parameterized by $T$ \added{and assumed to be non-zero only on a bounded interval}, and the function $\gamma$ is integrable with respect to $f_T$. Next, we define the \emph{averaged retrieval for a time window of duration $T$} as

\end{paracol}
\nointerlineskip
\begin{align}\label{windowedhmean}
\hat{h}_T(\x_\A,\x_\B,\omega) &:=\e{\hat{h}(\x_\A,\x_\B,\omega\,;\,T,\,\cdot\,)}_T\notag\\[.2cm]
& =\frac{2}{\rho c}\int_S\int_S  \big\langle \chi\chi\big\rangle_T(\x',\x'') \hat{F}^*(\x',\omega)\hat{F}(\x'',\omega)\hat{G}^*(\x,\x',\omega)\hat{G}(\x,\x'',\omega)d\x'd\x\,,
\end{align}
\begin{paracol}{2}
\switchcolumn
\noindent
where we assume that Fubini's theorem is satisfied and the {\it random windowing averaging kernel} is defined as 

\begin{equation}\label{chichi}
\big\langle \chi\chi\big\rangle_T(\x',\x''):= \big\langle\,\chi_{\left\{\norm{\x'-\x_n}\leq T\norm{\V}\right\}}(\x')\,\chi_{\left\{\norm{\x''-\x_n}\leq T\norm{\V}\right\}}(\x'')\big\rangle_T.
\end{equation}

\begin{figure}[H]
\includegraphics[width=0.74\textwidth]{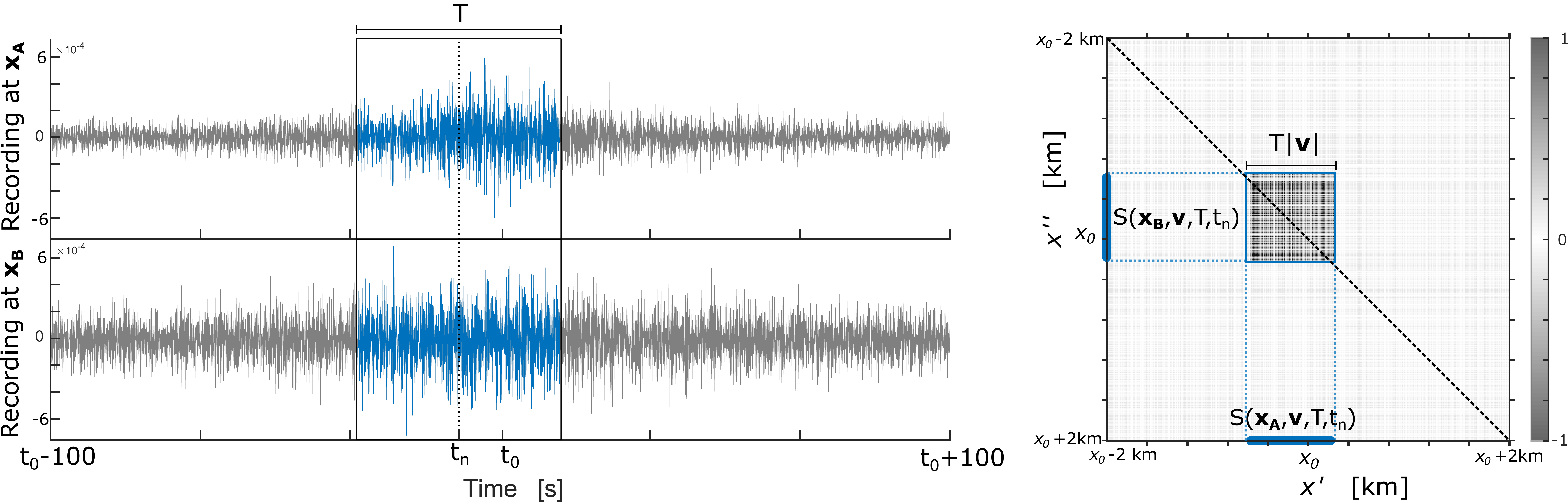}
\caption{{Illustration of the} 
 random windowing procedure for a single realization of a window of size $T$ in the temporal domain (\textbf{left}), and the corresponding window of length $T|{\bf v}|$ in the spatial domain (\textbf{right}) for $\V\neq0$. The panels on the left show the waveforms recorded at a pair of receivers generated by a moving noise source. The approximate time at which the moving source is collinear with the receiver pair is denoted by  $t_0$, and  the location of a window of duration $T$ (blue) is centered around the randomly drawn time $t_n$. The background in the inset on the right shows the spatial part of the source kernel $\langle\hat{F}^*(\x',\omega)\hat{F}(\x'',\omega)\rangle$ in Equation~(\ref{AverageAmbient}), where $\x_0$ corresponds to the spatial location of the moving source at time $t_0$. The boundary of the spatial window, centered at a random location $\x_n = t_n\norm{\V}$, and its projections are highlighted in blue and denoted by $S(\x,\V,T,t_n)$ for each receiver; the resulting square window (of size $T\norm{\V}$) is slightly off the diagonal. The parts of the source kernel excluded from this random window are grayed out.}\label{RandomWindFig}
\end{figure}

Please note that given the above formulation and the $\epsilon$-diagonal decomposition (\ref{hDecomp}), we~have \deleted{$\hat{h}_T(\x_\A,\x_\B,\omega)=\e{\hat{g}_\epsilon(\x_\A,\x_\B,\omega)}, \qquad \epsilon = T\norm{\V}.$}

\begin{equation}\label{htge}
\hat{h}_T(\x_\A,\x_\B,\omega)\propto\hat{g}_\epsilon(\x_\A,\x_\B,\omega), \qquad \epsilon = 2T\norm{\V}.
\end{equation}

Furthermore, similarly to standard averaging via the stacking of recordings as in \mbox{Equation~(\ref{AverageAmbient})}, the structure of the random windowing averaging kernel determines the quality of the retrieval of the  interferometric phase estimate from $\hat{h}_T(\x_\A,\x_\B,\omega)$. Note, in particular, that the random windowing averaging kernel $\big\langle \chi\chi\big\rangle_T(\x',\x'')$ is a symmetric function of the locations $\x', \x''\in S$, and is concentrated in the neighborhood of $\x'=\x''$. Note also that the source speed $\norm{\V}$ affects the properties of the random windowing averaging kernel. As suggested by the results illustrated in Figure~\ref{RandomWindFig} and confirmed by analytical estimates for a typical setup in Section~\ref{UniformtnExample}, the random windowing procedure has an analogous effect to the $\epsilon$-localization procedure in decomposition (\ref{meanhDecomp}) in the sense that interactions between incongruous source locations are implicitly restricted to within the time/space window of the recordings  before cross-correlating.  Hence, the contribution of the acausal crosstalk is mitigated. Moreover, if sufficient data are available, one can additionally apply the standard stacking operation to $\hat{h}_T(\x_\A,\x_\B,\omega)$ which yields the general retrieval

\end{paracol}
\nointerlineskip
\begin{align}\label{windowedh2means}
\e{\hat{h}_T(\x_\A,\x_\B,\omega)} & =\frac{2}{\rho c}\int_S\int_S   \e{\hat{F}^*(\x',\omega)\hat{F}(\x'',\omega)}\e{\,\chi_{\left\{\norm{\x'-\x_n}\leq T\norm{\V}\right\}}(\x')\,\chi_{\left\{\norm{\x''-\x_n}\leq T\norm{\V}\right\}}(\x'')}_T\notag\\[.2cm]
&\hspace{2cm}\times\,\hat{G}^*(\x,\x',\omega)\hat{G}(\x,\x'',\omega)\,d\x'd\x''\,,
\end{align}
\begin{paracol}{2}
\switchcolumn

\noindent
since the stacking $\langle \,\cdot\,\rangle$ and the window randomization $\langle \,\cdot\,\rangle_T$ are independent by design, and therefore they commute.  Importantly, the form of (\ref{windowedh2means}) implies that the random window average and the stacking average play, in principle, a similar role in the elimination of the crosstalk contributions in the ambient noise retrieval,  in the spirit of the $\epsilon$-diagonal decomposition (\ref{hDecomp}). For $T\ll1$ the random windowing kernel $\big\langle \chi\chi\big\rangle_T$ will concentrate the integrand in (\ref{windowedh2means}) to within  the $\mathcal{O}(T\norm{\V})$ neighborhood of $\x' = \x''$. A similar effect is achieved if enough data are available, and the noise sources  are such that the stacking average kernel $\big\langle \hat F^*\hat F\big\rangle$ localizes around $\x' = \x''$ (see Section~\ref{stacking_sec} and Figure~\ref{KernelMeanFig}). In principle, both operations can be applied concurrently. However, in contrast to the stacking average, the use of random windowing averaging relies on a single recording provided that a suitable value $T$ of the time window can be determined; derivation of such a  procedure described in Sections \ref{hTEstimation} and \ref{optT} is preceded by a simple example which  is aimed at elucidating the main properties of the random windowing average.

\subsection{Example: Uniformly Distributed Window Locations}\label{UniformtnExample}

Consider a default situation in which the temporal window location  $t_n$ is  uniformly distributed within the interval $D = \big[t_0-T,\;t_0 + T\big]$. We then have 

\begin{equation}
f_T(t_n) = \frac{1}{2T},  \qquad \big\langle t_n\big\rangle_T=t_0, \qquad  \textrm{Var}(t_n)  = \big\langle (t_n-t_0)^2\big\rangle_T= T^2/3.
\end{equation}

Then,  if we define the spatial source location at time $t_n$ as $\x_n = t_n\V$, we have   \mbox{$\big\langle \x_n\big\rangle_T = \V t_0 = \x_0$}. Applying the operator $\langle\,\cdot\,\rangle_T$ with the uniform density $f_T$ to the product of the indicator functions  in Equation~\deleted{\eqref{windowedhmean}}\added{\eqref{chichi}} corresponds to the  convolution of two boxcar functions of constant height one and  width $T\norm{\V}$, parameterized in terms of the source location discrepancy $|\x'-\x''|$, $\x',\x''\in S$. 

 A standard calculation shows that applying operator $\langle\,\cdot\,\rangle_T$ to the product of indicator functions in Equation~(\ref{windowedh2means}) yields a symmetric kernel in the random windowing averaging of the~form  

\end{paracol}
\nointerlineskip
\begin{align}\label{RandWindKernelMean}
\e{\,\chi_{\left\{\norm{\x'-\x_n}\leq T\norm{\V}\right\}}(\x')\,\chi_{\left\{\norm{\x''-\x_n}\leq T\norm{\V}\right\}}(\x'')}_T &= \big(\,\norm{\x'-\x''}+T\norm{\V}\,\big)\,\Theta \big(\,\norm{\x'-\x''}+T\norm{\V}\,\big)\notag\\[.2cm]
&\phantom{=}+\big(\,\norm{\x'-\x''}-T\norm{\V}\,\big)\,\Theta\big(\,\norm{\x'-\x''}-T\norm{\V}\,\big)\notag\\[.3cm]
&\phantom{=}-2\norm{\x'-\x''}\,\Theta\big(\,\norm{\x'-\x''}\,\big)\,,
\end{align}
\begin{paracol}{2}
\switchcolumn

\noindent
which represents the tent map supported on the interval $ -2T|\V|\leqslant \x'-\x''\leqslant 2T|\V|$ and centered at $\x'-\x''=0$, where $\Theta$ denotes the Heaviside function. An example of Equation~\eqref{RandWindKernelMean} is plotted in the right inset in Figure~\ref{RandWindKernelMeanFig}, where the kernel \eqref{RandWindKernelMean} is plotted in blue, and its approximation obtained via direct simulations is indicated by the dotted black~line.  

It is informative to compare the kernel $\big\langle \chi\chi\big\rangle_T$ of the random windowing average  in Equation~(\ref{RandWindKernelMean})  illustrated in Figure~\ref{RandWindKernelMeanFig} with the stacking average kernel  $\big\langle\hat{F}^*\hat{F}\big\rangle$ in Equation~(\ref{stack_av}) which is illustrated in Figure~\ref{KernelMeanFig}. Recall that both these kernels are present in the retrieval (\ref{windowedh2means}) and that they play, in principle, a similar role aimed at reducing the contribution from the off-diagonal, crosstalk terms into the retrieval. 
One can see that the random windowing procedure is akin to extraction of  the $\epsilon$-diagonal term  $\hat{g}_\epsilon$ in the decomposition  \eqref{hDecomp}, in the sense that for any pair of sources such that  $\norm{\x'-\x''}>2T\norm{\V}$  the random windowing kernel mutes the corresponding acausal contributions. For  uniformly distributed locations of the random windows around $t_0$, only the contributions satisfying $\norm{\x'-\x''}\leqslant 2T\norm{\V}$ are included in the resulting retrieval (\ref{windowedh2means}). Moreover, the analytical expression \eqref{RandWindKernelMean}  implies  that the width of the diagonal band induced by the random windowing procedure is directly proportional to the source speed $\norm{\V}$. This suggests that a progressively shorter time window size $T$ should be chosen for increasing train speeds to mitigate acausal arrivals associated with the crosstalk terms  in the ambient noise retrieval (\ref{hDecomp}).  

\subsection{Estimation of the Retrieval $ \hat{h}_T( \x_ \A, \x_ \B, \omega)$ for a Given Time Window Size $T$}\label{hTEstimation}
In this section, we describe a procedure for the estimation of the random windowing average $\langle\,\cdot\,\rangle_T$ with some fixed time window length $T>0$. A  systematic way of choosing the optimal time window size~$T_{opt}$ for any given single recording by means of estimating  the acausal energy present in each averaged retrieval $\hat{h}_{T}(\x_\A,\x_\B,\omega)$ in~Equation~(\ref{windowedhmean}) is discussed in {Section} 
 \ref{optT}.

\begin{figure}[H]
\includegraphics[width=0.74\textwidth]{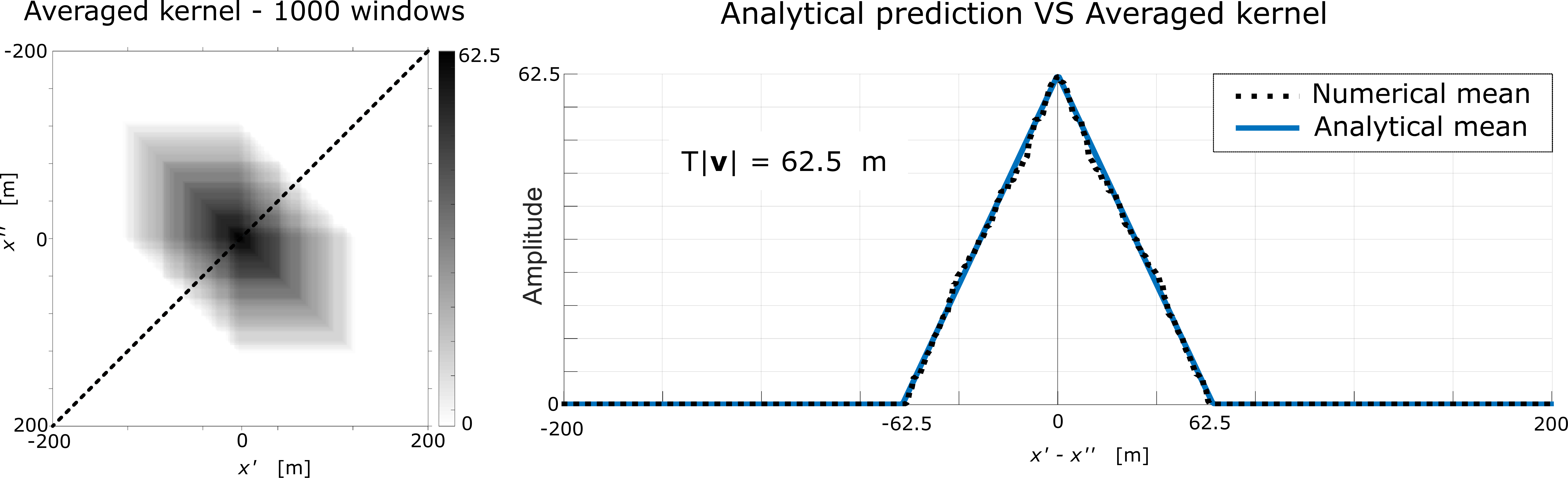}
\caption{{The structure} 
 of the kernel $\langle \chi\chi\rangle_T$ in Equation~(\ref{chichi}) of random windowing averaging (\ref{windowedhmean}) based on the uniform distribution of window locations. The kernel is given by a tent map (\ref{RandWindKernelMean})  with height $T\norm{\V}$ and width $2T\norm{\V}$, centered at $\x'=\x''$. The right panel shows the kernel parameterized in terms of the distance from the source locations $x'$ (for receiver $\x_\A$) and $x''$ (for receiver $\x_\B$) to the location of the source boundary with the receiver line \added{(see the parametrization used in Figure~\ref{KernelMeanFig})}; the analytical solution \eqref{RandWindKernelMean} is shown in blue and its approximation within the anti-diagonal section of the random windowing  kernel (see the left panel) obtained via direct simulations is indicated by the dotted black line.
The left panel shows the numerical approximation of the random window kernel based on 1000 windows for a single source train recording travelling at speed $\norm{\V} = 25$ m/s (90 km/h), with \mbox{$T= 2.5$ s}. Compare the structure of the above random windowing kernel $\langle \chi\chi\rangle_T$ with the stacking average kernel  $\langle\hat{F}^*\hat{F}\rangle$ illustrated in Figure~\ref{KernelMeanFig}. Both kernels are present in the general retrieval (\ref{windowedh2means}).}\label{RandWindKernelMeanFig}
\end{figure}

As outlined in Section \ref{rnd_wind_fr} and illustrated in Section \ref{UniformtnExample} the random windowing procedure will lead to an  improved phase retrieval through damping the contributions from the cross-correlation terms in Equation (\ref{windowedh2means}) provided that a suitable temporal window length $T>0$ is identified, so that the random  windowing kernel $\big\langle \chi\chi\big\rangle_T$ in Equation~\eqref{windowedhmean} effectively enforces  the recovery of the diagonal band term $\hat g_\epsilon$  in the decomposition~\eqref{hDecomp}.
In the case of standard stacking the choice of the spatial correlation length cut-off  $\epsilon$ can be inferred from the form of stacking average  kernel $\langle \hat F^*\hat F\rangle$ in~Equation~(\ref{stack_av}), assuming that the kernel is sufficiently localized. Clearly, the structure of the stacking average kernel cannot be determined from a single short recording or in situations when the stacking average kernel can be estimated but it is not  sufficiently localized (e.g., as in the example presented in Figure~\ref{DiagDecompDependentFig}).  On the other hand,  the random windowing average with an appropriately chosen $T$ can be applied to obtain the retrieval $\hat{h}_T(\x_\A,\x_\B,\omega)$ in the most challenging case where only a single (possibly short) recording of a source train is available. Moreover, as mentioned earlier and expressed via~Equation~(\ref{windowedh2means}), if enough recorded data are available, one can combine both averaging techniques to recover $\e{\hat{h}_T(\x_\A,\x_\B,\omega)}$. It is clear that the use of such a doubly averaged approach is more versatile than using only standard stacking in the absence of knowledge about the structure of the noise sources, and it will lead to an  improved retrieval, as long as at least one of the kernels is sufficiently concentrated around the set $\x'=\x''$, $\x',\x''\in S$.

The first step in the procedure for the empirical approximation of the random windowing average $\langle\,\cdot\,\rangle_T$ with some fixed time window length $T$ is to identify the approximate time in the recorded traces when the source train is collinear with the receiver pair (see for instance \cite{dales2020virtual} for one example of a method that achieves this using plane-wave beamforming). We chose a configuration in which the line through the receiver pair is perpendicular to the source boundary, as depicted in Figure 2. In such a geometry the Fresnel zone of the receiver pair is symmetric about the stationary phase point (see \cite{van2015retrieving}) on the source boundary but this is not essential to the random windowing method. To see this note that when the instantaneous source location is collinear with the receiver pair the signal recorded at the two receivers is generated at $\x' = \x''$, and the frequency recorded at each receiver, as well as the Doppler bias, are the same ($\theta_{\x_\A}(t') = \theta_{\x_\B}(t')$ in Equation~\eqref{doppler}). Thus, in this instantaneous configuration the effect of spurious contributions to the retrieval is minimized.  Finally, note that regardless of whether or not the noise source is in motion, the source location contained in the line through the receiver pair coincides with the stationary phase point (on the source boundary) which is always contained in the Fresnel zone of the receiver pair, unless the line through the receivers is parallel to the source boundary.  

The time tag at which the source is collinear with the receivers is denoted by  $t_0$ (see left panels of Figure \ref{RandomWindFig}) and the corresponding source location is denoted by  $\x_0 = \V t_0$. Please note that while $t_0$ is a known time tag, explicit knowledge of $\x_0$ is not necessary in practice. However, this notation will be useful when describing our method in the steps below. We note further that $\x_0$ will by construction be in the middle of the Fresnel zone of the receiver pair. Note also that for a given recording of a single source moving in one direction, the time tag $t_0$ remains fixed throughout the procedure.

\medskip
To estimate $\hat{h}_T(\x_\A,\x_\B,\omega)$ consider the sequence $n=1,\dots, N$, $N\gg1$, and  perform the following steps for the given pair of recordings $p(\x_\A,t)$ and $p(\x_\B,t)$:
\begin{enumerate}
\item[(i)] \emph{Generate a set of random window locations $\{t_n\}_{n=1}^N$} of  \deleted{size} \added{a fixed duration} $T$ by sampling  from the distribution associated with $f_T$ in Equation~(\ref{operatorT}), and  such that $\norm{t_n - t_0} < T$. There is flexibility in the choice of the probability distribution $f_T$ of $t_n$ provided that $\langle \,t_{\scaleobj{.8}{(\cdot)}}\,\rangle_T = t_0$ \added{and that $f_T$ is non-zero only on a bounded interval}.

\item[(ii)] \emph{For each $t_n$  extract the windowed recordings $p(\x_\A,t; T,t_n)$ and $p(\x_\B,t;T,t_n)$   of duration  $T$ and centered at $t_n$ from  $p(\x_\A,t)$ and $p(\x_\B,t)$}. This procedure is illustrated in the left panels of Figure \ref{RandomWindFig} and is performed on the recorded time series for each receiver. 

\item[(iii)] \textls[-25]{\emph{Perform interferometry using the windowed data} and derive a realization of $\hat{h}(\x_\A,\x_\B,\omega\,;\,T,t_n)$} according to Equation~(\ref{windowedhmean}), using the Fourier transforms of the windowed recordings from (ii).  
\item[(iv)] \emph{Repeat Steps (i)--(iii) for a large number $N$ of random window locations $t_n$} and take the arithmetic mean of $\{\hat{h}(\x_\A,\x_\B,\omega\,;\,T,t_n)\}_{n=1}^N$ in order to approximate the random windowing kernel $\big\langle \chi\chi\big\rangle_T$  given by Equation~(\ref{chichi}), and  to obtain the ambient noise retrieval $\hat{h}_T(\x_\A,\x_\B,\omega)$ in Equation~(\ref{windowedhmean}).
\end{enumerate}

Due to the law of large numbers, one has 

\begin{equation}\label{empiricalMean}
    \frac{1}{N}\sum_{n=1}^N \hat{h}(\x_\A,\x_\B,\omega\,;\,T,t_n)\underset{N\rightarrow\infty}{\longrightarrow}     \hat{h}_T(\x_\A,\x_\B,\omega),
\end{equation} 
\noindent
and the empirical mean of the sample $\{\hat{h}(\x_\A,\x_\B,\omega\,;\,T,t_n)\}_{n=1}^N$ converges, in the limit of  $N\rightarrow \infty$, to the expectation of $\hat{h}(\x_\A,\x_\B,\omega\,;\,T,\tau)$ as in Equation \eqref{operatorT} provided that $\tau$ has a finite variance (i.e., $\int_\mathbb{R}\tau^2f_T(\tau)d\tau<\infty$) and the locations $\{t_n\}_{n=1}^N$ are sampled independently from $f_T$. Finally, note that the steps outlined above do not require knowledge of the source speed $\V$, or for the source to be in motion, as long as it is possible to identify a suitable recorded time $t_0$ corresponding to the emission of energy from a source location collinear with the receiver pair.

\subsection{Procedure for Choosing the Optimal Time Window Size $T_{Opt}$ in the Random-Window Averaged Retrieval  $ \hat{h}_{T_{opt}}( \x_ \A, \x_ \B, \omega)$}\label{optT}

The procedure described in the previous section can be applied to any time window of size $0<T<\infty$. However, similarly to the general $\epsilon$-diagonal decomposition defined in Equation~(\ref{hDecomp}), a successful approximation of the phase of the inter-receiver Green's function depends on the choice of an optimal support of the random windowing kernel $\e{\chi\chi}_T$ in Equation~(\ref{chichi}). 
The steps outlined below provide a systematic way to choose an optimal time window size $T_{opt}$ for any given single recording by estimating  the acausal energy present in each averaged retrieval $\hat{h}_{T}(\x_\A,\x_\B,\omega)$ in~Equation~(\ref{windowedhmean}).

To determine $T_{opt}$ for a given pair of recordings $p(\x_\A,t)$ and $p(\x_\B,t)$, consider first a sufficiently large range of test values $T\in(T_{\min},T_{\max})$; this range will necessarily be limited by the length of the recordings available. Steps to find the optimal window size $T_\text{opt}$ for any fixed pair of recordings $p(\x_\A,t)$ and $p(\x_\B,t)$ are:
\begin{enumerate}
\item[(1)] Choose an ordered  collection of time window sizes  $\{T^{(j)}\}_{j=1}^J$ contained in the test range $T\in(T_{\min},T_{\max})$.
\item[(2)] Estimate the averaged retrieval $\hat{h}_{T^{(j)}}(\x_\A,\x_\B,\omega)$ for each time window size $\{T^{(j)}\}_{j=1}^J$  using the procedure described in Section \ref{hTEstimation}.
\item[(3)] Transform each estimate $\big\{\hat{h}_{T^{(j)}}(\x_\A,\x_\B,\omega)\big\}_{j=1}^J$ to the time domain to obtain the corresponding waveforms $\big\{h_{T^{(j)}}(\x_\A,\x_\B,t)\big\}_{j=1}^J$.
\item[(4)]  Order the waveforms $h_{T^{(j)}}(\x_\A,\x_\B,t)$ according to the  time window size $T^{(j)}$ (see bottom panel of Figure \ref{SpuriousEnergyAndStackFig}).
\item[(5)] Use the ordered family of waveforms  $h_{T^{(j)}}(\x_\A,\x_\B,t)$  to identify the time tag of the {\it feature of interest}, which is normally the first-arriving dominant wave, denoted by $t_I$. In Figure \ref{SpuriousEnergyAndStackFig}, $t_I = 0.4$ s and is highlighted in orange.
\item[(6)] Calculate the waveform energy of each trace $h_{T^{(j)}}(\x_\A,\x_\B,t)$ from time $t=0$ to a time slightly before $t_I$. This calculation quantifies the acausal energy present in each of the traces as a function of ${T^{(j)}}$. (In Figure \ref{SpuriousEnergyAndStackFig} the waveform energy is calculated for each trace between $t = 0$  and $t = 0.3$\,s. Please note that this curve would be different for different~recordings.)
\item[(7)] Compute  the spurious waveform energy calculated in Equation~(4) as a function of the time window sizes ${T^{(j)}}$ (see the top panel of Figure \ref{SpuriousEnergyAndStackFig}). 
\item[(8)] {Minimize the spurious energy with respect to ${T^{(j)}}$; the minimizer is denoted by $T_{opt}$.} 
\item[(9)] {Use the optimal window size $T_{opt}$ determined in Equation~(8) for the interferometric retrieval by following the steps described in Section \ref{hTEstimation}.} 
Compute the final/optimal ambient noise retrieval of the inter-receiver Greens function $\hat{h}_{T_{\text{opt}}}(\x_\A,\x_\B,\omega)$ given by Equation~(\ref{windowedhmean}) or its time-domain counterpart $h_{T_{\text{opt}}}(\x_\A,\x_\B,t)$.
\end{enumerate}
\vspace{-9pt}
\begin{figure}[H]
\includegraphics[width=0.47\textwidth]{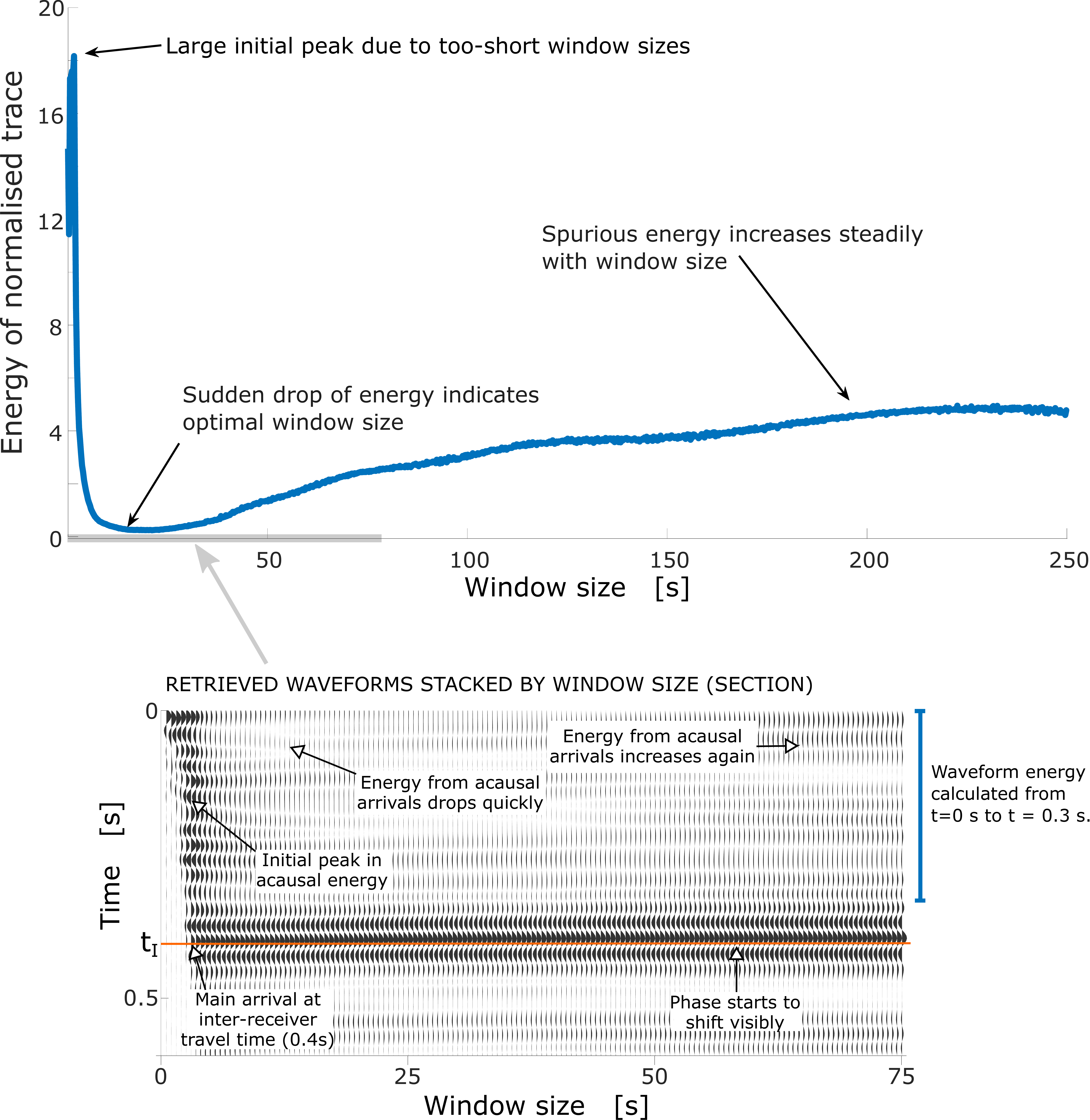}
\caption{Calculation of acausal energy curve for a recording of a single moving source. Values of $T$ tested ranged between 0.25 s and 250 s. The bottom panel shows the waveforms corresponding to the average random window retrieval for $T$ only in the range between 0.25 s and 75 s, for clarity. The waveforms are arranged by window size $T$. The arrival of interest is highlighted in orange at time $t_I=0.4$ s. The top panel shows the acausal energy curve in blue. The waveform energy was calculated for each waveform in the range from t = 0 to t = 0.3 s.  The acausal energy curve exhibits a clear initial peak for small values of $T$, followed by a sharp drop and then a slow increase with $T$. The minimum is achieved at $T_{\text{opt}}\approx 15$ s.}\label{SpuriousEnergyAndStackFig}
\end{figure}

Please note that the above procedure can easily be automated. For multiple source train recordings Steps (1)--(8) can be performed for each recording, noting that the optimal window size $T_{\text{opt}}$ will, in general, be different for each recording. Then, the standard stacking average can be applied to estimate $\langle\hat{h}_{T_{\text{opt}}}(\x_\A,\x_\B,\omega)\rangle$ and thus potentially improve the retrieval even further.

 \subsection{Results}\label{rw_res}
We now present some examples of the interferometric retrieval obtained via the random windowing method described in Section \ref{hTEstimation} which is carried out for the optimal time window size $T_{opt}$ determined according to the procedure described in Section \ref{optT}. \deleted{Although we have used a receiver pair oriented perpendicularly to the source-line, results are of essentially identical quality for orientations up to 30 degrees with respect to the source-line and are not included in a separate figure for the sake of brevity.}\added{We present results for a receiver pair that is perpendicularly oriented with respect to the source-line, as well as results for a receiver pair that is oriented at 30 degrees with respect to the source-line. The results are of essentially identical quality.}

The distribution of the window locations is chosen to be uniform, as in the example of Section~\ref{UniformtnExample}. The random windowing method was applied to the two configurations discussed in Section \ref{stacking_sec}. These configurations corresponded to (i) the case where the stacking average kernel $\e{\hat F^*\hat F}$ of the ambient noise is localized,  as in Figure \ref{DiagDecompFig} (left panels of \mbox{Figure~\ref{PhaseComparisonsFig}}), and (ii) the challenging scenario when $\e{\hat F^*\hat F}$ is not localized  and the spurious crosstalk arrivals do not decay (right panels of Figure~\ref{PhaseComparisonsFig}). The random windowing method leading to $\hat{h}_{T_{opt}}(\x_\A,\x_\B,\omega)$ is applied in all cases to a single short recording of a train of sources, and the stacking average is not applied to $\hat{h}_{T_{opt}}(\x_\A,\x_\B,\omega)$ unless otherwise stated. Figure \ref{PhaseComparisonsFig} shows phase estimates produced by the random windowing average in blue and the phase estimate based on $\hat{h}(\x_\A,\x_\B,\omega)$ in light gray for comparison. \added{Receivers oriented perpendicularly with respect to the source-line were used in the top row, whereas the bottom row of Figure \ref{PhaseComparisonsFig} shows the results for a receiver pair oriented at 30 degrees with respect to the source-line.} The phase estimated from individually recorded  sources, which allows us to calculate the reference retrieval using Equation \eqref{DirectInt}, is indicated in dashed black. We also show the phase of the inter-receiver Green's function retrieved through the standard stacking average over six source train recordings restricted to a one-minute-long window in a similar fashion to \cite{pinzon2021humming}. \added{For illustration, Figure \ref{WaveComparisonsFig} shows the waveforms corresponding to the phases depicted in the top-right panel of Figure \ref{PhaseComparisonsFig} (i.e., a correlated noise source with correlated phases, and perpendicularly oriented receivers) with normalized amplitudes. As expected, the waveform corresponding to the random windowing estimate (in blue) exhibits the best agreement with the reference waveform (in black dashes).} The phases estimated from the random windowing scheme agree well with the reference phase in all cases, showing that our method is effective in mitigating the undesired effects of correlation in the noise sources in ambient noise interferometry, especially where moving sources such as trains or traffic on a highway are concerned.

We note that  the workflow outlined in Section \ref{optT} and illustrated above is not restricted to ambient noise interferometry from correlated noise.  In fact, this methodology  can be applied to short recordings of uncorrelated noise to find an optimal window size that minimizes spurious arrivals. \added{In this scenario, the choice of $t_0$ as described in \mbox{Section~\ref{hTEstimation}} is more flexible as the source no longer exhibits coherent spatiotemporal correlation or a Doppler spread differential, as would be the case for a source in motion. Therefore, the choice of time tag $t_0$ may be arbitrary and the random windowing method can still be applied as described in Section \ref{optT}.} Figure \ref{RandomNoisePhasesFig} shows the phase estimated by the random windowing method when the ambient noise recording is uncorrelated. \added{The random windowing method was applied to a short (5 min) recording of uncorrelated ambient noise, producing satisfactory results.} Our approach has the potential to speed up convergence to the inter-receiver phase when applied to uncorrelated ambient {noise.} 

\clearpage
\end{paracol}
\nointerlineskip
\begin{figure}[H]
\widefigure
\includegraphics[width=0.9\textwidth]{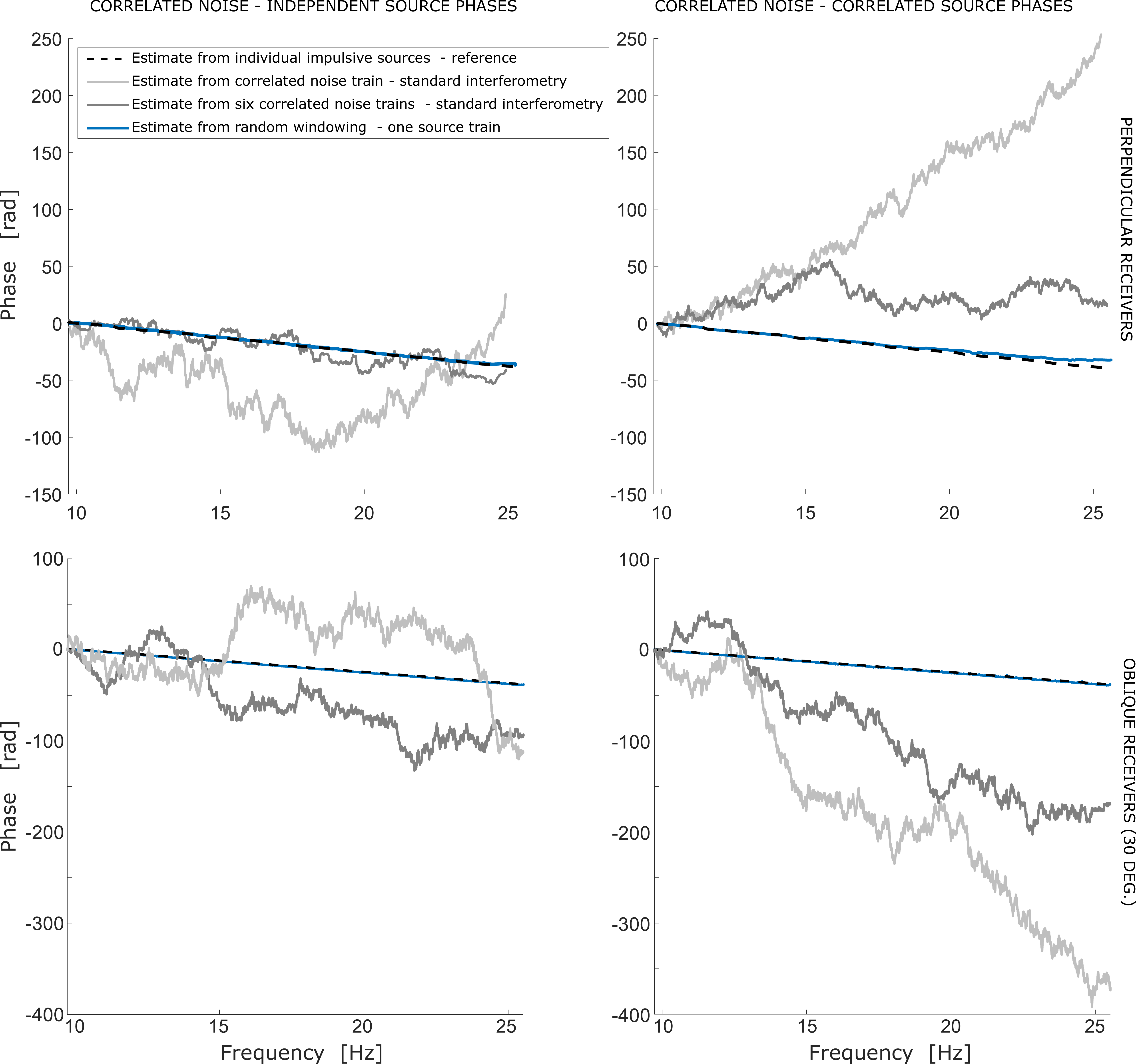}
\caption{\added{{Phase estimates} 
 obtained from the random windowing scheme applied to a short (5 min) recording of a single source using receivers that are perpendicularly oriented with respect to the source-line (\textbf{top}) and oriented at 30 degrees with respect to the source-line (\textbf{bottom}). The left column shows the results for correlated noise where source phases at different frequencies are independent (as in Figure~\ref{KernelMeanFig}). The phase estimated from discrete sources, in black dashes, is used for reference. The phases produced by the random windowing scheme are shown in blue and exhibit very good agreement with the reference phase. The phase of the retrieval using the standard method and no stacking is shown in light gray, and the phase of six stacked sources is shown in dark gray for comparison. The same phase estimates are shown in the right column for correlated noise with correlated source phases (as shown in Figure \ref{DiagDecompDependentFig}). The phase estimates from the random windowing scheme, in blue, are better than the retrieval obtained using the standard method, with slightly reduced accuracy at higher frequencies.}}\label{PhaseComparisonsFig}
\end{figure}
\begin{paracol}{2}
\switchcolumn

\begin{figure}[H]
\includegraphics[width=0.75\textwidth]{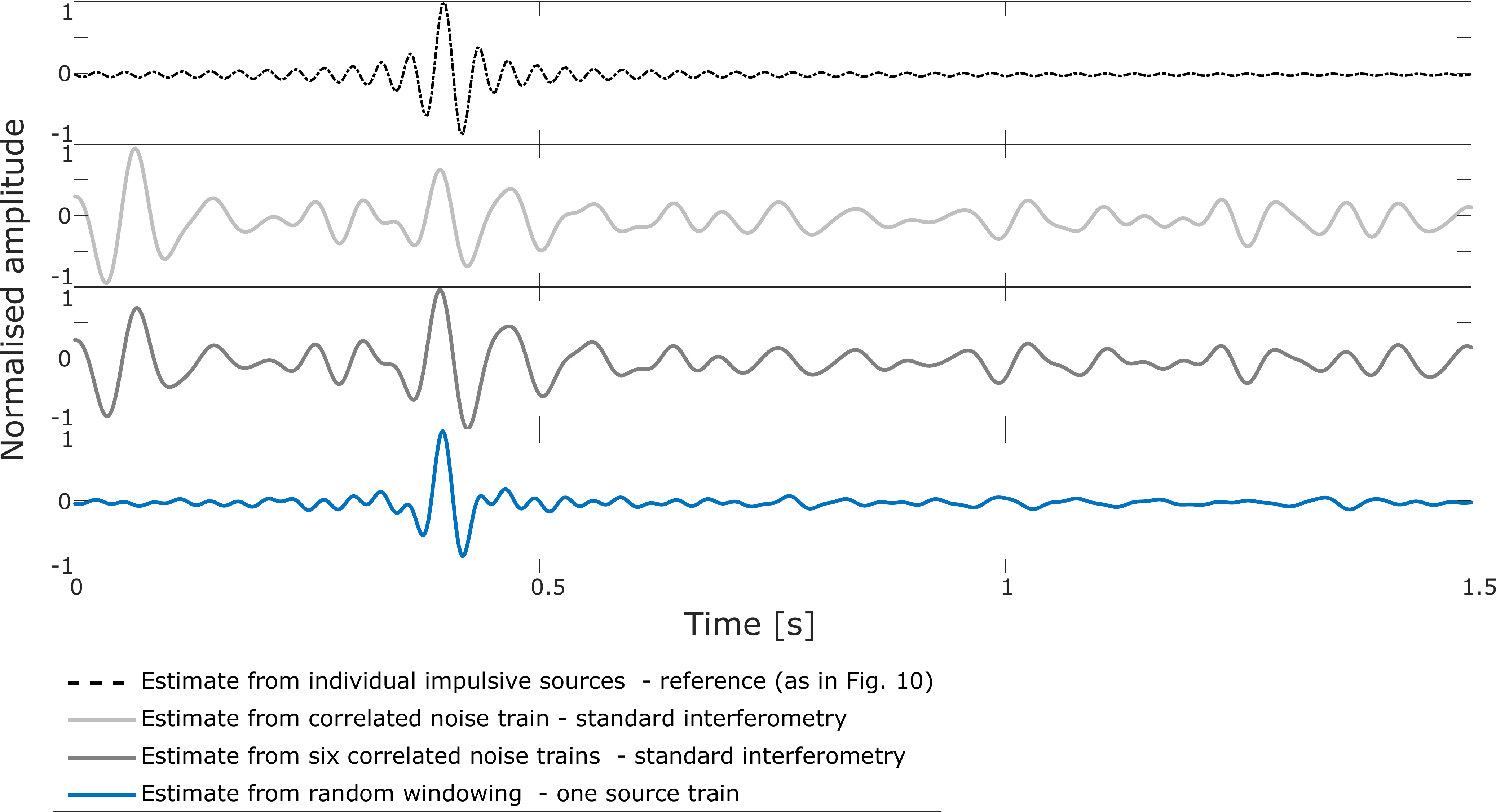}
\caption{\added{{Waveforms corresponding} 
 to the estimated phases shown in the top-right panel of \mbox{Figure \ref{PhaseComparisonsFig}}, for correlated noise with source phases recorded at a perpendicularly oriented pair of receivers 400 m apart, with the medium speed set at 1000 m/s. For reference, the estimate calculated from individual point sources is depicted in black dashes (\textbf{top}), exhibiting the inter-receiver arrival at 0.4 s. The estimate produced by standard interferometry without stacking is shown in light gray (\textbf{second from top}). The waveform resulting from standard stacking of six estimates is depicted in dark gray (\textbf{third from top}). The estimate obtained from the random windowing method applied to a short recording of a single source is shown in blue (\textbf{bottom}).}}\label{WaveComparisonsFig}
\end{figure}

\begin{figure}[H]
\includegraphics[width=0.75\textwidth]{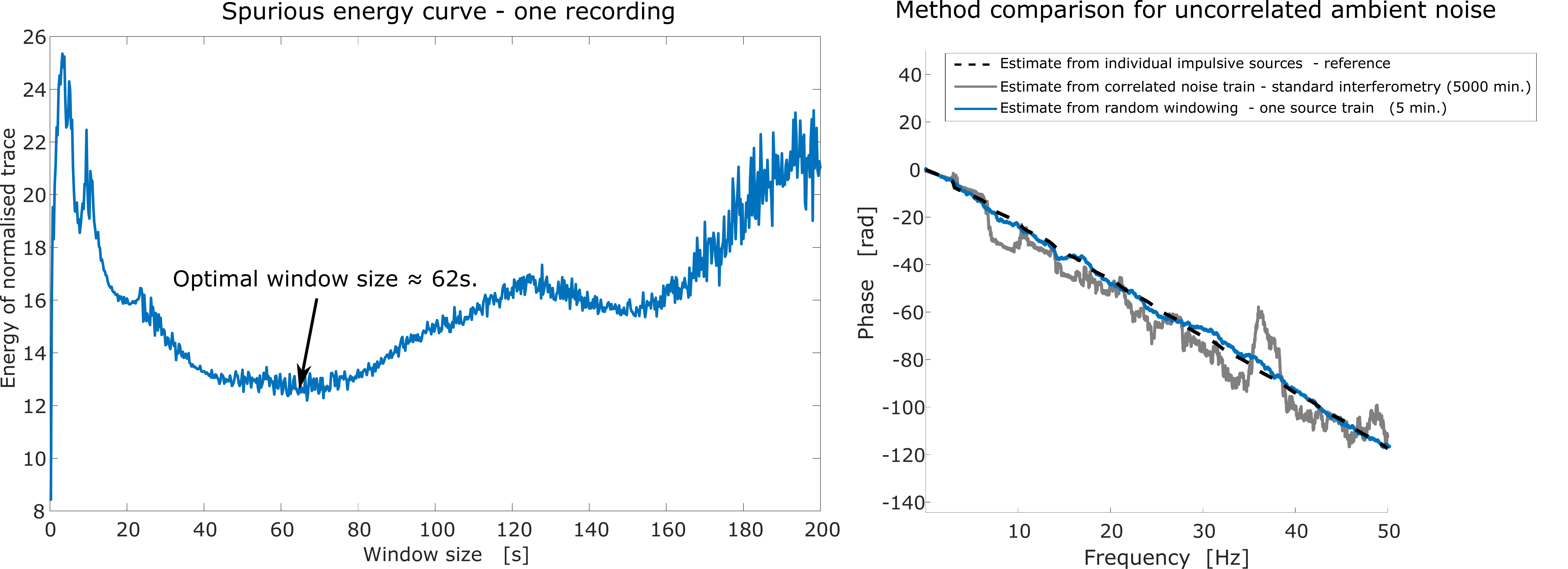}
\caption{{Random} 
 windowing method applied to uncorrelated ambient noise. The plot to the left shows the acausal energy curve estimated from one ambient noise recording. The plot to the right shows the resulting estimated phase in blue, which matches the reference phase in black dashes well. For comparison, the phase estimated from stacking 1000 simulated recordings of ambient noise five minutes long is shown in dark gray. We can see the stacking in this case will converge to the correct phase estimate as expected for truly uncorrelated ambient noise, but only after a very large number of ambient noise recordings are processed and stacked.}\label{RandomNoisePhasesFig}
\end{figure}

Finally, Figure \ref{OptimalWindowVSDistance} shows the mean optimal window sizes $T_{\text{opt}}$ as a function of the distance from the closest receiver to the source-line; in the illustrated case the  inter-receiver distance is  400\,m, and three different source speeds are considered. Recall that optimal window sizes $T_{\text{opt}}$ are, to some extent, recording-dependent due to the randomness in recorded signals, so their value will vary for different recordings  even if the receiver geometry remains fixed; hence, 30 samples for each considered closest receiver distance and  velocity were calculated and averaged to produce the curves in Figure \ref{OptimalWindowVSDistance}; the fluctuations in the curves diminish with the number of samples used. The relationship illustrated in \mbox{Figure \ref{OptimalWindowVSDistance}} indicates that  for a fixed source speed and a fixed  inter-receiver distance, the optimal window size  decreases with the distance of the receivers to the source boundary. Moreover, the optimal window size decreases for  increasing  source speeds, which agrees with physical intuition and with the form of the random windowing average \mbox{kernel~(\ref{RandWindKernelMean})} illustrated in Figure~\ref{RandWindKernelMeanFig}. Furthermore, as can be deduced from \mbox{Equations~\eqref{doppler} and~\eqref{tEmission}}, the spatial discrepancy of the implicit source locations (i.e., the difference $x' - x''$) will be smaller for decreasing inter-receiver distances. Consequently, the error due to the presence of the residual crosstalk terms in the retrieval decreases with decreasing inter-receiver distance, while larger inter-receiver distances require smaller values of the time window size $T_{\text{opt}}$ to mitigate this error. 
\begin{figure}[H]
\includegraphics[width=0.65\textwidth]{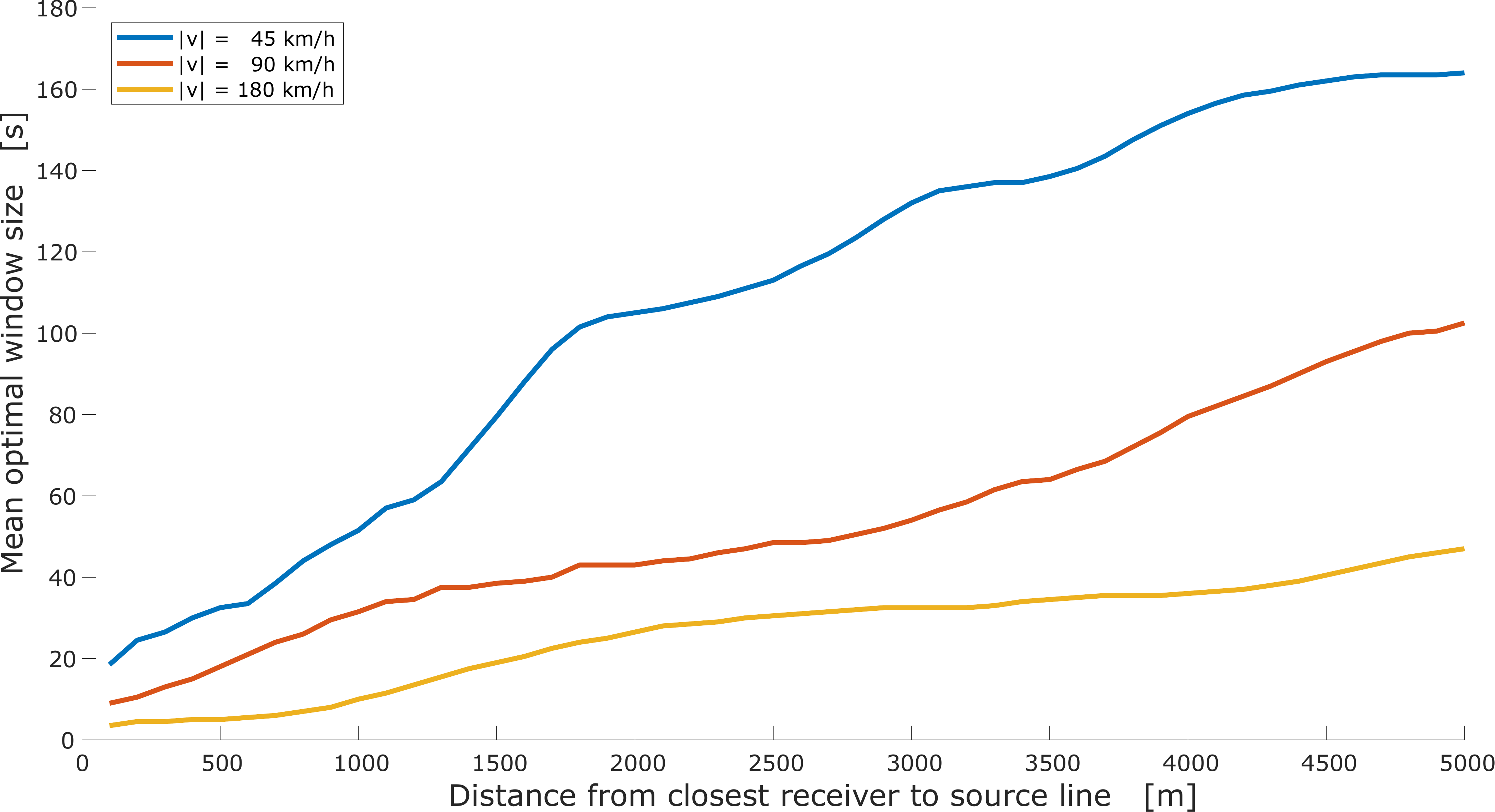}
\caption{Estimated optimal window size for the random windowing scheme, calculated for trains travelling at 45 km/h, 90 km/h and 180 km/h, in terms of the distance between the receivers and the road (using the closest receiver as reference) and averaged over 30 samples each. The receivers were 400 m apart from each other.}\label{OptimalWindowVSDistance}
\end{figure}

\section{Discussion}
The ambient noise interferometric retrieval based on the random windowing average (Sections~\ref{hTEstimation} and \ref{optT}) is capable of accurately estimating the phase of the inter-receiver Green's function based on a short pair of recordings  of  a correlated noise source (Section~\ref{rw_res}). As pointed out  in Equation~(\ref{htge}), the random windowing average  allows one to extract the $\epsilon$-diagonal term in decomposition (\ref{hDecomp}) of the ambient noise retrieval $\hat{h}(\x_\A,\x_\B,\omega)$  without the explicit knowledge of the stacking average kernel  $\e{\hat{F}^*\hat{F}}$ which need not be localized.   
For a given pair of recordings this new technique requires one to pre-process the spurious energy to determine the optimal time window size  $T_{opt}$; then, the optimal spatial window length can be used in decomposition (\ref{hDecomp}) to retrieve $\hat g_\epsilon$ with the optimal cut-off length $\epsilon = T_{opt}\norm{\V}$.  This procedure, illustrated in Figure \ref{SpuriousEnergyAndStackFig}, can be automated in practical applications.
Based on the derivations and the discussion in Sections \ref{stacking_sec} and \ref{newtech}, the optimal value $T_{opt}$ of the temporal window size decreases with the inter-receiver distance and with the distance between the source-line and the receiver pair, since the Doppler shift and the spatial incongruence become increasingly important. Moreover, the optimal value of the temporal window size decreases for increasing speeds of the moving noise source, as indicated in Equation \eqref{RandWindKernelMean}. These trends are  consistent with physical intuition, as faster source speeds have the compound effect of increasing the spatial coverage (and hence the prevalence of the crosstalk contributions to the ambient noise retrieval), and increasing the importance of the Doppler shift for the retrieval.

The  random windowing approach can be used in conjunction with the standard stacking average procedure as discussed in Section \ref{newtech}. This is especially tractable when estimating inter-receiver Green's functions from a small ensemble of source train recordings, where the random windowing procedure can be applied to each individual train recording before performing standard stacking to optimize the extraction of causal information. In Ref.~\cite{pinzon2021humming}, the distance from the receiver array to the rail track on which trains provide source energy is of the order of~5--10 km. This large distance combined with the stationary phase zone approach likely mitigates spurious effects arising from the Doppler effect induced by the trains' motion, since such a receiver array configuration and their localization methodology implicitly restrict the recordings to regions of the source boundary on which the train is roughly collinear with the receiver pairs. In fact, the choice of the optimal time window size in \cite{pinzon2021humming} in the stacked  retrieval is consistent with our findings. Incorporating our methodology in such a setup could allow for the retrieval of inter-receiver estimates from near-field recordings (closer to the rail tracks) maintaining the optimality of extraction of coherent information from each recorded trace.

\added{Sabra~\cite{sabra2010influence} concludes that under some circumstances the Doppler effect does not have a significant effect on the interferometric estimate of the inter-receiver Green's function. However, that work explicitly assumes that the ambient noise field is uncorrelated in time. To achieve this the sources considered were randomly and uniformly distributed in space around the receiver pair. The situation where the ambient noise field is statistically correlated is not considered. Moreover, to mitigate the differential Doppler effect, a stationary phase approximation is performed throughout, and the far-field assumption is made. These approaches are consistent with our findings, as well as with the random windowing method presented in this work.}

\textls[-15]{In interferometric studies from traffic noise, such as the one performed by \mbox{Behm et al.~\cite{behm2013love}}}, wave velocities are retrieved from a linear array of receivers which crosses  the main road perpendicularly. In this setup, the choice of the time window sizes could be further informed by our proposed framework, possibly using consistently shorter time windows for receiver pairs which are closer to the road. This interesting study points out that longer observation periods do not necessarily lead to better results, which indicates that  correlations in the underlying ambient noise are too significant for the stacked-averaged retrieval to be appropriate,  analogously to the case illustrated in Figure \ref{DiagDecompDependentFig}. Thus, integrating the random windowing approach could be beneficial by providing an informed method for  selecting temporal window sizes so that the effect of spurious arrivals is minimized, or even averaging over different time window sizes.
 Liu et al.~\cite{liu2021retrievability} conclude, using physically motivated  ray-path arguments, that the crosstalk can be negligible when performing interferometry using seismic noise generated by high-speed trains in the case of direct, scattered and refraction waves. Our approach to  interferometry from correlated noise sources links the quality of the retrieved phase of the inter-receiver Green's function from direct waves explicitly to the correlation structure of the noise source, as well as the receiver array geometry (inter-receiver distance and distance to the source-line) while further providing a methodology to mitigate the spurious effects induced by any of these causes. In the common case of sources in motion, we established a link between the spurious energy in the retrieved trace and the speed of the source, and the proposed methodology has the potential to mitigate spurious effects induced by the Doppler spread for this type of source. We further note that the receivers in \cite{liu2021retrievability} are very close to the location of the source train, confirming that correlated noise interferometry is applicable in the near-field of seismic noise generated by road traffic and trains.

We constrained our study to estimating the direct wave arrivals in two dimensions, which loosely resembles the case of surface waves in the three-dimensional setup. Similar derivations could be carried out for body waves that have been critically refracted in three dimensions along horizontal interfaces in the subsurface and then refract back up to the surface. Our methodology  could be used to image and monitor the subsurface using these non-physical refracted arrivals, as suggested by King and Curtis~\cite{king2011velocity} and Brenguier et al.~\cite{brenguier2019train}, respectively.

Finally, our method is optimized when the location of the source in motion can be identified from a location tracker or from the acquired data. The latter approach has been shown to be possible for example in \cite{brenguier2019train,pinzon2021humming}, but may not be feasible in an urban environment with a more complicated road network and multiple simultaneous sources. In those situations, one could apply directional decomposition of the wavefield into constituent energy from its different individual sources prior to using these methods, similarly to the methods employed by Maran\`o et al. \cite{marano2012seismic}. Although we have focused on the application of interferometry from noise sources in motion, representative of most known physical sources that contribute to ambient noise, the random windowing method can be used regardless of the underlying cause of statistical correlation in the ambient noise field.

\section{Conclusions}
We developed a novel, robust and general  procedure for accurate retrieval of the phase of the inter-receiver Green's function from a short recording of a single source of ambient noise, without the need to rely on the commonly used stacking average  procedure, and thus avoiding the potential pitfalls of standard ambient noise interferometry. 
The random windowing approach provides a versatile framework for ambient noise interferometry, and it should  prove useful  either as a standalone technique when only short recordings of correlated noise are available or in situations where the standard stacking operation fails due to significant correlations in the recorded signal and/or insufficiently long recordings of the seismic noise. The random windowing method mitigates the spurious crosstalk effects introduced by the presence of correlations in the ambient noise, regardless of the cause of the correlation. It includes the common case of correlation caused by a spatially migrating train of sources, and in that case it also potentially mitigates the bias introduced by the Doppler effect. This approach is general in the sense that it improves the quality of the phase of the inter-receiver Green's function without detriment to retrieval obtained using the standard stacking average methodology, and it applies to both correlated and uncorrelated ambient noise in both near and far-field configurations.  Thus, this unified  framework  offers a novel workflow that can  be deployed  for robust estimation of inter-receiver phases from ambient noise interferometry, regardless of the characteristics of the ambient noise and, in general, using less data than are required for standard interferometry that uses  the stacking average {approach. } 


\vspace{6pt} 



\authorcontributions{Conceptualization, D.A.-G., A.C. and M.B.; methodology, D.A.-G., A.C. and M.B.; software, D.A.-G.; validation, D.A.-G., A.C. and M.B.; formal analysis, D.A.-G., A.C. and M.B.; investigation, D.A.-G., A.C. and M.B.; resources, D.A.-G., A.C. and M.B.; data curation, \mbox{D.A.-G.}; writing---original draft preparation, D.A.-G.; writing---review and editing, D.A.-G., A.C. and M.B.; visualization, D.A.-G.; supervision, A.C. and M.B.; project administration, D.A.-G., A.C. and M.B.; funding acquisition, D.A.-G. All authors have read and agreed to the published version of the~manuscript.}

\funding{The authors wish to thank the Mexican Institute of Petroleum (IMP) for their support of Daniella Ayala-Garcia through their PCTRES funding programme.}

\institutionalreview{ Not applicable.
}

\informedconsent{ Not applicable.  

}

\dataavailability{ No data reported. 
}

\added{\acknowledgments{The authors wish to thank the editor and two anonymous reviewers for their comments, which helped to improve the manuscript.}}

\conflictsofinterest{The authors declare no conflict of interest. The funders had no role in the design of the study; in the collection, analyses, or interpretation of data; in the writing of the manuscript, or in the decision to publish the~results.}

\appendixtitles{yes} 
\appendixstart
\appendix
\section{Analytical Derivation of Moving Point Source Model}\label{pDetails}
Throughout this appendix, bold variables denote vectors whereas regular variables denote scalar quantities. We calculate the wavefield generated by a monochromatic point source that moves with constant speed $\V $ in a two-dimensional homogeneous medium, emitting sound at a constant angular frequency $\omTilde$. This wavefield is denoted by $p_\omTilde(t,\x,\V)$  when measured at time $t$ and at a receiver location $\x$. We may also refer to the constant source frequency as $\fTilde$ when we wish to express it in Hertz, so that $\omTilde = 2\pi\fTilde\,$.
We characterize the moving point source through the function

\begin{equation}\label{fSource}
s_{\omTilde}(t,\x,\V)=e^{-i\omTilde t}\delta(\x-\mathbf{v}t)\,.
\end{equation}

It is assumed without loss of generality that at time $t=0$ the monochromatic source is located at the origin. It is further assumed that the medium density $\rho$ and sound speed $c$ are constant. Then the wavefield, denoted by $p_{\omTilde}$, satisfies the linear wave equation

\begin{equation}\label{pWave}
\frac{1}{c^2}\frac{\partial^2 p_{\omTilde}}{\partial t^2} -\nabla^2 p_{\omTilde} = \rho\,\frac{\partial s_{\omTilde}}{\partial t} \,\
\end{equation}
where we restrict ourselves to real solutions.
To solve Equation \eqref{pWave} we introduce the abstract complex potential $\phi_{\omTilde}^\mathcal{C}(t,\x,\V)$ of the particle velocity, so that

\begin{equation}\label{potential}
p_{\omTilde}=\rho\,\mathfrak{R}\left\{\frac{\partial\phi_{\omTilde}^\mathcal{C}}{\partial t}\right\}\,.
\end{equation}

This potential satisfies the wave equation

\begin{equation}\label{wavePhi}
\frac{1}{c^2}\frac{\partial^2 \phi_{\omTilde}^\mathcal{C}}{\partial t^2}-\nabla^2\phi_{\omTilde}^\mathcal{C} = s_{\omTilde}\,,
\end{equation}
\noindent
which can be derived directly from Equation~\eqref{pWave} by integrating once with respect to time and cancelling a factor of $\rho$. Please note that if $\phi_{\omTilde}^\mathcal{C}$ is a solution to Equation \eqref{wavePhi}, then $p_{\omTilde}$ as defined by Equation \eqref{potential} will satisfy our original Equation \eqref{pWave}.

We can immediately write down the solution to Equation~\eqref{wavePhi} by convolving the source term Equation~\eqref{fSource} with the outgoing Green's function of the wave equation in time and space for a volume-injection source, given by 

\begin{equation}\label{GFWave}
G(t,\x)=\frac{\Theta\left(t-\frac{|\x|}{c}\right)}{2\pi\sqrt{t^2-\frac{|\x|^2}{c^2}}}\,,
\end{equation}
\noindent
where $\Theta$ denotes the Heaviside function. Hence, the complex potential $\phi_{\omTilde}^\mathcal{C}(t,\x,\V)$ can be expressed in integral form as

\begin{equation}\label{phiInt1}
\phi_{\omTilde}^\mathcal{C}(t,\x,\V)=\frac{1}{2\pi}\int_{-\infty}^\infty\int_{\mathbb{R}^2}e^{-i\omTilde t'} \delta(\x'-\mathbf{v}t')\frac{\Theta\left(t-t'-\frac{|\x-\x'|}{c}\right)}{\sqrt{(t-t')^2-\frac{|\x-\x'|^2}{c^2}}}\,d\x'\,dt'\,.
\end{equation}

Performing the integral in space is straightforward owing to the sifting property of the delta distribution, yielding

\begin{equation}\label{phiInt2}
\phi_{\omTilde}^\mathcal{C}(t,\x,\V)=\frac{1}{2\pi}\int_{-\infty}^\infty e^{-i\omTilde t'} \frac{\Theta\left(t-t'-\frac{|\x-\mathbf{v}t'|}{c}\right)}{\sqrt{(t-t')^2-\frac{|\x-\mathbf{v}t'|^2}{c^2}}}\,dt'\,.
\end{equation}

To perform the time integral in Equation \eqref{phiInt2}, define the auxiliary functions

\begin{equation}\label{q1}
q_1(t,\x,\V)=\frac{c^2t-\x\cdot\mathbf{v}}{c^2-|\V |^2}\,,
\end{equation}
\noindent
and

\begin{equation}\label{q2}
q_2(t,\x,\V)=q_1(t,\x,\V)^2-\frac{c^2t^2-|\x|^2}{c^2-|\V |^2}\,.
\end{equation}

Then, by the definition of the Heaviside function, we have

\begin{equation}\label{heaviside}
\Theta\left(t-t'-\frac{|\x-\mathbf{v}t'|}{c}\right)=
\begin{cases}
1&\quad\text{ if }t\leq \mathring{t}\,,\\
0&\quad\text{ if }t>\mathring{t}\,,
\end{cases}
\end{equation}
\noindent
where 

\begin{equation}\label{tEmission}
\mathring{t}=q_1(t,\x,\V)-\sqrt{q_2(t,\x,\V)}\,.
\end{equation}

It is useful highlight the physical interpretation of time $\mathring{t}$. This value is such that the argument of the Heaviside function in the left-hand side becomes zero or, equivalently,

\[c(t-\mathring{t}) = |\x-\mathbf{v}\mathring{t}|\,.\]

Please note that $t-\mathring{t}$ is the time it takes for energy to travel from the location of the source at time $\mathring{t}$ (given by $\mathbf{v}\mathring{t}$) to the receiver location $\x$. In other words, $\mathring{t}$ represents the time at which the wave arriving at time $t$ and location $\x$ was emitted. Equation \eqref{tEmission}, scaled by the source speed $\V$, is useful to calculate the location of the spatial window induced by a time window in Section \ref{newtech}.

Next, performing the time integral in Equation \eqref{phiInt2} yields the following solution for the potential:

\begin{equation}\label{phiMono}
\phi_{\omTilde}^\mathcal{C}(t,\x,\V)=\frac{ie^{-i\omTilde\,q_1(t,\x,\V)}}{4c\sqrt{c^2-|\V |^2}}\mathcal{H}_0^{(1)}\left(\omTilde\sqrt{q_2(t,\x,\V)}\right)\,.
\end{equation}
\noindent
where $\mathcal{H}_0^{(1)}$ denotes the first-kind Hankel function of order zero and $q_1$ and $q_2$ are as defined by Equations \eqref{q1} and \eqref{q2}
To simplify our expressions as possible, we define the non-dimensional parameter $M = \frac{|\V|}{c}$, which quantifies the ratio of the source speed with respect to the speed of sound in the medium. This parameter simplifies our analysis when considering the Doppler effect below. Furthermore, we place the origin of our reference system at the instantaneous location where the source is collinear with the perpendicular receiver pair. Hence we assume the receiver to be of the form $\x = (\,0\,,\,x_y\,)$, and that the source speed can be written as $\V = (\,v\,,\,0\,)$. Finally, we define for simplicity 

\begin{equation}\label{tx_eq}
t_\x = \frac{\norm{\x}}{c}\,,
\end{equation} 
\noindent
that is, the travel time from the reference origin described above to the receiver location $\x$.
Under these assumptions we can write the wavefield generated by a monochromatic moving point source as

\end{paracol}
\nointerlineskip
\appendix
\begin{equation}\label{pMonoReal}
\begin{split}
p_{\omTilde}(t,\x,M)&=\mathfrak{R}\left\{-\frac{i\rho\,\omTilde\,}{4c^2(1-M^2)^\frac32}\exp\left\{-i\,\frac{\omTilde}{1-M^2}\,t\right\}\phantom{\frac{M^2\,t}{\sqrt{M^2t^2+(1-M^2)\,t_\x^2}}\mathcal{H}_1^{(1)}\left(\frac{\omTilde}{1-M^2}\,\right)}\right.\\
&\\
&\phantom{\{\frac{-i\rho\,\,}{4c^2}}\times\left.\left[\frac{M^2\,t}{\sqrt{M^2t^2+(1-M^2)\,t_\x^2}}\mathcal{H}_1^{(1)}\left(\frac{\omTilde}{1-M^2}\,\sqrt{M^2t^2+(1-M^2)\,t_\x^2}\right)\right.\right.\\
&\\
&\left.\left.\phantom{\frac{M^2\,t}{\sqrt{M^2t^2+(1-M^2)\,t_\x^2}}\frac{M^2\,t}{\sqrt{M^2t^2}}}+\,iH_0^{(1)}\left(\frac{\omTilde}{1-M^2}\,\sqrt{M^2t^2+(1-M^2)\,t_\x^2}\right)\,\right]\right\}\,.
\end{split}
\end{equation}
\begin{paracol}{2}
\switchcolumn

Expressing the wavefield in this way allows us to see how its behavior depends on the source speed $|\V|$  and the distance to the road $\norm{\x}$, albeit always scaled by the medium speed $c$. It also allows us to see how the distance from the receiver to the road affects its behavior. See Figure \ref{MonoExampleFig} for a direct implementation of this equation in the time domain, as well as its numerically calculated frequency spectrum.

\begin{figure}[H]
\includegraphics[width =0.7\textwidth]{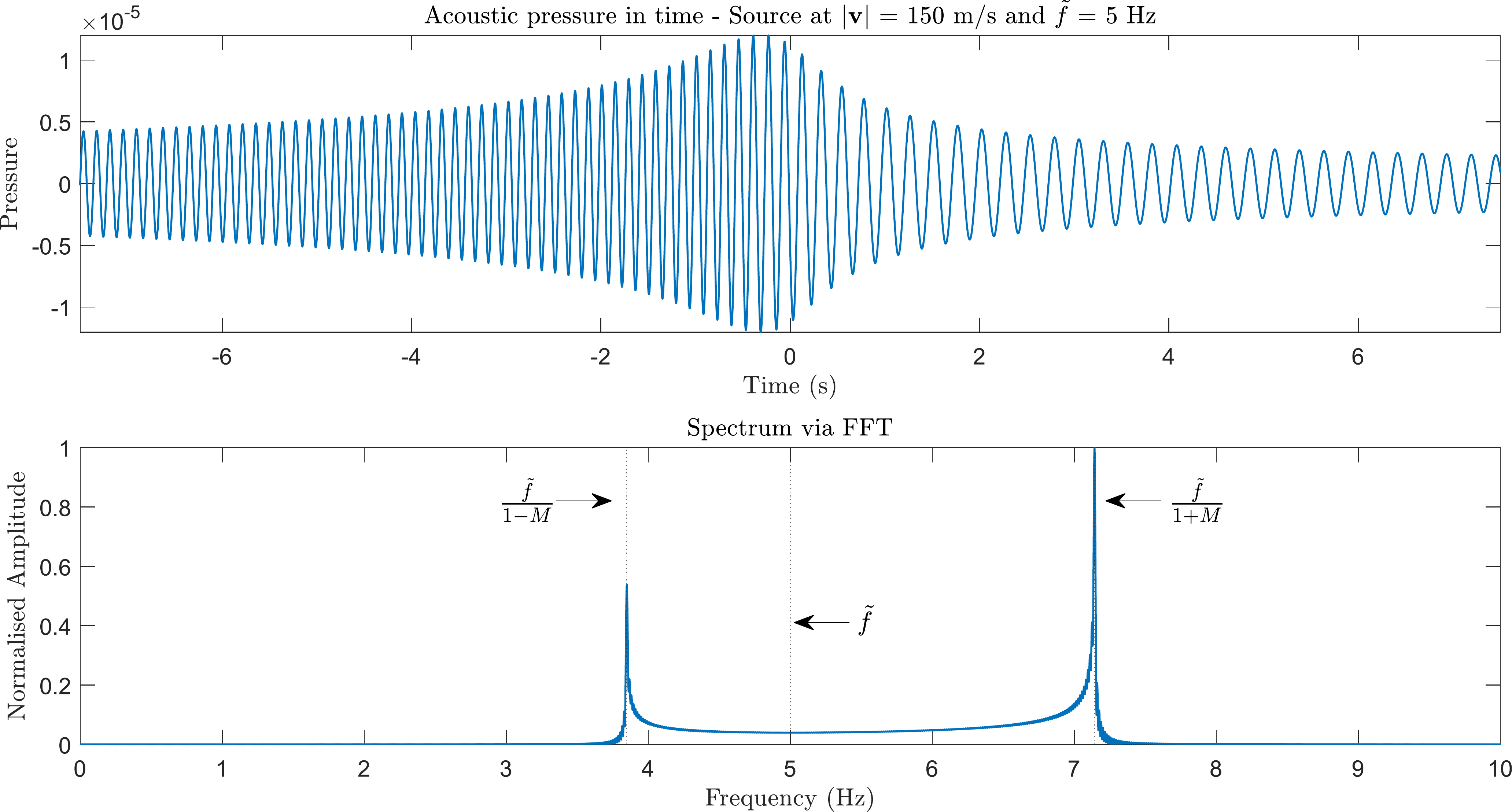}
\caption{{{Example of simulated monochromatic wavefield} 
 $p_\omTilde$.} Medium speed is $c = 500~\text{m/s}$ and medium density is set to $\rho= 1$. The receiver location is $\x = (0~\text{m}, 100~\text{m})$. The monochromatic source travels with speed $|\V| = 150\, \text{m/s}$ along the positive direction of the $x$-axis, emitting sound at $\fTilde = 5\, \text{Hz}$. It crosses the origin at time $t=0$ according to our parametrization. For illustration purposes, source parameters were chosen so that the Doppler effect could be plainly appreciated in the waveform (low emitting frequency $\fTilde$, relatively high value of $M$ at 0.3). The top panel shows the waveform in the time domain directly implemented from Equation \eqref{pMonoReal}. We can see how the perceived frequency increases or decreases when the source is moving towards or away from the receiver, respectively. The bottom panel shows the spectrum of this source, calculated numerically via \added{the fast Fourier transform} (FFT) in Matlab. The spectrum spreads between the frequency values predicted by the standard Doppler formula, $\frac{\fTilde}{1\pm M}$, highlighted in gray {dotted lines.} 
}\label{MonoExampleFig}
\end{figure}
To derive an exact frequency-domain expression for the interferometric retrieval we derive an analytical expression for the spectrum of Equation \eqref{pMonoReal}. Consider the following Fourier transform of the real part of the potential \eqref{phiMono}  

\begin{equation}\label{PotentialFourier}
\hat{\phi}_\omTilde(\omega,\x,\V )=\int_{-\infty}^\infty e^{-i\omega t}\mathcal{R}\left\{\,\phi_\omTilde^\mathcal{C}(t,\x,\V )\,\right\}\,dt\,.
\end{equation}

To perform this integration, we derive an integral representation for the spectrum of the potential by taking the real part of the integral representation \eqref{phiInt2}, where only the spatial integration has been performed, yielding

\begin{equation}\label{RealPotentialIntegral}
\mathcal{R}\left\{\,\phi_\omTilde^\mathcal{C}(t,\x,\V )\,\right\}=\int_{-\infty}^\infty\cos(\omTilde t')\frac{\Theta\left(t-t'-\frac{|\x-\mathbf{v}t'|}{c}\right)}{\sqrt{(t-t')^2-\frac{|\x-\mathbf{v}t'|^2}{c^2}}}\,dt'
\end{equation}

Substituting Equation \eqref{RealPotentialIntegral} into Equation \eqref{PotentialFourier} results after a few manipulations in the following expression for the spectrum of the wavefield generated by a moving point~source

\begin{equation}\label{phiMonoFReal}
\hat{\phi}_\omTilde(\omega,\x,\V ) =-\frac{i}{4} \int_{-\infty}^\infty  e^{-i\omega t'}\cos(\omTilde t')\mathcal{H}_0^{(2)}\left(\frac{\omega}{c}\,|\x-\V t'|\right)\,dt'\,.
\end{equation}
\noindent
where ${H}_0^{(2)}$ denotes the second-kind Hankel function of order zero.
Finally, to derive the spectrum of the wavefield $p_\omTilde$ Equation \eqref{potential} is Fourier transformed so that Equation~\eqref{phiMonoFReal} is simply multiplied by the corresponding frequency factor $i\omega$ that corresponds to a differentiation in the time domain, that is

\begin{equation}\label{pMonoF}
\hat{p}_{\tilde{\omega}}(\omega,\x,\V ) = \frac{\omega\rho}{4}\int_{-\infty}^\infty  e^{-i\omega t'}\cos(\omTilde t')\,\mathcal{H}_0^{(2)}\left(\,\frac{\omega}{c}\,|\x-\mathbf{v}t'|\right)\,dt'\,.
\end{equation}

Numerical implementation of the integral \eqref{pMonoF} agrees with the spectrum of the time-domain wavefield \eqref{pMonoReal} calculated via FFT. Moreover, the spectrum $\eqref{pMonoF}$ can be written in the standard form of a recording given by Equation \eqref{p}, where the source boundary $S$ is parameterized by $t'$ (via $\x'(t') = t'\V$)
and the source kernel is given explicitly by

\[\hat{F}(\x'(t'),\omega) = e^{-i\omega t'}\cos(\omTilde t') \]
\noindent
and the Green's function is indeed the Green's function of the Helmholtz operator for a volume-injection rate source, given by

\[ \hat{G}(\x,\x'(t'),\omega) = \frac{\omega\rho}{4} \mathcal{H}_0^{(2)}\left(\,\frac{\omega}{c}\,|\x-\x'(t')|\right)\,. \]

\subsection{Doppler Effect}
With frequency-domain analytical expressions available, the instantaneous frequency of the model can be evaluated. The complex model will be used. An analytical expression is derived for the instantaneous phase of the recorded source and then a derivative is taken to find the instantaneous frequency.

Consider the asymptotic approximation of the Hankel function. 

\begin{equation}\label{HankelAsymp}
\mathcal{H}_0^{2}(x)\sim e^{-i(x-\frac{\pi}{4})}\sqrt{\frac{2}{\pi x}}\,.
\end{equation}

The frequency regime and the source-receiver distance are already required to be sufficiently large for the monopole interferometry equation to be applied, which makes approximation~\eqref{HankelAsymp} reasonable. Using this approximation, the spectrum of the theoretical complex wavefield $\hat{p}_{\omTilde}^\mathcal{C}(\omega,\x,\V )$ (which is derived in a manner analogous to Equation \eqref{pMonoF}, taking the full complex potential) can be expressed as

\begin{equation}\label{pMonoFAsymp}
\hat{p}_{\omTilde}^\mathcal{C}(\omega,\x,\V ) \sim \frac{\omega\rho}{4}\int_{-\infty}^\infty e^{-i\left\{(\omega+\tilde{\omega})t'+\frac{\omega}{c}\,|\x-\mathbf{v}t'|-\frac{\pi}{4}\right\}}\sqrt{\frac{2c}{\pi\omega|\x-\mathbf{v}t'|}}\,dt'\,.
\end{equation}

The instantaneous phase recorded at a receiver location $\x$ will be denoted by $\text{ph}_{\x}(t')$ and given by 

\begin{equation}\label{phase}
\text{ph}_{\x}(t') = (\omega+\tilde{\omega})t'+\frac{\omega}{c}\,|\x-\mathbf{v}t'|-\frac{\pi}{4}\,,
\end{equation}
\noindent
and differentiating once in time yields 

\begin{equation}\label{phasedt}
 \frac{d}{dt'}\text{ph}_{\x}(t') = \omega + \tilde{\omega} - \frac{\omega}{c}\left(\frac{\x-\mathbf{v}t'}{|\x-\mathbf{v}t'|}\right)\cdot\mathbf{v}\,.
\end{equation}

The integral \eqref{pMonoFAsymp} is highly oscillatory and in accordance with the stationary phase approximation the main contributions correspond stationary points of the phase \eqref{phase}. The spectrum will be otherwise negligibly small by the Riemann-Lebesgue lemma. Hence the spectrum will be largely negligible except at the zeros of Equation~\eqref{phasedt}. Solving the resulting equation and using the fact that $\x-\mathbf{v}t'$ is the position of the source with respect to the receiver at time $t'$ results in the following expression for the instantaneous frequency recorded at location $\x$ and time $t'$

\begin{equation}\label{dopplerApp}
\breve{\omega}(\x,t') = \frac{\tilde{\omega}}{1-\frac{|\V |}{c}\cos\theta_{\x}(t')}\,.
\end{equation}

The instantaneous recorded frequency \eqref{dopplerApp} exhibits the Doppler shift induced by the movement of the source with respect to the receiver. Please note that for any time $t'$ the relative position angle $\cos\theta_{\x}(t')$ is bounded between $-\pi$ (when the source is infinitely far to the left of the receiver) and $\pi$ (when the source is infinitely far to the right), corresponding to the frequencies $\omTilde/1\pm M$ where $M=\norm{\V}/c$. Equation \eqref{dopplerApp} agrees with the well-known expressions for the recorded frequency when the receiver is directly in the path of the source, and is more general as the instantaneous frequency can be calculated for an arbitrary recording location $\x$. Moreover, it is straightforward to show that Equation~\eqref{dopplerApp} is larger than $\omTilde$ when the source is moving towards the receiver and smaller when the source is moving away, with the true emitted frequency $\omTilde$ recorded the moment the source-receiver ray-path perpendicular to the source boundary.
\subsection{Broadband Moving Source Model}
We construct a broadband source emitting noise on a discrete frequency domain $\omTilde\in\OmTilde$ by applying the principle of superposition to the monochromatic wavefields \eqref{pMonoF}, and imposing a random phase shift for each emitted frequency, i.e., we define the broadband wavefield

\begin{equation}\label{broadband_eq}
\hat{p}(\x,\omega) = \sum\limits_{\omTilde\in\OmTilde} c_\omTilde\hat{p}_\omTilde(\x,\omega)\,,
\end{equation} 
\noindent
where $c_\omTilde$ is an appropriate scaling coefficient that depends on how the spectrum $\OmTilde$ is constructed, and the dependence on the source speed $\V$ is left implicit. In this paper, it is assumed that the source spectrum can be represented in the form \eqref{broadband_eq}. A more general representation can be achieved by taking the Fourier transform of a wavefield of the form 

\begin{equation}
     p(\x,t) = \sum\limits_{\omTilde\in\OmTilde} c_\omTilde(t)p_\omTilde(\x,t)\,,
\end{equation}
\noindent
where the coefficients $c_\omTilde(t)$ are time dependent and random with their own correlation structure. Although we do not deal with this case in detail here, we point out such a setup would be necessary to derive a kernel that becomes localized for the case of a monochromatic source, as remarked under Equation \eqref{kernelMean}. 

Assuming that the source spectrum can be represented in Equation~\eqref{broadband_eq}, we have

\begin{equation}\label{pBroad}
\hat{p}(\x,\omega) = \frac{\omega\rho}{4}\sum\limits_{\omTilde\in\OmTilde}c_\omTilde \int_{-\infty}^\infty  e^{-i\omega t'} \cos(\omTilde t'+\theta_\omTilde)\,\mathcal{H}_0^{(2)}\left(\omega\,\frac{|\,\x-\mathbf{v}t'\,|}{c}\right)\,dt'\,,
\end{equation}
\noindent
where $\theta_\omTilde$ are a collection of random variables indexed by $\omTilde$.
The random quantity

\begin{equation}\label{generalSpectrum}
F(t';\theta_\OmTilde) = \sum\limits_{\omTilde\in\OmTilde} c_\omTilde  \cos(\omTilde t'+\theta_\omTilde)
\end{equation}
\noindent
characterizes the spectrum emitted by the source, as well as how the component frequencies interact with each other depending on the distribution of $\theta_\omTilde$ and their correlation structure. Equation \eqref{pBroad} can then be written more compactly as 

\begin{equation}\label{pBroadS}
\hat{p}(\x,\omega) = \frac{\omega\rho}{4}\int_{-\infty}^\infty F(t';\theta_\OmTilde)\,e^{-i\omega t'}\,\mathcal{H}_0^{(2)}\left(\omega\,\frac{|\,\x-\mathbf{v}t'\,|}{c}\right)\,dt'\,,
\end{equation}
\noindent
where it becomes clear that $\hat{p}$ is itself a random quantity.

\section{Analytical Kernel for a Band-Limited Source Train with Random-Phase Emitted Frequency Components}\label{UncorrelatedPhases}
The stacking average interferometric retrieval from the recordings of Equation~\eqref{pBroadS} at a receiver pair $\x_\A$ and $\x_\B$ can be written for a source travelling at speed $\V$ as 

\clearpage
\end{paracol}
\nointerlineskip
\begin{equation}\label{hApp}
\begin{split}
\e{\hat{h}(\x_\A,\x_\B,\V,\omega)}&=\frac{2}{\rho c}\e{\hat{p}^*(\x_\A,\omega)\hat{p}(\x_\B,\omega)}\\
&=\frac{\omega^2\rho}{8c}\int_{-\infty}^\infty\int_{-\infty}^\infty  \e{F^*(t';\theta_\OmTilde)F(t'';\theta_\OmTilde)}\,e^{-i\omega(t''- t')}\\
&\phantom{\frac{\omega^2\rho}{8c}\int_{-\infty}^\infty\int_{-\infty}^\infty }\times\,\mathcal{H}_0^{(1)}\left(\omega\,\frac{|\,\mathbf{A}-\mathbf{v}t'\,|}{c}\right)\,\mathcal{H}_0^{(2)}\left(\omega\,\frac{|\,\mathbf{B}-\mathbf{v}t''\,|}{c}\right)\,dt'\,dt''
\end{split}
\end{equation}
\begin{paracol}{2}
\switchcolumn

For an arbitrary discrete source spectrum $\OmTilde$ and associated source signature in the form of \eqref{generalSpectrum}, the generalized average source kernel in an ambient noise interferometric retrieval is given by

\end{paracol}
\nointerlineskip
\begin{align}
\e{F^*(t';\theta_\OmTilde)F^*(t'';\theta_\OmTilde)} =& \,\sum\limits_{\omTilde\in\OmTilde} c_\omTilde^2\Big(\cos(\omTilde t')\cos(\omTilde t'')\,\e{\cos^2(\theta_{\omTilde})}\notag\\[-.2cm]
&\hspace{1.5cm}-\sin(\omTilde (t'+t''))\,\e{\cos(\theta_{\omTilde})\sin(\theta_{\omTilde})}\notag\\[.1cm]
&\hspace{2.5cm}+\sin(\omTilde t')\sin(\omTilde t'')\,\e{\sin^2(\theta_{\omTilde})}\Big)\notag\\[.2cm]
&+\sum\limits_{\omTilde\in\OmTilde}\,\, \sum\limits_{\substack{\omTilde'\in \OmTilde\\ \omTilde\neq\omTilde'}} c_\omTilde c_{\omTilde'}\Big(\cos(\omTilde t')\cos(\omTilde' t'')\,\e{\cos(\theta_{\omTilde})\cos(\theta_{\omTilde'})}\notag\\[-.5cm]
&\hspace{2.8cm}  -\cos(\omTilde t')\sin(\omTilde' t'')\,\e{\cos(\theta_{\omTilde})\sin(\theta_{\omTilde'})}\notag\\[.1cm]
&\hspace{3.2cm}-\sin(\omTilde t')\cos(\omTilde' t'')\,\e{\sin(\theta_{\omTilde})\cos(\theta_{\omTilde'})}\notag\\[.1cm]
&\hspace{3.6cm}+\sin(\omTilde t')\sin(\omTilde' t'')\,\e{\sin(\theta_{\omTilde})\sin(\theta_{\omTilde'})}\Big),\label{ESS}
\end{align}
\begin{paracol}{2}
\switchcolumn

\noindent
\textls[-35]{where the double summation is split into diagonal terms and cross-terms $\omTilde\neq\omTilde'$. \mbox{Equation~\eqref{ESS}}} shows that the standard ensemble average operator depends on the phases of the emitted frequency components of the source, as well as their correlation.

For example, consider the case when the source spectrum is a regular partition of a bounded interval, that is $\OmTilde=\{\omTilde_j\}_{j=1}^{\NOm}$ for some finite spacing $\triangle\omTilde = \omTilde_{j+1}-\omTilde_j$, and associated random phase shifts $\theta_j$ for each frequency $\omTilde_j$. Moreover, assume that the random shifts are such that $\theta_j\sim\text{Unif}(0,2\pi)$ for all $j\in\{1,\,\ldots\,,\,\NOm\}$. Furthermore, assume that the phase shifts $\theta_j$ and $\theta_k$ are independent whenever $\omTilde_j\neq\omTilde_k$. These assumptions result in the following probability distributions for the random variables in~Equation~\eqref{ESS}

\[\sin(\theta)\,,\cos(\theta)\sim\text{ArcSin}(-1,1)\,,\] \[\sin(\theta)\cos(\theta)\sim\text{ArcSin}\left(-\textstyle\frac{1}{2},\frac{1}{2}\right)\,,\] 
\noindent
and

\[\sin(\theta)^2\,,\cos^2(\theta)\sim\text{ArcSin}(0,1)\,,\] 
\noindent
where $\text{ArcSin}$ denotes the arcsine probability distribution with known means. Moreover, assume that the coefficients $c_\omTilde$ in Equation~\eqref{generalSpectrum} do not depend on the individual frequencies $\omTilde$, and let this coefficient be denoted henceforth simply as $c_\OmTilde$. Then the random source kernel~\eqref{ESS} reduces to the deterministic expression

\begin{equation}\label{ESSUnif}
\begin{split}
\E{F^*(t';\theta_\OmTilde)F(t'';\theta_\OmTilde)} = c_\OmTilde^2\frac{1}{\NOm}\sum\limits_{j=0}^\NOm \cos(\,\omTilde_j (t''-t')\,).
\end{split}
\end{equation}

Next, one can deduce in a standard fashion~\cite{rudin1976principles} that

\begin{equation}
\E{F^*(t';\theta_\OmTilde)F(t'';\theta_\OmTilde)}    \underset{\underset{\NOm\triangle_\omTilde = const.}{\triangle_\omTilde\rightarrow0}}{\longrightarrow}\frac{c_\OmTilde^2}{\omTilde_{\max}-\omTilde_{\min}}\int_{\omTilde_{\min}}^{\omTilde_{\max}}\cos(\omTilde(t''-t'))d\omTilde\,,
\end{equation}
\noindent
where $\triangle_\omTilde=\omTilde_{j+1}-\omTilde_j$ and $\omTilde_{\max}-\omTilde_{\min}=\NOm\triangle_\omTilde$, which leads to

\begin{equation}\label{anSS}
\E{F^*(t';\theta_\OmTilde)F(t'';\theta_\OmTilde)} \approx\, c_\OmTilde^2\,\frac{\sin(\,\omTilde_{\max}(t''-t)\,)-\sin(\,\omTilde_{\min}(t''-t)\,)}{(\omTilde_{\max}-\omTilde_{\min})(t''-t')}\,.
\end{equation}

Finally, the right-hand side of Equation \eqref{anSS} can be expanded using well-known trigonometric identities to derive
\vspace{+6pt}
\end{paracol}
\nointerlineskip
\begin{equation}\label{EKernel}
\E{F^*(t';\theta_\OmTilde)F(t'';\theta_\OmTilde)} \approx c_\omTilde^2\cos\left(\frac{(\omTilde_{\max} +\omTilde_{\min})}{2}\,(t''-t')\right)\sinc\left(\frac{(\omTilde_{\max} -\omTilde_{\min})}{2}\,(t''-t')\right)
\end{equation}

\begin{paracol}{2}
\switchcolumn

The kernel Equation \eqref{EKernel} is parameterized in time. Under the assumption that the source is in motion, i.e., $\V\neq0$, Equation \eqref{EKernel} leads to expression \eqref{kernelMean}. To see this, let $\x'$ be the source location at time $t'$, and $\x''$ the source location at time $t''$, so that $\x' = (\,\norm{\V} t',\, 0\,)$ and $\x'' = (\,\norm{\V} t'',\, 0\,)$. Hence

\begin{equation}\label{timeSpaceLink_eq}
    t''- t'= \frac{\norm{\V}t''-\norm{\V}t'}{\norm{\V}} = \frac{x''-x'}{\norm{\V}}\,.
\end{equation}
\noindent
\added{where $x'=\norm{\V}t'$ and similarly $x''=\norm{\V}t''$.} Combining Equations \eqref{EKernel} and \eqref{timeSpaceLink_eq} yields an expression proportional to Equation \eqref{kernelMean}, \added{noting that the right-hand side of Equation \eqref{EKernel} is an even function (i.e., symmetric with respect to the sign of the argument)}.
\end{paracol}

\reftitle{References}





\begin{thebibliography}{999}

\bibitem[Campillo and Paul(2003)]{campillo2003long}
Campillo, M.; Paul, A.
\newblock Long-range correlations in the diffuse seismic coda.
\newblock {\em Science} {\bf 2003}, {\em 299},~547--549.

\bibitem[Wapenaar and Fokkema(2006)]{wapenaar2006green}
Wapenaar, K.; Fokkema, J.
\newblock Green’s function representations for seismic interferometry.
\newblock {\em Geophysics} {\bf 2006}, {\em 71},~SI33--SI46.

\bibitem[Curtis \em{et~al.}(2006)Curtis, Gerstoft, Sato, Snieder, and
  Wapenaar]{curtis2006seismic}
Curtis, A.; Gerstoft, P.; Sato, H.; Snieder, R.; Wapenaar, K.
\newblock Seismic interferometry---Turning noise into signal.
\newblock {\em  Lead. Edge} {\bf 2006}, {\em 25},~1082--1092.

\bibitem[Claerbout(1968)]{claerbout1968synthesis}
Claerbout, J.F.
\newblock Synthesis of a layered medium from its acoustic transmission
  response.
\newblock {\em Geophysics} {\bf 1968}, {\em 33},~264--269.

\bibitem[Rickett and Claerbout(1999)]{rickett1999acoustic}
Rickett, J.; Claerbout, J.
\newblock Acoustic daylight imaging via spectral factorization: Helioseismology
  and reservoir monitoring.
\newblock {\em  Lead. Edge} {\bf 1999}, {\em 18},~957--960.

\bibitem[Weaver and Lobkis(2001)]{weaver2001ultrasonics}
Weaver, R.L.; Lobkis, O.I.
\newblock Ultrasonics without a source: Thermal fluctuation correlations at MHz
  frequencies.
\newblock {\em Phys. Rev. Lett.} {\bf 2001}, {\em 87},~134301.

\bibitem[Derode \em{et~al.}(2003)Derode, Larose, Campillo, and
  Fink]{derode2003estimate}
Derode, A.; Larose, E.; Campillo, M.; Fink, M.
\newblock How to estimate the Green’s function of a heterogeneous medium
  between two passive sensors? Application to acoustic waves.
\newblock {\em Appl. Phys. Lett.} {\bf 2003}, {\em 83},~3054--3056.

\bibitem[Slob and Wapenaar(2007)]{slob2007electromagnetic}
Slob, E.; Wapenaar, K.
\newblock Electromagnetic Green's functions retrieval by cross-correlation and
  cross-convolution in media with losses.
\newblock {\em Geophys. Res. Lett.} {\bf 2007}, {\em 34}, {1--5.} 


\bibitem[Ruigrok \em{et~al.}(2008)Ruigrok, Draganov, and
  Wapenaar]{ruigrok2008global}
Ruigrok, E.; Draganov, D.; Wapenaar, K.
\newblock Global-scale seismic interferometry: Theory and numerical examples.
\newblock {\em Geophys. Prospect.} {\bf 2008}, {\em 56},~395--417.

\bibitem[Nishida \em{et~al.}(2009)Nishida, Montagner, and
  Kawakatsu]{nishida2009global}
Nishida, K.; Montagner, J.P.; Kawakatsu, H.
\newblock Global surface wave tomography using seismic hum.
\newblock {\em Science} {\bf 2009}, {\em 326},~112.

\bibitem[Shapiro \em{et~al.}(2005)Shapiro, Campillo, Stehly, and
  Ritzwoller]{shapiro2005high}
Shapiro, N.M.; Campillo, M.; Stehly, L.; Ritzwoller, M.H.
\newblock High-resolution surface-wave tomography from ambient seismic noise.
\newblock {\em Science} {\bf 2005}, {\em 307},~1615--1618.

\bibitem[Nishida \em{et~al.}(2008)Nishida, Kawakatsu, and
  Obara]{nishida2008three}
Nishida, K.; Kawakatsu, H.; Obara, K.
\newblock Three-dimensional crustal S wave velocity structure in Japan using
  microseismic data recorded by Hi-net tiltmeters.
\newblock {\em J. Geophys. Res. Solid Earth} {\bf 2008}, {\em
  113}, {1--22}. 

\bibitem[Arroucau \em{et~al.}(2010)Arroucau, Rawlinson, and
  Sambridge]{arroucau2010new}
Arroucau, P.; Rawlinson, N.; Sambridge, M.
\newblock New insight into Cainozoic sedimentary basins and Palaeozoic suture
  zones in southeast Australia from ambient noise surface wave tomography.
\newblock {\em Geophys. Res. Lett.} {\bf 2010}, {\em 37}, {1--6}. 

\bibitem[Bakulin and Calvert(2006)]{bakulin2006virtual}
Bakulin, A.; Calvert, R.
\newblock The virtual source method: Theory and case study.
\newblock {\em Geophysics} {\bf 2006}, {\em 71},~SI139--SI150.

\bibitem[Bakulin \em{et~al.}(2007)Bakulin, Mateeva, Mehta, Jorgensen,
  Ferrandis, Herhold, and Lopez]{bakulin2007virtual}
Bakulin, A.; Mateeva, A.; Mehta, K.; Jorgensen, P.; Ferrandis, J.; Herhold,
  I.S.; Lopez, J.
\newblock Virtual source applications to imaging and reservoir monitoring.
\newblock {\em  Lead. Edge} {\bf 2007}, {\em 26},~732--740.

\bibitem[Halliday and Curtis(2010)]{halliday2010interferometric}
Halliday, D.; Curtis, A.
\newblock An interferometric theory of source-receiver scattering and imaging.
\newblock {\em Geophysics} {\bf 2010}, {\em 75},~SA95--SA103.

\bibitem[Hong and Menke(2006)]{hong2006tomographic}
Hong, T.K.; Menke, W.
\newblock Tomographic investigation of the wear along the San Jacinto fault,
  southern California.
\newblock {\em Phys. Earth Planet. Inter.} {\bf 2006}, {\em
  155},~236--248.

\bibitem[Curtis \em{et~al.}(2009)Curtis, Nicolson, Halliday, Trampert, and
  Baptie]{curtis2009virtual}
Curtis, A.; Nicolson, H.; Halliday, D.; Trampert, J.; Baptie, B.
\newblock Virtual seismometers in the subsurface of the Earth from seismic
  interferometry.
\newblock {\em Nat. Geosci.} {\bf 2009}, {\em 2},~700--704.

\bibitem[Wapenaar(2004)]{wapenaar2004retrieving}
Wapenaar, K.
\newblock Retrieving the elastodynamic Green's function of an arbitrary
  inhomogeneous medium by cross correlation.
\newblock {\em Phys. Rev. Lett.} {\bf 2004}, {\em 93},~254301.

\bibitem[Curtis and Halliday(2010)]{curtis2010source}
Curtis, A.; Halliday, D.
\newblock Source-receiver wave field interferometry.
\newblock {\em Phys. Rev. E} {\bf 2010}, {\em 81},~046601.

\bibitem[Curtis \em{et~al.}(2012)Curtis, Behr, Entwistle, Galetti, Townend, and
  Bannister]{curtis2012benefit}
Curtis, A.; Behr, Y.; Entwistle, E.; Galetti, E.; Townend, J.; Bannister, S.
\newblock The benefit of hindsight in observational science: Retrospective
  seismological observations.
\newblock {\em Earth Planet. Sci. Lett.} {\bf 2012}, {\em
  345},~212--220.

\bibitem[Entwistle \em{et~al.}(2015)Entwistle, Curtis, Galetti, Baptie, and
  Meles]{entwistle2015constructing}
Entwistle, E.; Curtis, A.; Galetti, E.; Baptie, B.; Meles, G.
\newblock Constructing new seismograms from old earthquakes: Retrospective
  seismology at multiple length scales.
\newblock {\em J. Geophys. Res. Solid Earth} {\bf 2015}, {\em
  120},~2466--2490.

\bibitem[Chen and Saygin(2020)]{chen2020empirical}
Chen, Y.; Saygin, E.
\newblock Empirical Green's Function Retrieval Using Ambient Noise
  Source-Receiver Interferometry.
\newblock {\em J. Geophys. Res. Solid Earth} {\bf 2020}, {\em
  125}, {doi:10.1029/2019JB018261.}

\bibitem[Wapenaar \em{et~al.}(2011)Wapenaar, Van Der~Neut, Ruigrok, Draganov,
  Hunziker, Slob, Thorbecke, and Snieder]{wapenaar2011seismic}
Wapenaar, K.; Van Der~Neut, J.; Ruigrok, E.; Draganov, D.; Hunziker, J.; Slob,
  E.; Thorbecke, J.; Snieder, R.
\newblock Seismic interferometry by crosscorrelation and by multidimensional
  deconvolution: A systematic comparison.
\newblock {\em Geophys. J. Int.} {\bf 2011}, {\em
  185},~1335--1364.

\bibitem[Nicolson \em{et~al.}(2012)Nicolson, Curtis, Baptie, and
  Galetti]{nicolson2012seismic}
Nicolson, H.; Curtis, A.; Baptie, B.; Galetti, E.
\newblock Seismic interferometry and ambient noise tomography in the British
  Isles.
\newblock {\em Proc. Geol. Assoc.} {\bf 2012}, {\em
  123},~74--86.

\bibitem[Ardhuin \em{et~al.}(2011)Ardhuin, Stutzmann, Schimmel, and
  Mangeney]{ardhuin2011ocean}
Ardhuin, F.; Stutzmann, E.; Schimmel, M.; Mangeney, A.
\newblock Ocean wave sources of seismic noise.
\newblock {\em J. Geophys. Res. Ocean.} {\bf 2011}, {\em 116},~{C006952}.

\bibitem[Sabra \em{et~al.}(2005)Sabra, Roux, and Kuperman]{sabra2005arrival}
Sabra, K.G.; Roux, P.; Kuperman, W.
\newblock Arrival-time structure of the time-averaged ambient noise
  cross-correlation function in an oceanic waveguide.
\newblock {\em  J. Acoust. Soc. Am.} {\bf 2005},
  {\em 117},~164--174.

\bibitem[Halliday \em{et~al.}(2008a)Halliday, Curtis, and
  Kragh]{halliday2008seismic_a}
Halliday, D.; Curtis, A.; Kragh, E.
\newblock Seismic surface waves in a suburban environment: Active and passive
  interferometric methods.
\newblock {\em  Lead. Edge} {\bf 2008}, {\em 27},~210--218.

\bibitem[Nakata \em{et~al.}(2011)Nakata, Snieder, Tsuji, Larner, and
  Matsuoka]{nakata2011shear}
Nakata, N.; Snieder, R.; Tsuji, T.; Larner, K.; Matsuoka, T.
\newblock Shear wave imaging from traffic noise using seismic interferometry by
  cross-coherence.
\newblock {\em Geophysics} {\bf 2011}, {\em 76},~SA97--SA106.

\bibitem[Behm and Snieder(2013)]{behm2013love}
Behm, M.; Snieder, R.
\newblock Love waves from local traffic noise interferometry.
\newblock {\em  Lead. Edge} {\bf 2013}, {\em 32},~628--632.

\bibitem[Dales \em{et~al.}(2020)Dales, Pinzon-Ricon, Brenguier, Bou{\'e},
  Arndt, McBride, Lavou{\'e}, Bean, Beaupretre, Fayjaloun,
  et~al.]{dales2020virtual}
Dales, P.; Pinzon-Ricon, L.; Brenguier, F.; Bou{\'e}, P.; Arndt, N.; McBride,
  J.; Lavou{\'e}, F.; Bean, C.J.; Beaupretre, S.; Fayjaloun, R.; et al.
\newblock Virtual Sources of Body Waves from Noise Correlations in a Mineral
  Exploration Context.
\newblock {\em Seismol. Res. Lett.} {\bf 2020}, {\emph{91}, 2278--2286.} 

\bibitem[Brenguier \em{et~al.}(2019)Brenguier, Bou{\'e}, Ben-Zion, Vernon,
  Johnson, Mordret, Coutant, Share, Beauc{\'e}, Hollis,
  et~al.]{brenguier2019train}
Brenguier, F.; Bou{\'e}, P.; Ben-Zion, Y.; Vernon, F.; Johnson, C.; Mordret,
  A.; Coutant, O.; Share, P.E.; Beauc{\'e}, E.; Hollis, D.; et al.
\newblock Train traffic as a powerful noise source for monitoring active faults
  with seismic interferometry.
\newblock {\em Geophys. Res. Lett.} {\bf 2019}, {\em 46},~9529--9536.

\bibitem[Quiros \em{et~al.}(2016)Quiros, Brown, and Kim]{quiros2016seismic}
Quiros, D.A.; Brown, L.D.; Kim, D.
\newblock Seismic interferometry of railroad induced ground motions: Body and
  surface wave imaging.
\newblock {\em Geophys. Suppl. Mon. Not. R. Astron. Soc.} {\bf 2016}, {\em 205},~301--313.

\bibitem[Pinzon-Rincon \em{et~al.}(2021)Pinzon-Rincon, Lavou{\'e}, Mordret,
  Bou{\'e}, Brenguier, Dales, Ben-Zion, Vernon, Bean, and
  Hollis]{pinzon2021humming}
Pinzon-Rincon, L.; Lavou{\'e}, F.; Mordret, A.; Bou{\'e}, P.; Brenguier, F.;
  Dales, P.; Ben-Zion, Y.; Vernon, F.; Bean, C.J.; Hollis, D.
\newblock Humming Trains in Seismology: An Opportune Source for Probing the
  Shallow Crust.
\newblock {\em Seismol. Soc. Am.} {\bf 2021}, {\emph{92}, 623--635.} 

\bibitem[Liu \em{et~al.}(2021)Liu, Yue, Li, and Luo]{liu2021retrievability}
Liu, Y.; Yue, Y.; Li, Y.; Luo, Y.
\newblock On the Retrievability of Seismic Waves From High-Speed-Train-Induced
  Vibrations Using Seismic Interferometry.
\newblock {\em IEEE Geosci. Remote. Sens. Lett.} {\bf 2021}, {doi:10.1109/LGRS.2021.3050205.} 

\bibitem[Gerstoft \em{et~al.}(2006)Gerstoft, Sabra, Roux, Kuperman, and
  Fehler]{gerstoft2006green}
Gerstoft, P.; Sabra, K.G.; Roux, P.; Kuperman, W.; Fehler, M.C.
\newblock Green's functions extraction and surface-wave tomography from
  microseisms in southern California.
\newblock {\em Geophysics} {\bf 2006}, {\em 71},~SI23--SI31.

\bibitem[Curtis and Halliday(2010)]{curtis2010directional}
Curtis, A.; Halliday, D.
\newblock Directional balancing for seismic and general wavefield
  interferometry.
\newblock {\em Geophysics} {\bf 2010}, {\em 75},~SA1--SA14.

\bibitem[Fichtner \em{et~al.}(2016)Fichtner, Stehly, Ermert, and
  Boehm]{fichtner2016generalised}
Fichtner, A.; Stehly, L.; Ermert, L.; Boehm, C.
\newblock Generalised interferometry-I. Theory for inter-station correlations.
\newblock {\em Geophys. J. Int.} {\bf 2017}, {\emph{208}, 603--638.} 

\bibitem[Van~der Neut(2012)]{van2012interferometric}
Van~der Neut, J.R.
\newblock Interferometric Redatuming by Multidimensional Deconvolution. 2012.  {Available online:} \url{http://homepage.tudelft.nl/t4n4v/9_Theses_students/Neut.pdf}  (accessed {on 17th June 2021} 
). 

\bibitem[Van~Dalen \em{et~al.}(2015)Van~Dalen, Mikesell, Ruigrok, and
  Wapenaar]{van2015retrieving}
Van~Dalen, K.N.; Mikesell, T.D.; Ruigrok, E.N.; Wapenaar, K.
\newblock Retrieving surface waves from ambient seismic noise using seismic
  interferometry by multidimensional deconvolution.
\newblock {\em J. Geophys. Res. Solid Earth} {\bf 2015}, {\em
  120},~944--961.

\bibitem[Sabra(2010)]{sabra2010influence}
Sabra, K.G.
\newblock Influence of the noise sources motion on the estimated Green’s
  functions from ambient noise cross-correlations.
\newblock {\em  J. Acoust. Soc. Am.} {\bf 2010},
  {\em 127},~3577--3589.

\bibitem[Schuster(2009)]{schuster2009seismic}
Schuster, G.
\newblock {\em Seismic Interferometry}; Cambridge University Press: {Cambridge, UK,} 2009. 

\bibitem[Halliday and Curtis(2008b)]{halliday2008seismic_b}
Halliday, D.; Curtis, A.
\newblock Seismic interferometry, surface waves and source distribution.
\newblock {\em Geophys. J. Int.} {\bf 2008}, {\em
  175},~1067--1087.

\bibitem[Mehta \em{et~al.}(2007)Mehta, Bakulin, Sheiman, Calvert, and
  Snieder]{mehta2007improving}
Mehta, K.; Bakulin, A.; Sheiman, J.; Calvert, R.; Snieder, R.
\newblock Improving the virtual source method by wavefield separation.
\newblock \emph{Geophysics} {\bf 2007}, \emph{{72}}, {V79--V86}. 

\bibitem[Snieder(2004)]{snieder2004extracting}
Snieder, R.
\newblock Extracting the {G}reen’s function from the correlation of coda
  waves: A derivation based on stationary phase.
\newblock {\em Phys. Rev. E} {\bf 2004}, {\em 69},~046610.

\bibitem[Thorbecke and Wapenaar(2008)]{thorbecke2008analysis}
Thorbecke, J.; Wapenaar, K.
\newblock Analysis of spurious events in seismic interferometry. In {\em SEG
  Technical Program Expanded Abstracts 2008}; Society of Exploration
  Geophysicists: {Tulsa, OK, USA,} 2008; pp. 1415--1420. 

\bibitem[Van~Manen \em{et~al.}(2006)Van~Manen, Curtis, and
  Robertsson]{van2006interferometric}
Van~Manen, D.; Curtis, A.; Robertsson, J.
\newblock Interferometric modeling of wave propagation in inhomogeneous elastic
  media using time reversal and reciprocity. 
\newblock \emph{Geophysics} {\bf 2006}, \emph{71}, {SI47--SI60}.

\bibitem[Van~Dalen \em{et~al.}(2014)Van~Dalen, Wapenaar, and
  Halliday]{van2014surface}
Van~Dalen, K.N.; Wapenaar, K.; Halliday, D.F.
\newblock Surface wave retrieval in layered media using seismic interferometry
  by multidimensional deconvolution.
\newblock {\em Geophys. J. Int.} {\bf 2014}, {\em
  196},~230--242.

\bibitem[Wapenaar \em{et~al.}(2012)Wapenaar, van~der Neut, and
  Thorbecke]{wapenaar2012relation}
Wapenaar, K.; van~der Neut, J.; Thorbecke, J.
\newblock On the relation between seismic interferometry and the
  simultaneous-source method.
\newblock {\em Geophys. Prospect.} {\bf 2012}, {\em 60},~802--823.

\bibitem[King and Curtis(2011)]{king2011velocity}
King, S.; Curtis, A.
\newblock Velocity analysis using both reflections and refractions in seismic
  interferometry.
\newblock {\em Geophysics} {\bf 2011}, {\em 76},~\mbox{SA83--SA96}.

\bibitem[Maran{\`o} \em{et~al.}(2012)Maran{\`o}, Reller, Loeliger, and
  F{\"a}h]{marano2012seismic}
Maran{\`o}, S.; Reller, C.; Loeliger, H.A.; F{\"a}h, D.
\newblock Seismic waves estimation and wavefield decomposition: Application to
  ambient vibrations.
\newblock {\em Geophys. J. Int.} {\bf 2012}, {\em
  191},~175--188.

\bibitem[Rudin(1976)]{rudin1976principles}
Rudin, W.
\newblock {\em Principles of Mathematical Analysis}; McGraw-Hill: New York, NY, USA, 1976; Volume~3.

\end{thebibliography}
\end{document}